\documentclass{aa}  

\usepackage{graphicx}
\usepackage{subfigure}
\usepackage{txfonts}
\usepackage{color}
\usepackage{hyperref}

\newcommand{\SFR}{\dot{M}_\star}
\newcommand{\njop}{New J.~of.~Plasma~Phys.}

\newcommand{\specialcell}[2][c]{%
  \begin{tabular}[#1]{@{\vspace{-0.18cm}}c@{}}#2 \vspace{0.18cm}\end{tabular}}

\begin{document}

   \title{A model for the infrared-radio correlation of main sequence galaxies at GHz frequencies and its variation with redshift and stellar mass}

   \titlerunning{A model for the infrared-radio correlation}

   \author{J.~Schober \inst{1}  
          \and
          M.~T.~Sargent \inst{2, 3}   
          \and
          R.~S.~Klessen \inst{4,5}   
          \and
          D.~R.~G.~Schleicher \inst{6,7}  
          }

   \institute{Institute of Physics, Laboratory of Astrophysics, École Polytechnique Fédérale de Lausanne (EPFL), 1290 Sauverny, Switzerland,
             \email{jennifer.schober@epfl.ch}
             \and
             International Space Science Institute (ISSI), Hallerstrasse 6, CH-3012 Bern, Switzerland
             \and
             Astronomy Centre, Department of Physics and Astronomy, University of Sussex, Brighton BN1 9QH, UK
             \and
             Universität Heidelberg, Zentrum für Astronomie, Institut für theoretische Astrophysik, Albert-Ueberle-Straße 2, D-69120, Heidelberg, Germany
             \and            
             Universität Heidelberg, Interdisziplinäres Zentrum für Wissenschaftliches Rechnen, Im Neuenheimer Feld 205, D-69120 Heidelberg, Germany
             \and
             Departamento de Astronomía, Facultad Ciencias Físicas y Matemáticas, Universidad de Concepción, Av. Esteban Iturra s/n Barrio Universitario, Concepción, Chile
             \and
             Millennium Nucleus on Transversal Research and Technology to Explore Supermassive Black Holes (TITANS) 
             }

   \date{Received September 15, 1996; accepted March 16, 1997}

   \abstract
{The infrared-radio correlation (IRRC) of star-forming galaxies 
can be used to estimate their star formation rate (SFR) based on the 
radio continuum luminosity at MHz-GHz frequencies.
For its practical application in future deep radio surveys, it is crucial to know whether 
the IRRC persists at high redshift $z$. }
{Previous works have reported 
that 
the $1.4$~GHz IRRC correlation of star-forming
galaxies is nearly $z$-invariant up to $z \approx 4$, but depends strongly 
on the stellar mass $M_\star$. 
This should be taken into account for SFR calibrations based on radio luminosity.}
{To understand the physical cause 
behind the $M_\star$ dependence of the IRRC
and its properties at higher $z$, 
we constructed a phenomenological model for galactic radio emission.
Our model is based on a dynamo-generated magnetic field and a 
steady-state cosmic ray population. It includes a number of 
free parameters that determine the galaxy properties. 
To reduce the overall number of model parameters, we also employed
observed scaling relations.}
{We find that the resulting spread of the infrared-to-radio 
luminosity ratio, $q(z,M_\star)$, with respect to $M_\star$
is mostly determined by the scaling of the galactic radius with $M_\star$, 
while the absolute value of the $q(z,M_\star)$ curves decreases with more efficient 
conversion of supernova energy to magnetic fields and cosmic rays.
Additionally, decreasing the slope of the cosmic ray 
injection spectrum, $\alpha_\mathrm{CR}$, results in higher radio 
luminosity, decreasing the absolute values of the $q(z,M_\star)$ curves.
Within the uncertainty range of our model, the observed 
dependence of the IRRC on $M_\star$ and $z$ can be reproduced when the 
efficiency of supernova-driven turbulence is $5\%$, 
$10\%$ of the kinetic 
energy is converted into magnetic energy, and
$\alpha_\mathrm{CR} \approx 
3.0$.}
{For galaxies with intermediate to high ($M_\star\approx 
10^{9.5}-10^{11}~\mathrm{M}_\odot$) stellar masses, 
our model results in an IRRC that is nearly independent of $z$. 
For galaxies with lower masses ($M_\star\approx 10^{8.5}~\mathrm{M}_\odot$), 
we find that the IR-to-radio flux ratio increases with increasing redshift.
This matches the observational data in that mass bin which, however, only 
extends to $z\approx1.5$. 
The increase in the IR-to-radio flux ratio for
low-mass galaxies at $z\gtrsim1.5$ that is predicted by our model
could be tested with future deep radio observations.}


   \maketitle
%
%

\section{Introduction}

Local star-forming galaxies lie on the ``infrared-radio correlation'' 
(IRRC), which spans at least 
three orders in magnitude of radio continuum 
luminosity. 
It was first observed at a rest frame frequency of $1.4$ 
GHz \citep[e.g.,][]{VanDerKruit1971,VanDerKruit1973, HelouEtAl1985,deJongEtAl1985,HelouEtAl1985,YunEtAl2001}, 
and has increasingly also been studied all the way 
down to the MHz-regime with the advent of LOFAR 
\citep[e.g.,][]{SmithEtAl2021,BonatoEtAl2021,McCheyneEtAl2022}.
At 1.4 GHz, the IRRC is commonly parametrized in terms of the IR-to-radio luminosity ratio, namely:\ 
\begin{eqnarray}
  q = \log_{10} \left( \frac{L_\mathrm{IR}}{3.75\times 10^{12} ~\mathrm{W}}\right)   
      - \log_{10} \left(\frac{L_{1.4}}{\mathrm{W}~ \mathrm{Hz}^{-1} } \right),
\label{eq_q}
\end{eqnarray}
with $L_{1.4}$ being the rest-frame 1.4 GHz radio 
continuum luminosity, and $L_\mathrm{IR}$ the dust 
continuum luminosity integrated over the rest-frame 
wavelength region $8-1000$ $\mu$m in the infrared. 
The parameter $q$ is constant as long as the 
correlation is linear.

The origin of the IRRC lies within the physics 
of cosmic rays, magnetic fields, and interstellar dust,
which are intertwined through star formation activity. 
Young stars heat their surrounding dust that re-emits thermal 
continuum in the IR (measured quantitatively via the 
luminosity $L_\mathrm{IR}$ over the wavelength range 
8-1000\,$\mu$m) and the supernova explosions of short-lived 
massive stars produce CRs that emit non-thermal radio continuum 
(the intrinsic brightness of which is quantified, 
e.g., at 1.4\,GHz, via the luminosity $L_{1.4}$)
in the form of synchrotron radiation \citep{Draine2011,KlessenGlover2016}.
One of the first quantitative models of the IRRC was proposed by 
\citet{Voelk1989}, who assumes that cosmic ray electrons lose all
their energy via synchrotron radiation within the galaxy and that 
all UV photons emitted by young stars are absorbed by dust and then
re-emitted in the infrared \citep{Draine2003}.
However, this calorimeter model lacks to explain certain observed properties of
galaxies, for instance:~it predicts overly steep radio spectra
\citep[see, e.g.,][and references within]{VollmerEtAl2022}.
Various refined models explaining the IRRC have been subsequently 
proposed \citep{MurphyEtAl2006, SchleicherBeck2013, SchoberEtAl2016},
including extensions to dwarf galaxies \citep{SchleicherBeck2016}.
A common key ingredient of models of the
IRRC is star formation
and, therefore, it opens up a pathway 
for calibrating the 
star formation rate (SFR) with radio observations 
\citep{Condon1992,Bell2003,MurphyEtAl2011,SchoberEtAl2017,MolnarEtAl2021},
as long as an active
galactic nuclei (AGN) contribution to the radio flux 
can be excluded. 

The question of whether current calibrations of GHz radio continuum emission 
as an SFR tracer in low-redshift galaxies remain valid out to high redshifts 
is still open \citep{SchoberEtAl2017,SmithEtAl2021}. 
The minimum requirements for the correlation to uphold 
in young galaxies are:
(i) a sufficient cosmic ray (CR) population; 
(ii) strong magnetic fields; and
(iii) the presence of interstellar dust.
While the CR production rate increases with
the increasing SFR and the accompanying increase in supernova shocks
at high $z$, CR electrons suffer 
from increasing energy losses via inverse Compton scattering 
with the cosmic microwave background. 
\citet{SchleicherBeck2013} proposed that this leads to 
a breakdown of the correlation at high $z$. 
Requirement (ii) of a strong magnetization of the 
interstellar medium (ISM) should be satisfied 
by efficient turbulent dynamo amplification 
\citep{SchoberEtAl2013,RiederTeyssier2016,RiederTeyssier2017,Martin-AlvarezEtAl2021,LiuEtAl2022}.
Regarding requirement (iii), a breakdown of the IRRC due to a lack
of interstellar dust can likely be excluded for 
the currently observationally accessible galaxies. 
Observations at {(sub-)millimetre}
wavelengths have provided convincing evidence of rapid dust enrichment 
in high-$z$ galaxies \citep{InamiEtAl2022,SommovigoEtAl2022}. 
Efficient dust production explaining these observations has been suggested by various ISM models 
\citep{ManciniEtAl2015, Michalowski2015,MarassiEtAl2019,GrazianiEtAl2020}.

On observational grounds, the potential evolution of the IRRC 
over cosmic timescales is still much discussed. 
Several studies report a mild or moderate evolution of the correlation
with redshift
\citep{SeymourEtAl2009,IvisonEtAl2010a,MichalowskiEtAl2010a,MichalowskiEtAl2010b,IvisonEtAl2010b,MagnelliEtAl2015,BasuEtAl2015,DelhaizeEtAl2017,CalistroRiveraEtAl2017}.
Other authors, however, have found no significant evidence for evolution
\citep{Garrett2002,AppletonEtAl2004,JarvisEtAl2010,SargentEtAl2010,BourneEtAl2011}
and have instead considered other interpretations, such as observational biases
or an evolution of the dust temperature \citep{SmithEtAl2014}.
In a recent study, \citet{DelvecchioEtAl2020} report that the infrared-radio correlation of star-forming
galaxies is strongly $M_\star$-dependent but nearly $z$-invariant up to $z \approx 4$.

Here, we present a model of the FIR-radio correlation that uses 
$M_\star$ and $z$ as the basis for setting the underlying free 
parameters of our model, using observationally established 
scaling relations and/or constraints from theory.
We aim to find the conditions under which the observations 
by \citet{DelvecchioEtAl2020} can be reproduced
and to make predictions for higher redshift.
The paper is organized as follows: In Sect.~\ref{sec_model_radio}, we
present our semi-analytical model for galactic radio emission. 
The results for different parameters of the model are presented in 
Sect.~\ref{sec_results} and compared to the data of \citet{DelvecchioEtAl2020}.
We draw our conclusions in Sect.~\ref{sec_conclusion}.

\section{Model for galactic emission}
\label{sec_model_radio}

\begin{figure*}[t]
\centering
  \subfigure{\includegraphics[width=\textwidth]{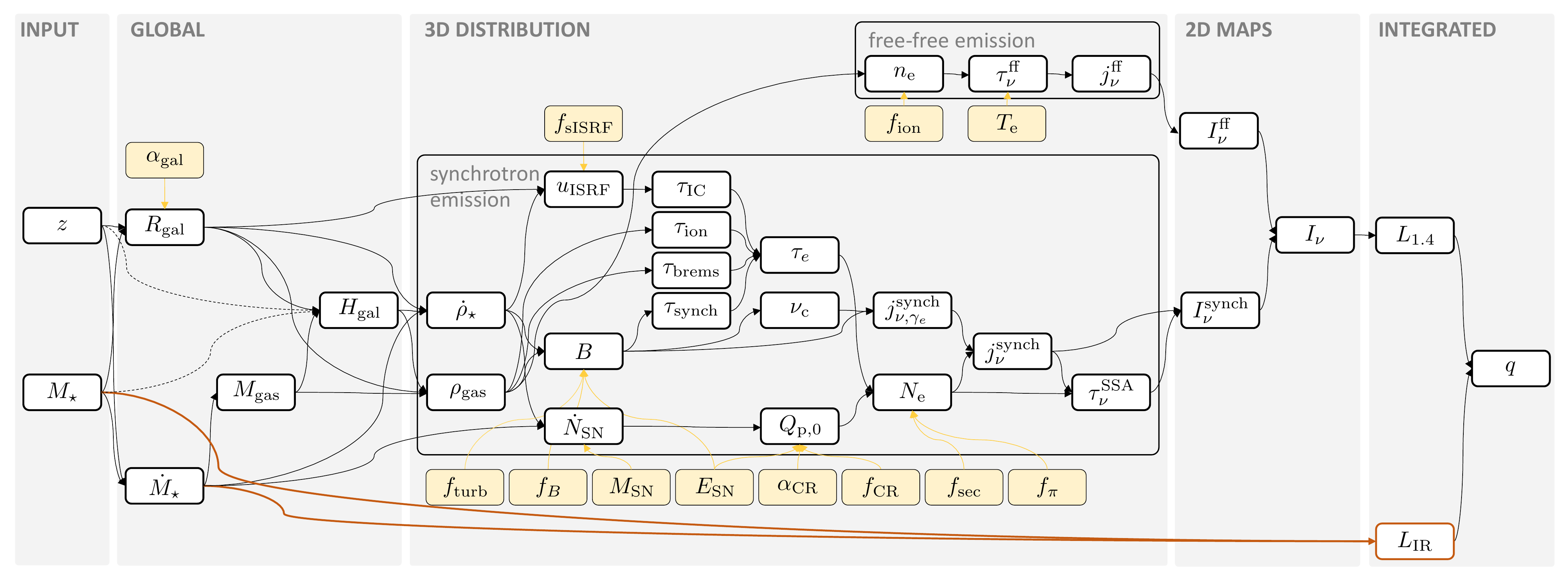}}
\caption{Overview of the model for the radio and far-infrared emission with which
the $q$ parameter can be calculated.
The input parameters, redshift $z$ and total stellar mass $M_\star$, are shown on the left
and the final output is the $q$ parameter that combines 
$1.4$ GHz luminosity $L_{1.4}$ and the rest-frame
$8-1000$ $\mu$m luminosity $L_\mathrm{IR}$.
For the galactic scale height, $H_\mathrm{gal}$, different scenarios are considered which is indicated by 
the dashed lines; see Sect.~\ref{sec_RgalHgal}.
Free parameters of the model are highlighted by yellow boxes and
their default values are listed in Table~\ref{table_model_parameters}.
See Sect.~\ref{sec_model_radio} for details of the model.
}
\label{fig_Model_anatomy2}
\end{figure*} 

To calculate the radio luminosity, we used a
semi-analytical galaxy model 
that is based on the following steps. 
1) The input is the stellar mass, $M_\star$, and redshift, $z$. 
2) The global galaxy properties are calculated from observational 
correlations, which include the gas mass, $M_\mathrm{gas}$, the star 
formation rate, $\SFR$, 
and galactic radius and scale height, $R_\mathrm{gal}$ 
and $H_\mathrm{gal}$, respectively.
3) On a 3D grid, the galaxy is modeled with an exponential 
gas density profile, using $R_\mathrm{gal}$ and $H_\mathrm{gal}$ 
from step 2. 
The total $M_\mathrm{gas}$ from step 2 is used for the normalization of the gas density. 
Likewise, the distribution of the star formation rate density is 
modeled and normalized by $\SFR$. 
A cosmic ray population is implemented based on a steady-state assumption
and the magnetic energy distribution is implemented based on a saturated 
small-scale dynamo.  
The synchrotron and free-free emission are calculated in each grid cell. 
4) Absorption processes are taken into 
account to construct 2D maps for observations with different lines of sight.
5) Integration over the 2D maps yields the total
radio luminosity which is used to calculate $q$.

These steps are summarized in Fig.~\ref{fig_Model_anatomy2}
and the default values of the free parameters of the model are presented in Table~\ref{table_model_parameters}.
The dependence on the free parameters is systematically explored
and different series of models are introduced in 
Table~\ref{tab_models}.

\begin{table}
\begin{center}
      \begin{tabular}{p{4.9cm}|p{0.8cm}|p{1.9cm}}
      \hline  \hline  
      Parameter                         & Name                  & Default value            \\
      \hline
      Power law scaling of $R_\mathrm{gal}$  with $M_\star$   & $\alpha_\mathrm{gal}$               & $0.05$\\
      SN energy                              & $E_\mathrm{SN}$       & $10^{51}~\mathrm{erg}$ \\  
      Typical mass of SN progenitor star    & $M_\mathrm{SN}$       & $22.37~M_\odot$ \\
      Fraction of SN energy that is converted to turbulence & $f_\mathrm{turb}$       & $0.05$               \\  
      Fraction of SN energy that is converted to CR protons & $f_\mathrm{CR}$       & $0.04$               \\  
      Efficiency of the dynamo               & $f_B$       & $0.1$ \\
      Slope of CR injection spectrum        & $\alpha_\mathrm{CR}$                & $3.0$ \\
      Fraction of CR protons that decay into pions   & $f_\pi$       & $0.2$ \\
      Fraction of secondary CR electrons   & $f_\mathrm{sec}$       & $0.8$ \\
      Ionization fraction                   & $f_\mathrm{ion}$      & $0.1$ \\
      Free electron temperature                   & $T_e$      & $10^4~\mathrm{K}$ \\
      ISRF, stellar contribution            & $f_\mathrm{sISRF}$     & $1$ \\
      \hline  \hline  
    \end{tabular}
\end{center}
\caption{Values of the parameters, as presented in Fig.~\ref{fig_Model_anatomy2}, that
are used in the reference model.
In this table the acronyms ``SN'' (supernova), ``CR'' (cosmic rays), and
``ISRF'' (interstellar radiation field) are used.
We note that not all of these parameters are independent.}
\label{table_model_parameters}
\end{table}

\subsection{Global observational correlations}
\label{sec_global}

In this paper, we focus on mass-selected star-forming galaxies 
on the galaxy main sequence. 
We do not attempt to model starburst 
galaxies with high efficiencies and high specific SFRs, which 
would require different values for the gas or SFR 
density, for instance, compared to the ones adopted in the following.

To reduce the number of free parameters in our model, we set the 
global properties of our model galaxies in accordance with 
observational scaling relations from the literature for a given 
$M_\star$ and $z$ (see Secs.~\ref{sec_SFR}-\ref{sec_RgalHgal} below).
These scaling relations have been derived using galaxy samples with 
stellar masses typically in the range $M_{\star}\,>\,10^{9}\,M_{\odot}$ 
and drawn from studies probing a redshift range up to $z\,{\approx}$\,3 or 4. 
This approach is thus well constrained in the redshift and mass range of 
the \citet{DelvecchioEtAl2020} study of the joint mass- and redshift dependence 
of the IRRC, which we aim to match here. 
Extrapolations to higher redshift will, however, necessarily be more 
speculative.

\subsubsection{Total SFR}
\label{sec_SFR}

For a given $M_\star$ and $z$, the total star formation rate  
of a galaxy, $\SFR$, is calculated from the observational calibration of
\citet{SchreiberEtAl2015}:
\begin{eqnarray}
  \mathrm{log}_{10}\left[\frac{\SFR(M_\star,z)}{M_\odot/\mathrm{yr}}\right] &=&  m - m_0 + a_0 r  \nonumber \\
  ~ & & -  a_1 \left[\mathrm{max}(0, m - m_1 - a_2 r) \right]^2,
\label{eq_SFR}
\end{eqnarray}    
with $r\equiv\mathrm{log}_{10}(1+z)$ and 
$m\equiv\mathrm{log}_{10}\left[M_\star/(10^9~M_\odot)\right]$ and the 
constants $m_0 = 0.5\pm 0.07$, $m_1 = 0.36\pm0.3$, $a_0 = 1.5\pm0.15$, $a_1 = 0.3\pm0.08$, and $a_2 = 2.5\pm0.6$.
Equation~\ref{eq_SFR} is based 
on a Salpeter initial mass function (IMF). 
By multiplying both $M_\star$ and $\SFR$ by a factor of $1.7$,
we can convert Eq.~\ref{eq_SFR} to make it consistent with a
Chabrier IMF, so that all parts of our model are based on the same IMF.

\subsubsection{Total gas mass}

Following \citet{SargentEtAl2014}, we estimated the total gas 
mass, $M_\mathrm{gas}$, of a star-forming galaxy 
based on its SFR 
(which, in turn, follows from its stellar mass via 
Eq.~\ref{eq_SFR}) as:
\begin{eqnarray}
 \mathrm{log}_{10}\left[\frac{M_\mathrm{gas}(M_\star,z)}{M_\odot}\right] =  
    \alpha_{2,\mathrm{SFR}} + \beta_2 \, \mathrm{log}_{10}\left[\frac{\SFR (M_\star,z)}{M_\odot \mathrm{yr}^{-1}}\right].
\label{eq_Mgas}
\end{eqnarray}   
Here, $\alpha_{2,\mathrm{SFR}}= 9.22\pm0.02$ and 
$\beta_2= 0.81\pm0.03$\footnote{We note that the classic 
Kennicutt-Schmidt relation \citep{Kennicutt1998,KennicuttEvans2012} 
relates surface densities from the regime of normal galaxies to 
starburst galaxies, with an empirical power law index of 1.4. 
In contrast, Eq.~\ref{eq_Mgas} represents the galaxy-integrated 
star-formation law for main-sequence galaxies only 
\citep[e.g.,][]{DaddiEtAl2010, SargentEtAl2012}. 
For this population, the relation between gas mass 
and total SFR has an index of $1/\beta_2\approx 1.23$.}.

\subsubsection{Galactic radius and scale height}
\label{sec_RgalHgal}

For a given stellar mass and redshift, the
characteristic half-light radius of the galaxy is \citep{Jimenez-AndradeEtAl2019} as follows:
\begin{eqnarray}
  R_\mathrm{gal}(M_\star,z) = \frac{R_\mathrm{gal,0}}{(1+z)^{\alpha_z}} \left(\frac{M_\star}{M_{\star,0}}\right)^{\alpha_\mathrm{gal}},
\label{eq_Rgal}
\end{eqnarray}    
with $R_\mathrm{gal,0}= (2100\pm200)~\mathrm{pc}$, $\alpha_z=0.26\pm0.08$, 
and $M_{\star,0}=(6.08 \pm1.14)\times10^{10} ~M_\odot$ 
being used as the default values \citep{LicquiaNewman2015}.
We note that this size scale was derived for massive main-sequence 
galaxies over the range $0<z<2$.
Based on the observed mild dependence on stellar mass, as reported in
\citet{Jimenez-AndradeEtAl2021}, we also introduced a scaling with
the power law index, $\alpha_\mathrm{gal}$. 
As a default, we used $\alpha_\mathrm{gal}=0.05$.
A weak dependence of $R_\mathrm{gal}$ on $M_{\star}$ is also found 
for the stellar component of star-forming galaxies \citep[e.g.,][]{MowlaEtAl2019}.
Equation~\ref{eq_Rgal} was obtained in the radio continuum,
namely, it is a directly relevant quantity here, since 
it is an estimate of the spatial extent of the star 
formation.

Following \citet{ThompsonEtAl2005}, we assume that the galactic scale 
height depends on the radius as:
\begin{eqnarray}
  H_\mathrm{gal}(M_\star,z) =  \frac{M_\mathrm{gas}}{M_\mathrm{gas} + M_\star} \frac{Q}{2^{3/2}} R_\mathrm{gal}(M_\star,z),
\label{eq_Hgal}
\end{eqnarray}   
where we use a Toomre parameter of $Q=1$.
Since Eq.~\ref{eq_Hgal} is obtained under the assumption of a
Salpeter IMF, we correct $M_\star$ by a factor of $1.7$ to adjust to a Chabrier 
IMF.
Additionally, we present scenarios based on two 
other models: $H_\mathrm{gal} \propto R_\mathrm{gal}(M_\star,z)$ and
$H_\mathrm{gal} \propto (1+z)$.

\begin{table*}
\begin{center}
\caption{\label{Table_models} Different series of models. 
We note that $f_B^{1/2}f_\mathrm{turb}^{1/3}$ determines the
strength of the magnetic field and, therefore, the synchrotron 
emission and $f_\mathrm{ion}$ regulates the free-free emission.
Synchrotron emission also depends on the number of CR electrons 
which is subject to the number of CR protons (controlled by the 
value of $f_\pi f_\mathrm{CR}/f_\mathrm{sec}$) and 
the slope of the injection spectrum $\alpha_\mathrm{CR}$, 
as well as on the loss rates, which depend, e.g.,~on the ISRF 
(controlled by the value of $f_\mathrm{sISRF}$).}
  \begin{tabular}{l|llllll}
   \hline  \hline  
   ~      & $H_\mathrm{gal}(M_\star, z)$   & $\alpha_\mathrm{gal}$  & $f_B^{1/2}f_\mathrm{turb}^{1/3}$    & $f_\mathrm{ion}$     & $f_\pi f_\mathrm{CR}/f_\mathrm{sec}$   & $\alpha_\mathrm{CR}$    \\
   \hline  
   %
   Reference   &   $\frac{M_\mathrm{gas}}{M_\mathrm{gas} + M_\star} \frac{Q}{2^{3/2}} R_\mathrm{gal}(M_\star,z)$ & $0.05$  &  $0.026$   & $0.1$    &  $0.025$    & $3.0$  \\
   \hline  
       Series A   &   $\frac{M_\mathrm{gas}}{M_\mathrm{gas} + M_\star} \frac{Q}{2^{3/2}} R_\mathrm{gal}(M_\star,z)$ & $0.05$ &  \specialcell{$0.013$ \\ $0.026$ \\ $0.5$}  & \specialcell{$0.05$ \\ $0.1$ \\$0.2$}    &  $0.025$    & $3.0$  \\
   \hline  
   %
   Series B  &   $\frac{M_\mathrm{gas}}{M_\mathrm{gas} + M_\star} \frac{Q}{2^{3/2}} R_\mathrm{gal}(M_\star,z)$ & $0.05$ &  $0.026$   & $0.1$   &  \specialcell{$0.015$ \\ $0.025$ \\ $0.050$}    & \specialcell{$2.8$ \\ $3.0$ \\ $3.2$}  \\
   \hline  
   %
   Series C   &   $\frac{M_\mathrm{gas}}{M_\mathrm{gas} + M_\star} \frac{Q}{2^{3/2}} R_\mathrm{gal}(M_\star,z)$ & \specialcell{$0.0$ \\ $0.1$} &    $0.026$  & $0.1$    &  $0.025$   & $3.0$  \\
   \hline  
   %
   Series D   & \specialcell{$0.1 ~R_\mathrm{gal}(M_\star,z)$ \\ $0.2 ~R_\mathrm{gal}(M_\star,z)$\\$200~\mathrm{pc}~(1+z)$ \\ $400~\mathrm{pc}~(1+z)$ }  & $0.05$    &  $0.026$   & $0.1$    &  $0.025$   & $3.0$ \\
   \hline  
   \end{tabular}
   \label{tab_models}
\end{center}
\end{table*}

\subsection{3D galaxy model}

Based on the global galaxy parameters estimated in the previous section, we
set up a three-dimensional (3D)  model of the gas density, the cosmic ray 
distribution, and other relevant parameters. 
This allows us to calculate the galactic synchrotron emission including 
absorption processes that occur along 
a given line of sight. 

The model is constructed on a grid in a box with length $L$. 
The volume of a single grid cell $V_\mathrm{cell}=L^3/\mathcal{R}^3$ 
depends on the resolution $\mathcal{R}$. 
We typically use $\mathcal{R}=100$ and $L$ is 
adjusted to the size of the
galaxy (as default we use $L= 10~R_\mathrm{gal}$). We refer to Appendix~\ref{app_res} for a discussion on convergence.

\subsubsection{Gas distribution}

The gas density distribution is set up as a
double exponential profile:
\begin{eqnarray}
  \rho_\mathrm{gas}(x_1,x_2,x_3) & =& \rho_\mathrm{gas,c} \left[ \mathrm{exp}\left(-\frac{(x_1^2+x_2^2)^{1/2}}{R_\mathrm{gal}} \right) \mathrm{exp}\left(-\frac{x_3}{H_\mathrm{gal}} \right)\right],\nonumber \\ 
\label{eq_rhogas}
\end{eqnarray}    
where the normalization $\rho_\mathrm{gas,c}$ is chosen such that 
\begin{eqnarray}
  M_\mathrm{gas} = \int \rho_\mathrm{gas} (x_1,x_2,x_3) ~\mathrm{d}x_1\mathrm{d}x_2\mathrm{d}x_3
,\end{eqnarray}   
and $R_\mathrm{gal}$ and $H_\mathrm{gal}$ 
are taken from
the expressions presented in Sect.~\ref{sec_RgalHgal}.
The gas mass is related to the star formation rate and stellar mass (see Eq.~\ref{eq_Mgas}).
The gas number density is calculated as
\begin{eqnarray}
  n_\mathrm{gas} (x_1,x_2,x_3)  = \frac{\rho_\mathrm{gas}(x_1,x_2,x_3)}{1.75 m_\mathrm{p}},
\label{eq_ngas}
\end{eqnarray}    
where we assume that the gas is composed of atoms with average mass $1.75 m_\mathrm{p}$ 
(for a primordial mixture of gas with $75\%$ hydrogen and $25\%$ helium).
In Fig.~\ref{app_RadProf} in the appendix, the radial dependence of 
 $n_\mathrm{gas}$ is presented for different values of $M_\star$ and $z$.

\begin{figure}[t!]
\centering
  \includegraphics[width=0.45\textwidth]{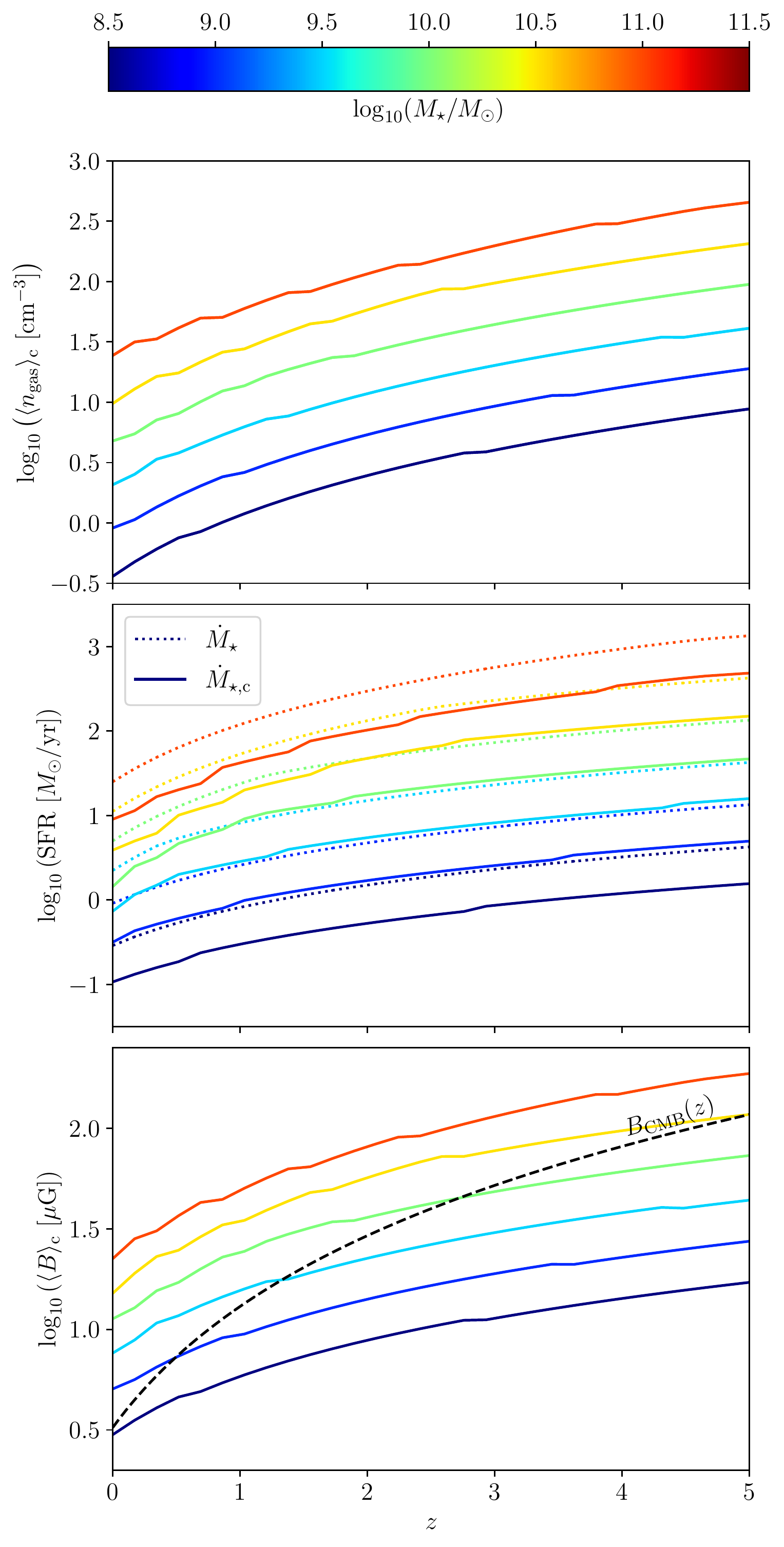}
\caption{
Input relations for the reference galaxy model (see Table~\ref{Table_models}).
\textit{Upper panel:} Mean density in the central region of the 
galaxy obtained from integrating Eq.~\ref{eq_ngas}.
\textit{Middle panel:} Total SFR from Eq. \ref{eq_SFR} shown 
by dotted lines and the SFR in the core region of the galaxy obtained 
from integrating Eq.~\ref{eq_rhoSFR} shown by solid lines.
\textit{Lower panel:} Mean magnetic field strength in the core region of 
the galaxy obtained from integrating Eq.~\ref{eq_B}
with the default value $f_B=0.1$.
}
\label{fig_model_input}
\end{figure}

In Fig.~\ref{fig_model_input}, the typical volume average of the
density is presented for the parameter range 
of interest. 
Since the volume average depends on the size of the
box, $L$, we show the average value in 
the central region of the galaxy instead.
We define this central average of the gas density as
$\langle n_\mathrm{gas}\rangle_\mathrm{c} = \int_{-R_\mathrm{gal}}^{R_\mathrm{gal}}\int_{-R_\mathrm{gal}}^{R_\mathrm{gal}}\int_{-H_\mathrm{gal}}^{H_\mathrm{gal}} n_\mathrm{gas}~\mathrm{d}x_1\mathrm{d}x_2\mathrm{d}x_3/V_\mathrm{c}$,
with $V_\mathrm{c} = \int_{-R_\mathrm{gal}}^{R_\mathrm{gal}}\int_{-R_\mathrm{gal}}^{R_\mathrm{gal}}\int_{-H_\mathrm{gal}}^{H_\mathrm{gal}} \mathrm{d}x_1\mathrm{d}x_2\mathrm{d}x_3$.

\subsubsection{Star formation and supernova rate}

Supernova (SN) explosions are responsible for producing 
two key ingredients of galactic synchrotron emission:
magnetic fields and CRs.
Therefore, we need to model the spatial distribution of the SN rate
which depends on the galactic star formation activity.
To this end, we make use of the empirical Kennicutt-Schmitt 
law \citep{Kennicutt1998,KennicuttEvans2012}, which relates 
the SFR surface density to the gas surface density.
Moving from surface densities to volume densities, we calculate
the local SFR density:
\begin{eqnarray}
  \dot{\rho}_\star (x_1,x_2,x_3) &=& \dot{\rho}_{\star,\mathrm{c}} \left[ \mathrm{exp}\left(-\frac{(x_1^2+x_2^2)^{1/2}}{R_\mathrm{gal}} \right) \mathrm{exp}\left(-\frac{x_3}{H_\mathrm{gal}} \right)\right]^{n_\mathrm{SFR}}, \nonumber \\
\label{eq_rhoSFR}
\end{eqnarray}   
in each grid cell.
In Eq.~\ref{eq_rhoSFR}, the same profile as the one of the 
gas density in Eq.~\ref{eq_rhogas}
is used but with   $n_\mathrm{SFR}$ as the exponent. 
The observed Kennicutt-Schmidt relation
for surface densities has
$n_\mathrm{SFR}=1.4$. 
However, we note that a relation between $\dot{\rho}_\star$ and $\rho_\mathrm{gas}$ 
can also be derived theoretically when assuming that 
the SFR is proportional to the gas density over the
dynamical timescale.
This yields $\dot{\rho}_\star \approx \epsilon_\mathrm{SFR} G^{1/2} \rho_\mathrm{gas}^{3/2}$, 
where $\epsilon_\mathrm{SFR}$ is the star formation
efficiency and $G$ is the gravitational constant. 
If $\epsilon_\mathrm{SFR}$ is independent of $\rho_\mathrm{gas}$ then
$n_\mathrm{SFR}=3/2$ \citep{SchayeDallaVecchia2008}, which we use in our model 
\footnote{We note that the value of the index $n_\mathrm{SFR}=3/2$ 
differs a corresponding value appearing in Eq.~\ref{eq_Mgas} 
that is $1/\beta_2\approx1.23$.
Since Eq. \ref{eq_Mgas} is an empirical relation that relates 
masses and Eq.~\ref{eq_rhoSFR} is based on mass density, a difference 
between $n_\mathrm{SFR}$ and $1/\beta_2$ 
should not be considered to be an inconsistency.}.
The normalization in Eq.~\ref{eq_rhoSFR}, $\dot{\rho}_{\star,\mathrm{c}}$, is chosen such that 
\begin{eqnarray}
  \SFR  = \int_0^L \int_0^L \int_0^L \dot{\rho}_\star (x_1,x_2,x_3) ~\mathrm{d}x_1\mathrm{d}x_2\mathrm{d}x_3.
\end{eqnarray}   
The global SFR, $\SFR$, and the SFR in the core region,
$\dot{M}_{\star,\mathrm{c}} = \int_{-R_\mathrm{gal}}^{R_\mathrm{gal}}\int_{-R_\mathrm{gal}}^{R_\mathrm{gal}}\int_{-H_\mathrm{gal}}^{H_\mathrm{gal}} \dot{\rho}_\star~\mathrm{d}x_1\mathrm{d}x_2\mathrm{d}x_3$,
are presented in Fig.~\ref{fig_model_input}.

Assuming a Chabrier initial mass function and a minimum mass
of stars that result in a SN of $8~M_\odot$, 
the supernova rate in each grid cell is: 
\begin{eqnarray}
  \dot{N}_\mathrm{SN} (x_1,x_2,x_3)  = 0.23 \frac{ \dot{\rho}_\star (x_1,x_2,x_3) V_\mathrm{cell}}{\overline{M}_\mathrm{SN}},
\end{eqnarray}  
where $\overline{M}_\mathrm{SN} \approx 22.37~M_\odot$ is the 
average mass of a star that ends in a supernova.
The corresponding SN rate density in a grid cell is given by
\begin{eqnarray}
  \dot{n}_\mathrm{SN} (x_1,x_2,x_3)  =  \frac{ \dot{N}_\mathrm{SN} (x_1,x_2,x_3)}{V_\mathrm{cell}}.
\end{eqnarray}

\subsubsection{Magnetic field strength}

A key assumption in this work is that magnetic 
fields are produced on short timescales by a turbulent dynamo
\citep{Kazantsev1968,KulsrudAnderson1992, BrandenburgSubramanian2005}. 
This mechanism converts turbulent kinetic energy to magnetic 
energy exponentially over time, 
until the magnetic energy becomes comparable to the kinetic energy and a 
back reaction on the velocity field sets in. 
The characteristic timescale of the small-scale turbulent 
dynamo is the eddy turnover time
on the viscous spatial scale determined by the Reynolds 
number in the system 
\citep{RogachevskiiKleeorin1997,SchoberEtAl2012.1,SchoberEtAl2012.3}.

In the context of galaxy formation, 
a seed magnetic field
is first amplified by compression during
the gravitational collapse. 
The end of the collapse phase can be defined as
the moment when the gas density has become approximately 
constant. 
Then the main mechanism of magnetic field amplification is 
the small-scale dynamo that is sourced by turbulence due to 
accretion and SN explosions.
Since galaxies have very high Reynolds numbers ($\approx 10^{10}-10^{12}$), 
the small-scale dynamo timescale is very short ($\approx 10^{-2}-10^{-1}~\mathrm{Myr}$)
compared to other dynamical timescales. 
Therefore, once the small-scale dynamo is activated, the magnetic field quickly 
reaches the equipartition value already in young galaxies 
\citep{SchoberEtAl2013, SchleicherEtAl2013}.
This magnetic field evolution is being seen in cosmological 
simulations that solve the magnetohydrodynamical equations
and have sufficient resolution
\citep{RiederTeyssier2016, RiederTeyssier2017, Martin-AlvarezEtAl2018}.
In the case described by \citet{Martin-AlvarezEtAl2018}, the collapse phase of the
galaxy ends at $z\approx 12$ and the 
timescale of the subsequent small-scale dynamo is on the order of 2 Gyrs. 
The timescale, however, is much longer than what is expected in 
real galaxies due to the limited resolution of cosmological simulations.
\footnote{The growth rate of the small-scale dynamo depends
on the Reynolds number $\mathrm{Re}$
which is on the order of $\mathrm{Re}\approx10^{10}-10^{12}$ for young galaxies. 
In simulations such high values of $\mathrm{Re}$ can never be achieved
due to limited resolution.
For example, the runs presented in \citet{Martin-AlvarezEtAl2018} 
reach $\mathrm{Re}\approx 15-160$. 
Therefore, the dynamo timescale in simulations is always overestimated
by orders of magnitudes.}

The energy source of the small-scale dynamo is turbulence. 
Here, we assume that the main drivers of turbulence are supernova explosions
in the ISM \citep{MacLowKlessen2004}. 
Under steady state the turbulent velocity $v_\mathrm{turb}$ can be estimated from 
a balance between the loss rate $1/2 \rho_\mathrm{gas} v_\mathrm{turb}^2/(H_\mathrm{gal}/v_\mathrm{turb})$ 
and the energy injection rate $\dot{n_\mathrm{SN}} f_\mathrm{turb} E_\mathrm{SN}$. 
Here, $f_\mathrm{turb} E_\mathrm{SN}$ 
is the fraction of supernova energy converted into turbulence 
with the default values of $f_\mathrm{turb}=0.05$ and $E_\mathrm{SN} = 10^{51}~\mathrm{erg}$.
Solving for $v_\mathrm{turb}$ gives the following result:\ 
\begin{equation}
  v_\mathrm{turb}(x_1,x_2,x_3) = \left(2 \dot{n_\mathrm{SN}} f_\mathrm{turb} E_\mathrm{SN} H_\mathrm{gal} \rho_\mathrm{gas}^{-1}\right)^{1/3}.
\label{eq_vturb}
\end{equation}
Assuming that at saturation, the magnetic energy reaches a fraction $f_B$ 
of the turbulent kinetic energy, the magnetic field strength is then:
\begin{eqnarray}
  B(x_1,x_2,x_3) & = & \left[f_B ~ 8\pi ~\frac{1}{2} \rho_\mathrm{gas} \left(v_\mathrm{turb}\right)^2\right]^{1/2} \nonumber \\
                 & = &  \left( 4\pi f_B \right)^{1/2}   \left(2 \dot{n_\mathrm{SN}} f_\mathrm{turb} E_\mathrm{SN} H_\mathrm{gal} \right)^{1/3} \rho_\mathrm{gas}^{1/6}.
\label{eq_B}
\end{eqnarray}
Since $\dot{n}_\mathrm{SN}\propto\dot{\rho}_\star$ in our model, 
Eq.~\ref{eq_B} leads to $B \propto \dot{n}_\mathrm{SN}^{1/3} \propto \dot{\rho}_\star^{1/3}$.
This is close to the observed scaling of $B \propto \dot\Sigma_\star^{0.3}$ 
in various types of galaxies 
\citep{NiklasBeck1997,ChyzyEtAl2011,BasuEtAl2017}, where
$\dot\Sigma_\star$ is the SFR surface density.

MHD dynamo theory makes predictions 
for the saturation level, $f_B$. 
In the most ideal case, namely, of  incompressible (or Kolmogorov) turbulence,
$f_B$ reaches its maximum value of $\approx0.4$.
However, in the SN-driven case, turbulence is typically compressive 
with Mach numbers reaching values on the order of $10$.
In that case, direct numerical simulations (DNS) with driven turbulence result in saturation 
levels of $f_B\approx0.01$
\citep{FederrathEtAl2011b,FederrathEtAl2014b, SetaFederrath2021}
and $f_B\approx0.05$ in SN-driven simulations of the ISM \citep{GentEtAl2021}.
On the other hand, observations indicate that 
a large faction of the interstellar medium, in particular, 
the warm ionized medium, 
has relatively low Mach numbers, implying sub- to transonic turbulence
\citep{BurkhartEtAl2010,GaenslerEtAl2011,KoleyRoy2019}.
Therefore, we use $f_B\approx0.1$ as the default value in our model.
However, we explore the
dependence of the results on $f_B$ in Sect.~\ref{sec_synch_ff}. 
The mean magnetic field strength in the central region of the galaxy, defined as
$\langle B\rangle_\mathrm{c} = \int_{-R_\mathrm{gal}}^{R_\mathrm{gal}}\int_{-R_\mathrm{gal}}^{R_\mathrm{gal}}\int_{-H_\mathrm{gal}}^{H_\mathrm{gal}} B~\mathrm{d}x_1\mathrm{d}x_2\mathrm{d}x_3/V_\mathrm{c}$,
is shown in the bottom panel of Fig.~\ref{fig_model_input}.
The resulting field strength in the central part of the modeled galaxies 
is, at low redshifts, larger than the CMB equivalent field strength
$B_\mathrm{CMB}(z) \approx 3.25(1 + z)^2~\mu\mathrm{G}$.
At a higher $z$, the value of $\langle B\rangle_\mathrm{c}$ 
eventually drops below $B_\mathrm{CMB}$.

\subsubsection{Cosmic ray electrons}

In every grid cell, the cosmic ray (CR) energy spectrum is modeled.
We assume that CR protons are produced in diffusive shock
acceleration within supernova shock fronts
and that CR electrons are composed of both primaries 
and secondaries from decays of CR protons. 
Following \citet{LackiBeck2013}, we further assume that CR electrons 
are in a steady state.
The full CR implementation is sketched in the following.

The energy spectrum of injected protons is \citep{Bell1978a,Bell1978b}
\begin{equation}
  Q_\mathrm{p}(E_\mathrm{p}) = Q_{\mathrm{p},0}~\left(\frac{E_\mathrm{p}}{m_\mathrm{p} c^2}\right)^{-\alpha_\mathrm{CR}},
\label{eq_Q}
\end{equation}
with the normalization $Q_{\mathrm{p},0}$ and the injection power law index $\alpha_\mathrm{CR}$.
Integration from lowest energy to infinity gives the total energy injected as protons, namely:
\begin{eqnarray}
  \int_{E_{\mathrm{p},0}}^\infty Q_\mathrm{p}(E_\mathrm{p})~\mathrm{d}E_\mathrm{p} &=& Q_{\mathrm{p},0} ~m_\mathrm{p} c^2~ \frac{1}{\alpha_\mathrm{CR}-1} \left(\frac{E_{\mathrm{p},0}}{m_\mathrm{p} c^2}\right)^{1-\alpha_\mathrm{CR}} \nonumber \\
                                                                                   &=& Q_{\mathrm{p},0} ~m_\mathrm{p} c^2~ \frac{1}{\alpha_\mathrm{CR}-1} \gamma_{\mathrm{p},0}^{1-\alpha_\mathrm{CR}},
\label{eq_Qint}
\end{eqnarray}
where $\gamma_{\mathrm{p},0}=1~\mathrm{GeV}/(m_\mathrm{p} c^2) \approx 1$.
The normalization of Eq.~\ref{eq_Qint} is obtained by comparison with
the energy injected by supernovae,
$f_\mathrm{CR} \dot{N}_\mathrm{SN} E_\mathrm{SN}$.
Using $f_\mathrm{CR} = 0.1$ as the fraction of SN energy that is 
converted into CRs and a characteristic energy per SN of $E_\mathrm{SN} = 10^{51}~\mathrm{erg}$ 
yields
\begin{equation}
  Q_{\mathrm{p},0} = \frac{f_\mathrm{CR} E_\mathrm{SN} \dot{N}_\mathrm{SN} (\alpha_\mathrm{CR}-1)}{m_\mathrm{p} c^2 ~ \gamma_\mathrm{p,0}^{1-\alpha_\mathrm{CR}}}.
\label{eq_Qp0}
\end{equation}
The CR protons produce secondary CR electrons with a spectrum
\begin{equation}
  Q_\mathrm{e,sec}(E_\mathrm{e,sec}) = \frac{f_\pi}{6} \left(\frac{E_\mathrm{p}}{E_\mathrm{e,sec}}\right)^2 Q_\mathrm{p}(E_{\mathrm{p}}),
\end{equation}
which, in $\gamma=E/(mc^2)$ space, is: 
\begin{equation}
  Q_\mathrm{e,sec}(E_\mathrm{e,sec}) ~m_\mathrm{p} c^2= \frac{f_\pi}{6} \left(\frac{E_\mathrm{p}}{E_\mathrm{e,sec}}\right)^2 Q_\mathrm{p}(E_{\mathrm{p}}) ~m_\mathrm{e} c^2.
\end{equation}
The factor $f_\pi$ is the fraction of CR protons that decay
into pions and we assume $f_\pi = 0.2$ in the reference model.
It follows \citep[with $E_{\mathrm{p}} =20~E_{\mathrm{sec}}$,][]{LackiBeck2013}:
\begin{eqnarray}
  Q_\mathrm{e,sec}(E_\mathrm{e,sec}) &=& \frac{f_\pi}{6} \left(\frac{E_\mathrm{p}}{E_\mathrm{e,sec}}\right)^2 Q_\mathrm{p}(E_{\mathrm{p}}) \frac{m_\mathrm{e}}{m_\mathrm{p}}  \nonumber \\
                                     &=& \frac{f_\pi}{6} 20^2 Q_{\mathrm{p},0}\left(\frac{20 \gamma_e m_\mathrm{e} c^2}{m_\mathrm{p} c^2}\right)^{-\alpha_\mathrm{CR}} \frac{m_\mathrm{e}}{m_\mathrm{p}} \nonumber \\
                                     &=& \frac{f_\pi}{6} 20^{2-\alpha_\mathrm{CR}} Q_{\mathrm{p},0}\gamma_e^{-\alpha_\mathrm{CR}} \left(\frac{m_\mathrm{e}}{m_\mathrm{p}}\right)^{1-\alpha_\mathrm{CR}}, 
\end{eqnarray}
where $\gamma_e$ is the Lorentz factor of the electrons.
The total injection rate of CR electrons (primaries and secondaries) is related to the secondary rate $Q_\mathrm{e,sec}$
by the factor $f_\mathrm{sec}$:
\begin{equation}
  Q_{e} = \frac{Q_\mathrm{e,sec}}{f_\mathrm{sec}}.
\end{equation}
We use $f_\mathrm{sec}= 0.8$ as a default value \citep{LackiBeck2013}. 
For steady state, the total number of cosmic ray electrons can be estimated as:
\begin{eqnarray}
  N_\mathrm{e}(\gamma_e) &=& \frac{Q_{e}(\gamma_e)\tau_{e} (\gamma_e)}{\alpha_\mathrm{CR}-1}  \nonumber \\
         &=&  \frac{20^{2-\alpha_\mathrm{CR}}}{6 (\alpha_\mathrm{CR}-1)} \frac{f_\pi}{f_\mathrm{sec}} \left(\frac{m_\mathrm{e}}{m_\mathrm{p}}\right)^{1-\alpha_\mathrm{CR}} ~Q_\mathrm{p,0} \gamma_e^{-\alpha_\mathrm{CR}}  \tau_e(\gamma_e).
\label{eq_Ne}
\end{eqnarray}
The CR cooling time $\tau_{e}$ is determined by different 
loss channels, including ionization ($\tau_\mathrm{ion}$),
bremsstrahlung ($\tau_\mathrm{brems}$), inverse Compton scattering ($\tau_\mathrm{IC}$), 
and synchrotron losses ($\tau_\mathrm{synch}$),
\begin{equation}
  \tau_{e} = \left(\tau_\mathrm{ion}^{-1} + \tau_\mathrm{brems}^{-1} + \tau_\mathrm{IC}^{-1} + \tau_\mathrm{synch}^{-1} \right)^{-1},
\label{te}
\end{equation}
with 
\begin{eqnarray}
  \tau_\mathrm{ion} & = & \frac{\gamma_\mathrm{e}}{2.7~c~\sigma_\mathrm{T}~(6.85 + 0.5~\mathrm{ln}\gamma_\mathrm{e})~n_\mathrm{gas}} \label{tau_ion}, \\
  \tau_\mathrm{brems} & = & 
  \frac{1}{4 \alpha_\mathrm{em} r_0^2 c n_\mathrm{ion} \beta_e} \left[\mathrm{ln}(2\gamma_e) -\frac{1}{3}\right]^{-1}
  \label{tau_brems}, \\
  \tau_\mathrm{IC} & = & \frac{3~m_\mathrm{e}~c}{4~\sigma_\mathrm{T}~u_\mathrm{ISRF}~\gamma_\mathrm{e}} \label{tau_IC}, \\
  \tau_\mathrm{synch} & = & \frac{3~m_\mathrm{e}~c}{4~\sigma_\mathrm{T}~u_B~\gamma_\mathrm{e}}. \label{tau_synch}
\end{eqnarray}
In the expression for $\tau_\mathrm{brems}$, $\alpha_\mathrm{em}$ and $r_0$ are the fine-structure constant and the electron radius, respectively,
$\beta_e=(1 - \gamma_e^{-2})^{1/2}$, and 
$n_\mathrm{ion}$ is the ion density, for which we assume $n_\mathrm{ion} = f_\mathrm{ion} n_\mathrm{gas}$ 
with $f_\mathrm{ion}$ being the ionization degree
 \citep{BlumenthalGould1970}. 
Further, in the expressions for $\tau_\mathrm{IC}$ and $\tau_\mathrm{synch}$, $\sigma_\mathrm{T}$ is the Thomson cross-section for an electron and
$u_\mathrm{ISRF}$ and $u_B=B^2/(8\pi)$ are the energy density of the interstellar radiation field
and the magnetic energy density, respectively.
The model of $u_\mathrm{ISRF}$ includes both, contributions from the stellar population and the
cosmic microwave background (see Appendix~\ref{app_ISRF} for details).

The number of CR electrons, modeled here via 
Eq.~\ref{eq_Ne}, determines the resulting 
radio emission of the galaxy. 
It includes several free parameters. 
In particular, $N_e$ is proportional to the ratio
$f_\pi f_\mathrm{CR}/f_\mathrm{sec}$
but it also depends on the slope of the injection
spectrum $\alpha_\mathrm{CR}$.
Analytical models for diffusive shock acceleration (DSA)
predict values of $\alpha_\mathrm{CR}=2.0$ 
for non-relativistic gas and $\alpha_\mathrm{CR}=2.5$ 
for relativistic gas \citep{Bell1978b,BlandfordEichler1987}. 
Detailed models of supernova shock fronts result  
$\alpha_\mathrm{CR}= 2.1 - 3$ \citep{BogdanVolk1983},
whereas more recent studies of diffusive shock acceleration 
based on PIC/hybrid approach
predict $\alpha_\mathrm{CR} \approx 1.5$ for Mach numbers $\gtrsim$ 10 
and non-relativistic ions \citep{CaprioliSpitkovsky2014}.
In our reference model, a steeper value of 
$\alpha_\mathrm{CR}=3$ is used to reproduce the observed values
of $q(z)$.
A more detailed cosmic ray model of cosmic ray propagation,
including diffusion and losses via galactic winds,
could allow for values of $\alpha_\mathrm{CR}$ that 
are closer to the DSA predictions.

\subsection{Radio emission}

\begin{figure}
\centering
  \subfigure{\includegraphics[width=0.45\textwidth]{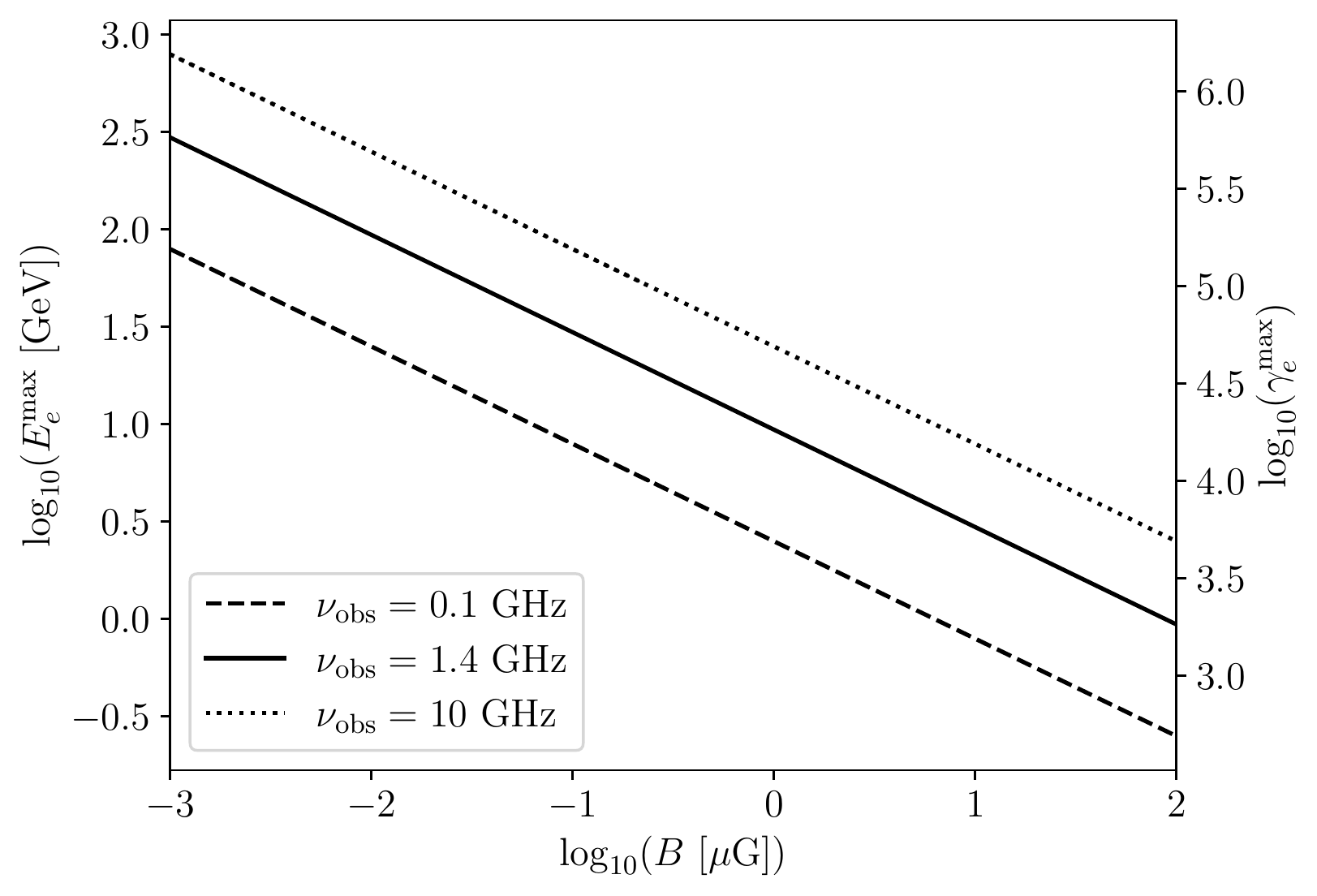}}
\caption{Energy (left ordinate) and Lorentz factor (right ordinate) 
of CR electrons that contribute mostly to the synchrotron emission at 
different observing frequencies $\nu_\mathrm{obs}$
and different magnetic field strengths $B$;
see Eq.~\ref{eq_Emax}.}
\label{fig_Emax_B}
\end{figure}

\begin{figure*}
\centering
  \subfigure{\includegraphics[width=0.85\textwidth]{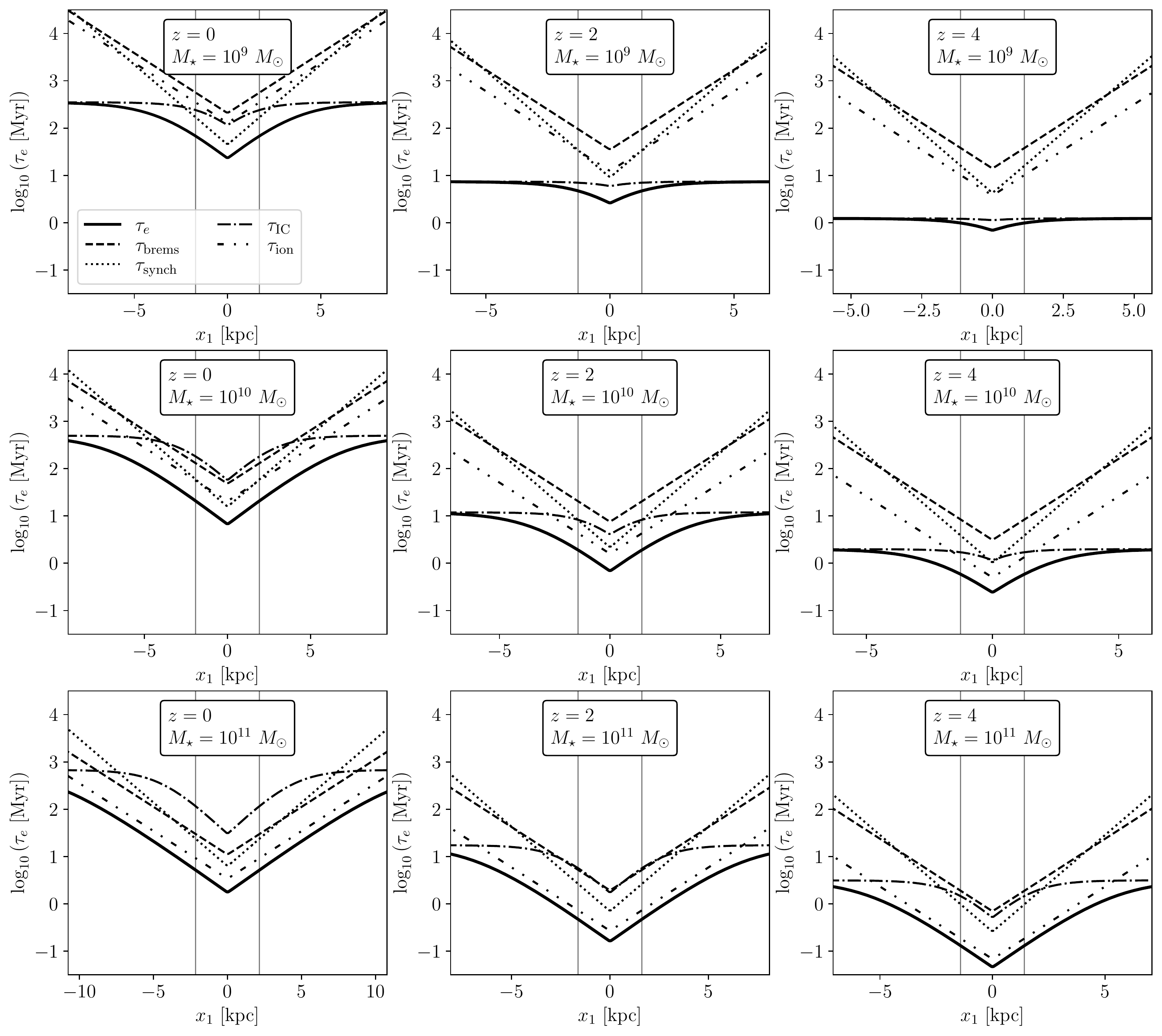}}
\caption{Cooling timescales of CR electrons in a slice through the center 
of galaxies with different stellar masses, $M_\star$, (top to bottom panels) 
and different redshifts, $z$ (left to right panels).
The value of $\gamma_e$ used in this plot corresponds to the 
one at which CR electrons contribute mostly to the synchrotron emission
at $1.4~\mathrm{GHz}$,
i.e.,~at which $\nu_\mathrm{c} = \nu_\mathrm{obs} = 1.4~\mathrm{GHz}$;
see Eq.~\ref{eq_gammamax}.
The total timescale, $\tau_e$, (solid lines) has contributions from
bremsstrahlung ($\tau_\mathrm{brems}$, dashed lines),
synchrotron losses ($\tau_\mathrm{synch}$, dotted lines),
inverse Compton losses ($\tau_\mathrm{IC}$, dashed-dotted lines),
and ionization losses ($\tau_\mathrm{ion}$, dashed-dotted-dotted lines).
Grey vertical lines indicate the characteristic radius $R_\mathrm{gal}$,
i.e.,~they enclose $2R_\mathrm{gal}$.
}
\label{fig_tauCR}
\end{figure*}

\subsubsection{Synchrotron emission and self-absorption}

Synchrotron emission (in [erg s$^{-1}$ Hz$^{-1}$]) 
of a single particle is calculated via: 
\begin{eqnarray}
  j^\mathrm{synch}_{\nu,\gamma_e} = \frac{\sqrt{3}~e^3~ B}{m_e c^2}~\frac{\nu}{\nu_\mathrm{c}}~ \int_{\nu/\nu_\mathrm{c}}^\infty K_{5/3}(\xi)~\mathrm{d}\xi, 
\label{eq_singlesynch}
\end{eqnarray}
where $K_{5/3}(\xi)$ is the modified Bessel function of the second kind \citep{BlumenthalGould1970}.
This spectrum peaks roughly at the critical frequency, 
\begin{eqnarray}
  \nu_\mathrm{c}(\gamma_e) = \frac{3 \gamma_e^2~e~B}{4\pi~c~m_\mathrm{e}}.
\end{eqnarray}

For a distribution of electrons with different energies according to Eq.~\ref{eq_Ne}, 
the spectral emission per grid cell 
(in [erg s$^{-1}$ Hz$^{-1}$ cm$^{-3}$]) is determined by \citep{BlumenthalGould1970}
\begin{eqnarray}
  j^\mathrm{synch}_\nu & = & \int_{\gamma_{\mathrm{e},0}}^\infty j^\mathrm{synch}_{\nu,\gamma_e} \frac{N_\mathrm{e}(\gamma_e)}{\delta x^3} ~\mathrm{d}\gamma_e  \int N(\alpha) (\mathrm{sin}(\alpha))^{(\alpha_\mathrm{CR}+1)/2}~\mathrm{d}\Omega_\alpha. \nonumber \\
\label{eq_jsynchnu}
\end{eqnarray}
The last integral in Eq.~\ref{eq_jsynchnu} over the pitch angle $\alpha$ is
on the order of $10$ for $\alpha_\mathrm{CR} = 3.0$ and if isotropy is assumed, namely,~the distribution of pitch angles is $N(\alpha)\approx 1$.

The numerical integration of Eq.~\ref{eq_jsynchnu} is expensive and therefore we 
use the following two approximations. 
First, instead of performing the expensive numerical integration of the Bessel function
in Eq.~\ref{eq_singlesynch}, we follow \citet{AharonianEtAl2010} and use
\begin{eqnarray}
   a \int_a^\infty K_{5/3}(\xi)~\mathrm{d}\xi  
       &\approx& 2.15 a^{1/3}~(1+3.06 a)^{1/6}  \nonumber \\
       && \times \frac{1+0.884a^{2/3} + 0.471a^{4/3}}{1+1.64a^{2/3} + 0.974a^{4/3}}~ e^{-a}.
\end{eqnarray}
Second, it is useful to note that the function 
$j^\mathrm{synch}_{\nu,\gamma_e}$ peaks at the 
characteristic frequency $\nu_\mathrm{c}(\gamma_e)$. 
This implies that synchrotron radiation
at a given frequency $\nu_\mathrm{obs}$ is emitted by cosmic ray electrons with a specific 
Lorentz factor of
\begin{eqnarray}
  \gamma_{e}^\mathrm{max} = \left( \frac{4\pi~c~m_\mathrm{e} \nu_\mathrm{obs}}{3~ e~B }\right)^{1/2} 
\label{eq_gammamax}
\end{eqnarray}
or, in terms of energy,
\begin{eqnarray}
  E_{e}^\mathrm{max} = \left( \frac{4\pi~c~m_\mathrm{e} \nu_\mathrm{obs}}{3~ e~B }\right)^{1/2} m_e c^2.
\label{eq_Emax}
\end{eqnarray}
The resulting value of $E_{e}^\mathrm{max}$ is shown for 
$\nu_\mathrm{obs}=1.4~\mathrm{GHz}$ 
as a function of magnetic field strength
in Fig.~\ref{fig_Emax_B}.
To save computational time, the integration over $\gamma_e$ in Eq.~\ref{eq_jsynchnu} 
is performed in the interval 
$[\mathrm{max}(1,0.01~\gamma_{e}^\mathrm{max}), 100~\gamma_{e}^\mathrm{max}]$.

Knowing the characteristic energy of the CRs that emit synchrotron radiation at $\nu_\mathrm{obs}$
allows us to determine the most relevant cooling processes of the emission. 
The numerical value of the cosmic ray cooling timescale, $\tau_{e}$,
and its different contributions are shown in 
Fig.~\ref{fig_tauCR} for $\gamma_{e}=\gamma_{e}^\mathrm{max}$.
For the default values of the free parameters, the 
dominant process determining the population CR electrons that emit 
synchrotron radiation at $\nu_\mathrm{obs}=1.4~\mathrm{GHz}$  at $z=0$
are synchrotron (for low and intermediate $M_\star$) and ionization (for the highest $M_\star$ considered in this study) in
the center of the galaxy and inverse Compton (with the CMB) in the outer, very 
low gas density regions of the simulation box. 
With increasing $z$, inverse Compton becomes the dominant cooling process even in 
more central parts of the galaxy. 
The reason for this is the CMB radiation field which increases
proportional to $(1+z)^4$.

If absorption is ignored, the total synchrotron emission is
obtained by
\begin{eqnarray}
  L^\mathrm{synch,0}_\nu =  \int \int \int j^\mathrm{synch}_\nu~\mathrm{d}x_1\mathrm{d}x_2\mathrm{d}x_3.
\end{eqnarray}
However, at low frequencies and/or high gas density, absorption of 
synchrotron photons can become significant.
The synchrotron self-absorption (SSA)
coefficient is given by \citep{Schlickeiser2002},
\begin{eqnarray}
  \kappa^{\mathrm{SSA}}_\nu &=& - \frac{c^2}{8 \pi m_e c^2 \nu^2} 
  \int_1^\infty j^\mathrm{synch}_\nu \gamma_e^2 \frac{\mathrm{d}}{\mathrm{d}\gamma_e} \left[\frac{N_e(\gamma_e)}{\gamma_e^2}\right] ~\mathrm{d}\gamma_e, 
\label{eq_kappaSSA}
\end{eqnarray}
and leads to a corresponding optical depth of
\begin{eqnarray}
  \tau^{\mathrm{SSA}}_\nu(s_0, s_1) &=&  \int_{s_0}^{s_1}  \kappa^{\mathrm{SSA}}_\nu~\mathrm{d}s.
\label{eq_tauSSA}
\end{eqnarray}
We note that the optical depth depends on the thickness of 
the layer of synchrotron emitting CRs considered. 
If the most distant emission occurs at position $s_0$, 
then Eq.~\ref{eq_tauSSA} gives the value of $\tau^{\mathrm{SSA}}_\nu$
at position $s_1$.
Another absorption channel is free-free absorption which is discussed in the subsequent section.

\subsubsection{Free-free emission and absorption}
\label{sec_freefree}

The free-free emission adds a contribution to the total radio flux
with the emissivity calculated as follows
\citep{RybickiLightman1986}:
\begin{eqnarray}
  j^\mathrm{ff}_\nu &=&  \frac{2^5 \pi e^6}{3 m_e c^3} \left(\frac{2\pi}{3 k_\mathrm{B} m_e}\right)^{1/2} 
  T_\mathrm{e}^{-1/2} n_e n_\mathrm{ion} Z^2 \mathrm{e}^{- h \nu/(k_\mathrm{B} T_e)} \overline{g}^\mathrm{ff}_\nu. 
\label{eq_Lff}
\end{eqnarray}
The velocity-averaged Gaunt factor is given by \citep{NovikovThorne1973}:
\begin{eqnarray}
  \overline{g}^\mathrm{ff}_\nu = \frac{\sqrt{3}}{\pi} \mathrm{ln} \left[ \frac{1}{4\xi^{5/2}Z} \left(\frac{k_\mathrm{B} T_e}{ h \nu}\right) \left(\frac{k_\mathrm{B} T_e}{ 13.6~\mathrm{eV}}\right)^{0.5} \right],
\label{eq_gff}
\end{eqnarray}
with $\xi\approx 1.781$.
We assume that the ion and electron densities are comparable, 
that is,~$n_\mathrm{ion}\approx n_e = f_\mathrm{ion} n_\mathrm{gas}$ and $Z=1$.
As the default value for the ionization degree, we use $f_\mathrm{ion}=0.1$ and we assume an electron temperature 
of $T_e=10^4~\mathrm{K}$.

With the free-free absorption coefficient, 
\begin{eqnarray}
  \kappa^\mathrm{ff}_\nu &=& \frac{4 e^6}{3 m_e h c} \left(\frac{2\pi}{3 k_\mathrm{B} m_e}\right)^{1/2} 
  T_\mathrm{e}^{-1/2} n_e n_\mathrm{ion} Z^2 \nu^{-3} 
  \left(1- \mathrm{e}^{- h \nu/(k_\mathrm{B} T_e)}\right) \overline{g}_\mathrm{ff}, \nonumber \\
\label{eq_kappaff}
\end{eqnarray}
the optical depth along a line of sight between $s_0$ and $s_1$ can be calculated as
\begin{eqnarray}
  \tau^\mathrm{ff}_\nu(s_0, s_1) &=& \int_{s_0}^{s_1} \kappa^\mathrm{ff}_\nu ~ \mathrm{d}s.
\label{eq_tauff}
\end{eqnarray}

The intensity at a slice of the box at the position $s_1$ along a
line of sight is then given as
\begin{eqnarray}
  I^\mathrm{ff}_\nu(s_0,s_1) &=& \frac{1}{4\pi} \frac{j^\mathrm{ff}_\nu}{\kappa^\mathrm{ff}_\nu} \left(1 - \mathrm{e}^{-\tau^\mathrm{ff}_\nu(s_0,s_1)}\right),
\end{eqnarray}
and the luminosity can be obtained by integrating over the slice, 
namely, in the case of a face-on galaxy, 
via
\begin{eqnarray}
  L^\mathrm{ff}_\nu &=& \int_0^L \int_0^L I^\mathrm{ff}_\nu(0,L) ~\mathrm{d}x_1\mathrm{d}x_2,
\label{eq_Lnuff_fo}
\end{eqnarray}
where $L$ is the size of the numerical domain and the integration 
variable in Eq.~\ref{eq_tauff} is $s=x_3$. 
For an edge-on galaxy, the line of sight 
is parallel to the $x_1$ direction, namely,~$s=x_1$,
and the total free-free luminosity is:
\begin{eqnarray}
  L^\mathrm{ff}_\nu &=& \int_0^L \int_0^L I^\mathrm{ff}_\nu(0,L) ~\mathrm{d}x_2\mathrm{d}x_3.
\label{eq_Lnuff_eo}
\end{eqnarray}

\subsubsection{Total radio luminosity}

The total radio emission is composed of both free-free and synchrotron radiation
with relative contributions that depend on the frequency range. 
Furthermore, the geometry of the emitting galaxy plays a role in the sense that
the line of sight of the observation becomes relevant when including absorption 
processes. 
In this work, we consider the two extreme cases, face-on and edge-on observations, to
bracket reality.

Following \citet{WerhahnEtAl2021}, we use 
$\tau_\nu = \tau^\mathrm{ff}_\nu + \tau^\mathrm{SSA}_\nu$ to calculate the absorption of
synchrotron intensity at a position $s_1$ as:
\begin{eqnarray}
  I^\mathrm{synch}_\nu(s_0,s_1) &=&  \int_{s_0}^{s_1} j^\mathrm{synch}_\nu~e^{-\tau_\nu(s,s_1)}~\mathrm{d}s,
\label{eq_Isynch}
\end{eqnarray}
where $j^\mathrm{synch}_\nu$ is given in Eq.~\ref{eq_jsynchnu}.

Again, the total synchrotron luminosity depends on 
whether the galaxy is observed face-on, for which 
\begin{eqnarray}
  L^\mathrm{synch}_\nu &=& \int_0^L \int_0^L I^\mathrm{synch}_\nu(0,L) ~\mathrm{d}x_1\mathrm{d}x_2,
\label{eq_Lnusynch_fo}
\end{eqnarray}
with $s=x_3$ or edge-on, for which
\begin{eqnarray}
  L^\mathrm{synch}_\nu &=& \int_0^L \int_0^L I^\mathrm{synch}_\nu(0,L) ~\mathrm{d}x_2\mathrm{d}x_3
\label{eq_Lnusynch_eo}
\end{eqnarray}
with $s=x_1$ in Eq.~\ref{eq_Isynch}.

The total radio luminosity is
 \begin{eqnarray}
  L_\nu = L^\mathrm{synch}_\nu + L^\mathrm{ff}_\nu,
\label{eq_Lnu}
\end{eqnarray}
with $L^\mathrm{synch}_\nu$ and $L^\mathrm{ff}_\nu$
given by Eqs.~\ref{eq_Lnusynch_fo} and \ref{eq_Lnuff_fo} for
face-on observation and
Eqs.~\ref{eq_Lnusynch_eo} and \ref{eq_Lnuff_eo} for
edge-on observation, respectively.

\subsection{$q$ parameter}

To account for the stellar mass-dependent balance between 
obscured and unobscured SFR, when calculating the infrared luminosity,
we adopt the calibration of \citet{BernhardEtAl2014}:
\begin{eqnarray}
  L_\mathrm{IR}  = \frac{\SFR}{K_\mathrm{UV}} \frac{10^\mathrm{IRX}}{1+\frac{K_\mathrm{IR}}{K_\mathrm{UV}} 10^\mathrm{IRX}},
\label{eq_LFIR}
\end{eqnarray}
where 
$K_\mathrm{IR}= 1.7\times 10^{-10}~M_\odot \mathrm{yr}^{-1} L_\odot^{-1}$,
$K_\mathrm{UV}= 2.8\times 10^{-10}~M_\odot \mathrm{yr}^{-1} L_\odot^{-1}$,
and
\begin{eqnarray}
 \mathrm{IRX}= \alpha \left( \mathrm{log}\left(\frac{M_\star}{M_\odot}\right) -10.35\right) + \mathrm{IRX}_0
,\end{eqnarray}
with
$\alpha=0.71$ and $\mathrm{IRX}_0 =1.32$.
The calibration given in Eq.~\ref{eq_LFIR} is based on
a self-consistent of model the redshift evolution of the
UV and IR luminosity functions, which takes into account that the 
fraction of escaping UV photons increases with increasing redshift \citep[e.g.,][]{BurgarellaEtAl2013}.

The comparison between IR and radio luminosity is typically
expressed in terms of the logarithm of their ratio.
This value is named $q$ and has been defined in 
Eq.~\ref{eq_q}.
The luminosities in this definition are in the rest frame.

\section{Results and comparison with observations}
\label{sec_results}

\subsection{Radio emission and IR-radio correlation in the reference model}

In this section, we present results for the model 
that describes the observed dependence of the
IR-radio correlation on stellar mass best. 
The parameters of this reference model are 
given in Table~\ref{table_model_parameters}. 

\begin{figure}
\centering
   \subfigure{\includegraphics[width=0.45\textwidth]{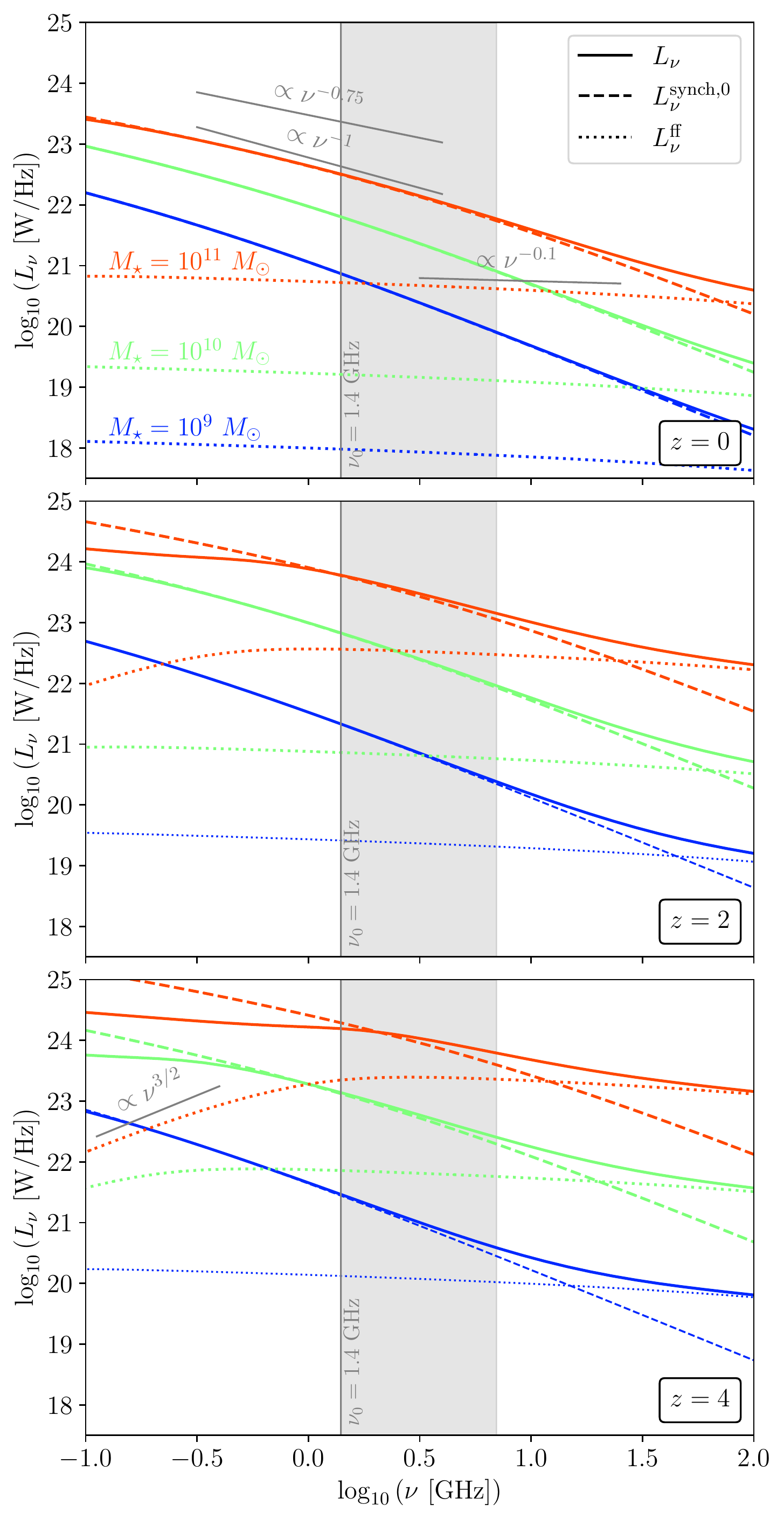}}
\caption{
Rest-frame spectral luminosities at different redshifts
for galaxies that are observed face-on.
Different colors indicate results from models with
different stellar masses: $M_\star=10^9~M_\odot$ (blue), 
$M_\star=10^{10}~M_\odot$ (green), and $M_\star=10^{11}~M_\odot$ (red).
Different linestyles refer to different emission types:
the total radio emission $L_\nu$ (solid lines) depends on
the free-free luminosity $L^\mathrm{ff}_\nu$ (dotted lines) 
and the unabsorbed synchrotron emission $L^\mathrm{synch,0}_\nu$ (dashed lines).
The grey vertical line represents $\nu_\mathrm{obs}=1.4~\mathrm{GHz}$
and the grey region indicates the range for redshifting
between $z=0-4$ (i.e.,~from left to right).
}
\label{fig_spec}
\end{figure} 

Radio spectra obtained for the reference model of
galaxies with stellar masses of $10^{9}M_\odot$, $10^{10}M_\odot$, and $10^{11}M_\odot$
that are observed face-on 
are shown in Fig.~\ref{fig_spec}.
The total radio emission $L_\nu$ includes contributions
from both free-free emission, $L^\mathrm{ff}_\nu$, 
the intrinsic synchrotron emission, $L^\mathrm{synch,0}_\nu$.
The absorption processes included in our model
affect the radio spectra mostly at low frequencies. 
Here, the free-free emission increases at very 
low frequencies of $\propto \nu^{3/2}$ 
(for most cases, this range is below the minimum frequency, 
as shown in Fig.~\ref{fig_spec})
and at high frequencies it decreases at $\propto \nu^{-0.1}$.
The spectrum of $L^\mathrm{synch,0}_\nu$ is close to
a power law, but mildly curved due to cosmic ray cooling processes, 
over the frequency range presented in Fig.~\ref{fig_spec}.
Such mildly curved synchrotron spectra have also 
been seen in the observations of \citet{KleinEtAl2018}.

\begin{figure}
\centering
   \subfigure{\includegraphics[width=0.45\textwidth]{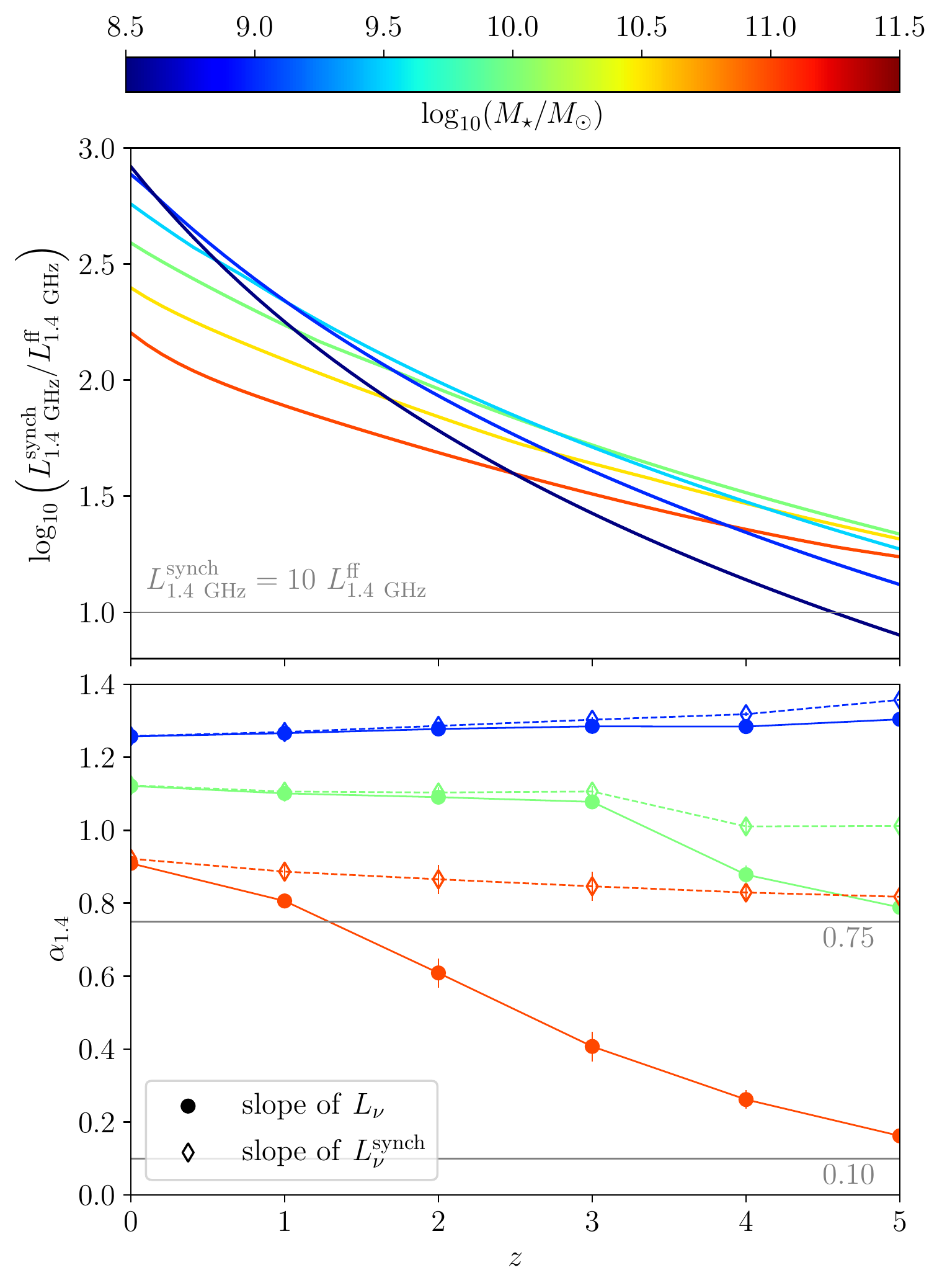}}
\caption{Analysis of radio spectra at $\nu=1.4~\mathrm{GHz}$.
The top panel shows the ratio of non-thermal 
($L^\mathrm{synch}_{1.4~\mathrm{GHz}}$) and thermal emission
($L^\mathrm{ff}_{1.4~\mathrm{GHz}}$) as a function of redshift.
The gray horizontal line indicates a thermal fraction 
of 10\% which is typically observed in local galaxies.
The bottom panel shows the result of a fit of the total radio
luminosity ($L_\nu$) and the synchrotron component 
($L^\mathrm{synch}_\nu$) to a function $f(\nu)\propto \nu^{-\alpha_{1.4}}$.
The fit is performed around $\nu=1.4~\mathrm{GHz}$.
}
\label{fig_slope14_z}
\end{figure}

For $z=0$, synchrotron emission at $\nu=1.4~\mathrm{GHz}$ dominates 
over free-free emission for all stellar masses
by at least two orders of magnitude.
Absorption affects the spectrum only for 
$\nu\lesssim 0.02~\mathrm{GHz}$ for~the case of
$M_\star=10^{10}~M_\odot$,
for instance, which is below the frequency range presented in Fig.~\ref{fig_spec}.
However, for $z=4,$ the slope of $L^\mathrm{synch,0}_\nu$ is affected
by absorption at $\nu\lesssim 30~\mathrm{GHz}$ for $M_\star=10^{10}~M_\odot$.
At high radio frequencies, the free-free emission dominates the spectra. 
For a galaxy with $M_\star=10^{10}~M_\odot$
this crossover from synchrotron to free-free dominance occurs at $\approx 30~\mathrm{GHz}$ at $z=0$ 
(see the upper panel of Fig.~\ref{fig_spec}). 
With increasing $M_\star$ and increasing $z$, the crossover point moves to lower $\nu$. 
In the KINGFISHER sample of nearby galaxies \citep{TabatabaeiEtAl2017}
the transition from synchrotron-dominated emission to free-free-dominated emission
occurs, on average around $\approx 10~\mathrm{GHz}$, yet the sample includes several 
galaxies with crossover frequencies that are comparable to our model.
The sample of local luminous infrared galaxies presented by \citep{DeyEtAl2022} 
includes several galaxies with crossover frequencies of $\gtrsim30~\mathrm{GHz}$.
Overall, our model predicts that the free-free contribution increases strongly with $z$ and 
starts affecting the slope of the total spectrum at
$\nu=1.4~\mathrm{GHz}$ already at $z\approx4$.
In the top panel of Fig.~\ref{fig_slope14_z}, the ratio of 
non-thermal ($L^\mathrm{synch,0}_\nu$) to thermal ($L^\mathrm{ff}_\nu$) 
luminosity at $\nu=1.4~\mathrm{GHz}$ is presented, demonstrating the clear dominance of $L^\mathrm{synch,0}_\nu$ 
at low $z$ for all $M_\star$ considered and the increase
in the non-thermal contribution with increasing $z$.
Therefore, especially at higher redshifts, 
one needs to be careful when shifting the observed spectrum 
to the rest frame for extracting 
the value of $L_{1.4}$ due to the change in the spectral slope.  

The slope of $L^\mathrm{synch,0}_\nu$ resulting from our model 
depends mostly on the slope of the proton injection spectrum $\alpha_\mathrm{CR}$. 
It can be estimated\footnote{The spectral synchrotron emission $j_{\nu}^\mathrm{synch}$ is given in 
Eq.~\ref{eq_jsynchnu} and is proportional to the emission $j_{\nu,\gamma_e}^\mathrm{synch}$ that corresponds to
electrons with a fixed Lorentz factor $\gamma_e$ and the number of CR electrons at a 
given energy, $N_e(\gamma_e)$, times $\gamma_e$.
From Eq.~\ref{eq_singlesynch}, which holds for a homogeneous magnetic field, 
we see that $j_{\nu,\gamma_e}^\mathrm{synch}\propto v/\gamma_e^2$ and from 
Eq.~\ref{eq_Ne} that $N_e(\gamma_e)\propto \gamma_e^{-\alpha_\mathrm{CR}}$, 
if the total cooling timescale does not depend strongly on $\gamma_e$.
Therefore, $j_{\nu}^\mathrm{synch} \propto \nu/\gamma_e^{\alpha_\mathrm{CR}+1}$.
Using Eq.~\ref{eq_gammamax} for the estimate of the energy that 
contributes mostly to the emission at a given frequency gives
$\gamma_e\propto\nu^{1/2}$ and therefore $j_{\nu}^\mathrm{synch} \propto \nu^{-(\alpha_\mathrm{CR}-1)/2}$.
For an inhomogeneous magnetic field
deviations from this scaling can be expected.}
as $-(\alpha_\mathrm{CR}-1)/2 = -1.0$ for $\alpha_\mathrm{CR}=3.0$, although
there can be effects on the slope from the electron energy losses 
that contribute differently at different frequencies. 
However, the reference model results in a scaling of synchrotron emission 
close to $\propto \nu^{-1.0}$ as indicated in Fig.~\ref{fig_spec}. 
Therefore, the resulting spectrum $L_\nu$ at $\nu=1.4~\mathrm{GHz}$ 
is steeper than the assumption 
used in the analysis of \citet{DelvecchioEtAl2020}, which is 
$L_\nu\propto \nu^{-0.75}$.
We show the results of power-law fits 
(with a fitting range of $0.43-4.5~\mathrm{GHz}$)
of the model spectra to 
both $L^\mathrm{synch,0}_\nu$ and $L_\nu$ 
at $\nu=1.4~\mathrm{GHz}$ in the lower panel of Fig.~\ref{fig_slope14_z}.
There, a clear dependence of the power law index $\alpha_{1.4}$
on $M_\star$ is demonstrated.  
We note that a dependence of the slope of radio spectra on the 
SFR, which in our model is coupled to 
$M_\star$,
has also been seen in observations \citep{TabatabaeiEtAl2017,KleinEtAl2018}.
Furthermore, Fig.~\ref{fig_slope14_z} shows the significant flattening 
of the radio spectra at $\nu=1.4~\mathrm{GHz}$
in high-mass galaxies due to the increase in the thermal contribution.
We note, however, that the opposite trend has been reported 
from observations 
\citep{AnEtAl2021,HeesenEtAl2022}, where radio spectra 
become steeper with increasing SFR, namely, increasing $M_\star$.
The discrepancy could be caused by the fact that we do 
not model the escape of CR electrons through galactic winds.
Another caveat of our reference model is that it produces 
a thermal fraction at $z=0$ at least an order of 
magnitude smaller than the observed fraction in local 
galaxies of $\approx10~\%$ \citep{TabatabaeiEtAl2017}. More details on this are shown in the
upper panel of Fig.~\ref{fig_slope14_z}. 
With increasing redshift, however, 
the reference model produces higher thermal fractions.

\begin{figure}
\centering
  \subfigure{\includegraphics[width=0.45\textwidth]{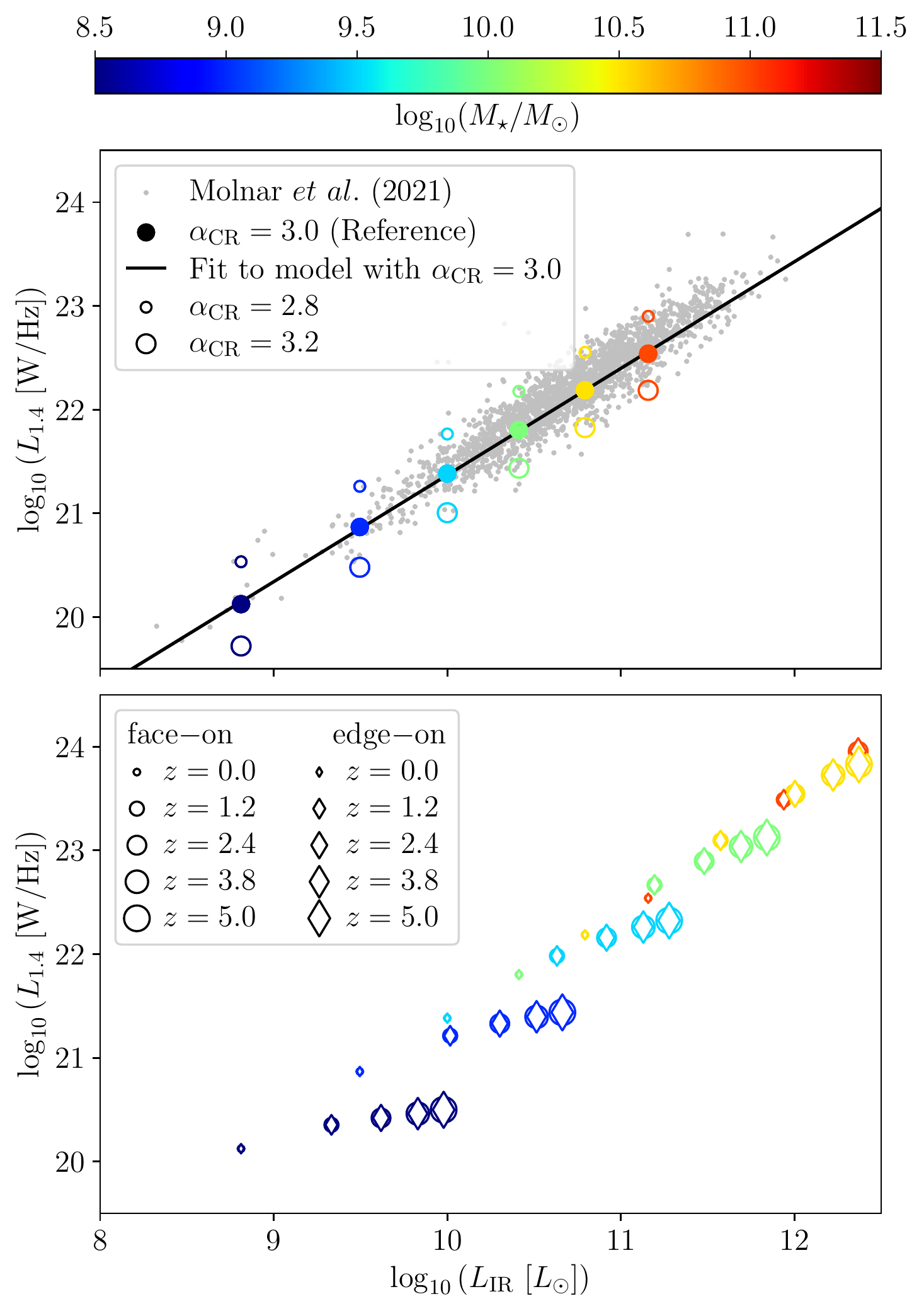}}
\caption{
Radio luminosity $L_{1.4}$ as a function of the IR luminosity 
$L_\mathrm{IR}$.
\textit{Top:}
 Solid dots show the result using the
default parameters of the model (with $\alpha_\mathrm{CR}=3.0$). 
The different colors are models with different 
stellar masses $M_\star$, according to the color 
bar.
The data in the background are local observations
of \citet{MolnarEtAl2021}.
The lines show the results of different fits to the 
reference model (see Eq.~\ref{eq_fit}).
The model predictions for different cosmic ray
spectral indexes, $\alpha_\mathrm{CR}=2.8$ and 
$\alpha_\mathrm{CR}=3.2$, are shown for comparison.
\textit{Bottom:}
Model predictions for $L_{1.4}$ and
$L_\mathrm{IR}$ for $\alpha_\mathrm{CR}=3.0$
at different redshifts as indicated but the marker size 
(smallest to biggest: $z=0-5$).
Different symbols indicate different treatments of 
absorption (circles: face-on observation,
diamonds: edge-on observation).
}
\label{fig_Lradio_LFIR}
\end{figure}

Next, we compare the radio emission at 
$\nu_\mathrm{obs}=1.4~\mathrm{GHz}$ and to the
IR emission obtained from Eq.~\ref{eq_LFIR}.
In the top panel of Fig.~\ref{fig_Lradio_LFIR}, 
we present the output of the 
reference model and compare it to the 
\citet{MolnarEtAl2021} data set of local sources. 
As indicated by the color of the large dots,
the model predicts low mass galaxies to be located
in the low luminosity end of the 
observed correlation and high mass galaxies in the 
intermediate/high luminosity end. 
The trend of the observed correlation as well as the absolute
value are in good agreement with our model at $z=0$.
The extensive catalog of low-redshift star-forming galaxies
presented by \citet{MolnarEtAl2021} indeed reveals 
a nonlinear IRRC with a slope of $1.11\pm0.01$. 
When fitting our model at $z=0$ with the following function:
\begin{eqnarray}
   \log_{10} \left(\frac{L_{1.4}}{\mathrm{W}~ \mathrm{Hz}^{-1} } \right) = m~\log_{10} \left( \frac{L_\mathrm{IR}}{3.75\times 10^{12} ~\mathrm{W}}\right) - b  
\label{eq_fit}
,\end{eqnarray}
we find a slope of $m=1.03\pm0.01$ and $b=2.92\pm0.10$.
We note that for $m=1$, $b=q$.
The correlation resulting from our model is, therefore, 
similar to the observed correlation,
namely:\ it is close to linear.
The fit to the model is presented as a black line in
the top panel of Fig.~\ref{fig_Lradio_LFIR}. 
In Appendix \ref{app_nonlin}, we present fits to the 
function (\ref{eq_fit}) for the modelled IRRC at 
$0\leq z \leq 5$.
To illustrate the dependence of the model
on the free parameters, the results for changing $\alpha_\mathrm{CR}$, 
while using the reference values for all other free parameters,
are presented in Fig.~\ref{fig_Lradio_LFIR}.  
We find that varying $\alpha_\mathrm{CR}$ between $2.8$ and
$3.2$ captures the entire observed scatter of the IRRC. 
However, changing other free parameters, like $f_B$ and $\alpha_\mathrm{CR}$,
has a similar effect on the resulting IRRC.
The systematics of the free parameters will be analyzed in detail
in Sect.~\ref{subsec_parameters}.

In the bottom panel of Fig.~\ref{fig_Lradio_LFIR}, 
we show the evolution of the luminosities with
redshift, as indicated by the increasing size of the markers. 
With increasing redshift, both 
$L_{1.4}$ and $L_\mathrm{IR}$ increase in our model, but the 
correlation remains conserved.
Even at the highest redshift considered in this plot, 
$z=5$, only small deviations from the correlation can be 
seen for high-mass galaxies (see the symbols of the largest size).
Here, the correlation becomes more nonlinear; for details, we refer to the discussion
on nonlinearity in Appendix~\ref{app_nonlin}.
Furthermore, there is no dependence on the line of sight for the
redshift range presented in Fig.~\ref{fig_Lradio_LFIR}.
This is not surprising because the only way differences between 
face-on and edge-on cases can arise in our current model is through 
free-free and synchrotron self-absorption 
processes, namely, opacity differences arising from different
path lengths.
Stronger variations with respect to the line of sight can be
expected in models that account for cosmic ray transport and 
ordered large-scale magnetic fields,
which is beyond the scope of the present study. 
Since there seems to be no 
dependence on galaxy orientation up to $z=5,$ we 
use the face-on assumption as the default unless indicated otherwise. 

\begin{figure}
\centering
  \subfigure{\includegraphics[width=0.45\textwidth]{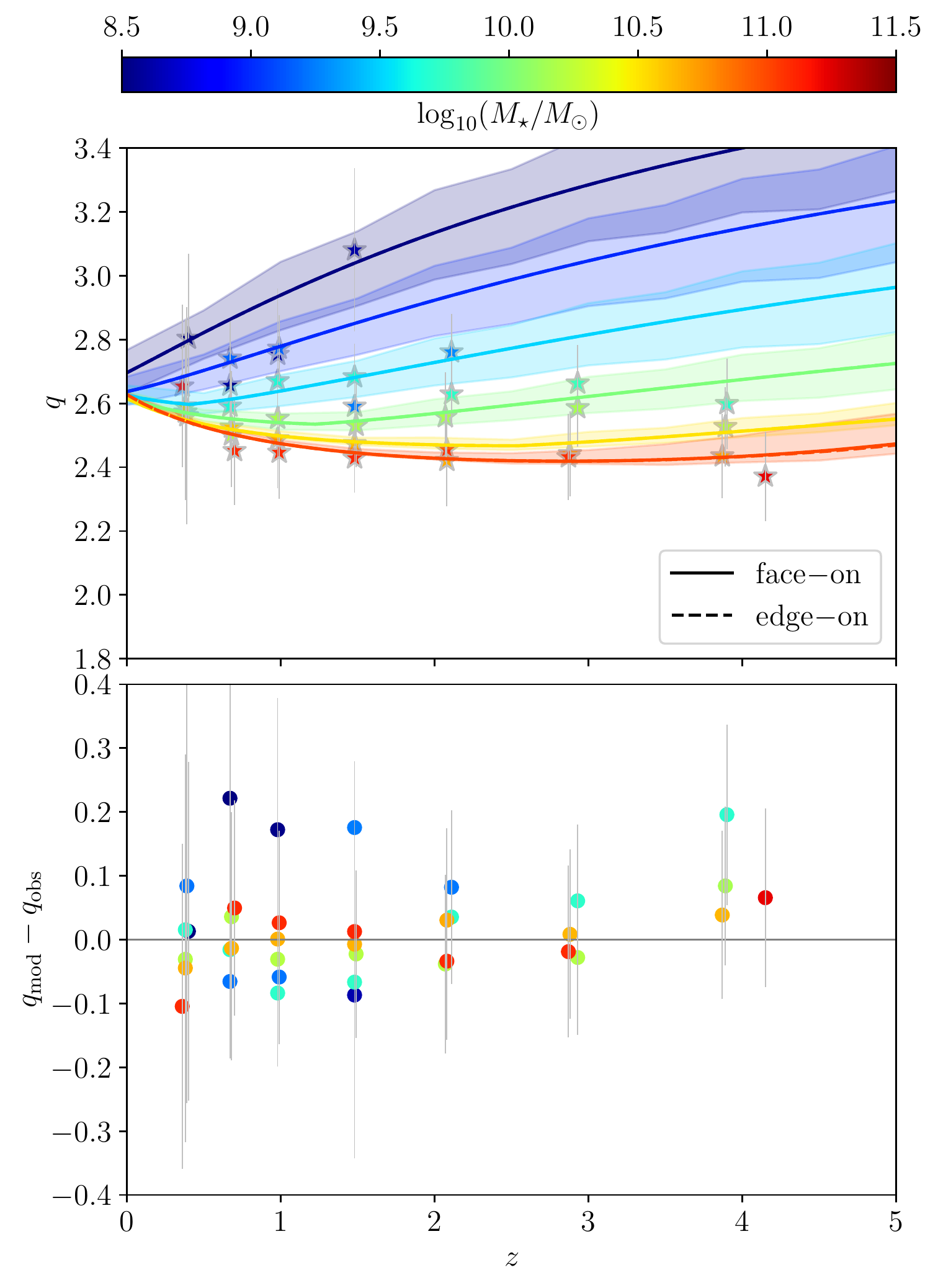}}
\caption{
 $q$ parameter as defined in Eq.~\ref{eq_q} versus~redshift~$z$.
\textit{Top:}
Lines show the results of our reference galaxy model and
the stars with grey contours are observations 
\citep[AGN-corrected values from][]{DelvecchioEtAl2020}.
Colors refer to different stellar masses $M_\star$
as indicated by the color bar.
For solid lines the galaxies are observed face-on and for
dashed lines edge-on.
Shaded regions indicate the standard deviation.
\textit{Bottom:} Difference between the observed values of
$q_\mathrm{obs}$ and the corresponding model predictions $q_\mathrm{mod}$.
The errors include only the uncertainties of the observational data.
}
\label{fig_q_z}
\end{figure}

In Fig.~\ref{fig_q_z}, the redshift evolution of $q$ for the
reference model is directly compared to the 
AGN-corrected observations of \citet{DelvecchioEtAl2020}.
Generally, there is an agreement between model and observation:
With the exception of galaxies with low $M_\star$, the
value of $q$ is nearly independent of redshift, but there
is a spread of $q$ for different masses. 
As for the observed data set, our reference model results in 
a spread of $q$ from $q\approx 2.4$ for $M_\star \approx10^{11}~M_\odot$
up to, for $z\lesssim2$, $q\approx 3.2$ for $M_\star\approx 10^{8.5}~M_\odot$.
We note that in the  $z=0-1 $ range, the spread of $q$ vanishes for masses of
$M_\star \gtrsim 10^{10}~M_\odot$ which is also seen in the observation.
Towards the highest $z$ shown in Fig.~\ref{fig_q_z} the spread
between different masses becomes larger.
We note that the 1-$\sigma$ uncertainty regions 
(as indicated by the shaded regions in the top panel of Fig.~\ref{fig_q_z}), caused by uncertainties in the input empirical relations,
are barely overlapping.
Even at low stellar masses, where the error bands are the
largest, we find a systematic trend of $q$ with $M_\star$.
This strengthens our conclusion that the IRRC is indeed dependent on
the stellar mass (see also the comparison to $M_\star$-agnostic observations of the IRRC in Appendix \ref{sec_otherobs}).
Finally, it is worth mentioning that our model predicts different
redshift dependencies of $q$ for different values of $M_\star$. 
This is in contrast to the fitting formula used in \citet{DelvecchioEtAl2020}, 
who report the same redshift dependence of $(1+z)^{-0.023}$ for the mass bins
$10.5<\mathrm{log}_{10}(M_\star/M_\odot)<11$ and $11<\mathrm{log}_{10}(M_\star/M_\odot)<12$.
Deeper radio and IR observations, and complete input galaxy samples, 
capable of homogeneously probing all stellar masses 
$M_{\star}\,{\geq}\,10^9\,M_{\odot}$ over a larger redshift range,
are required to establish the true level of complexity present 
in the mass dependence of the IRRC out to higher redshift.

\subsection{Dependence model parameters}
\label{subsec_parameters}

So far, we have discussed the results of the reference model
for which all free parameters have been fixed to values that are listed 
in Table~\ref{table_model_parameters} and that 
produce $q(z,M_\star)$ curves that are in good agreement with the observed IRRC. 
However, some of the free parameters in our model are not well 
constrained and a systematic exploration of the parameter space 
is required. 
The goal of this section is to analyze the sensitivity of the 
model to the various free parameters and to identify the most 
crucial dependencies.

\subsubsection{Ratio of synchrotron to free-free emission}
\label{sec_synch_ff}

\begin{figure}[t]
\centering
  \includegraphics[width=0.5\textwidth]{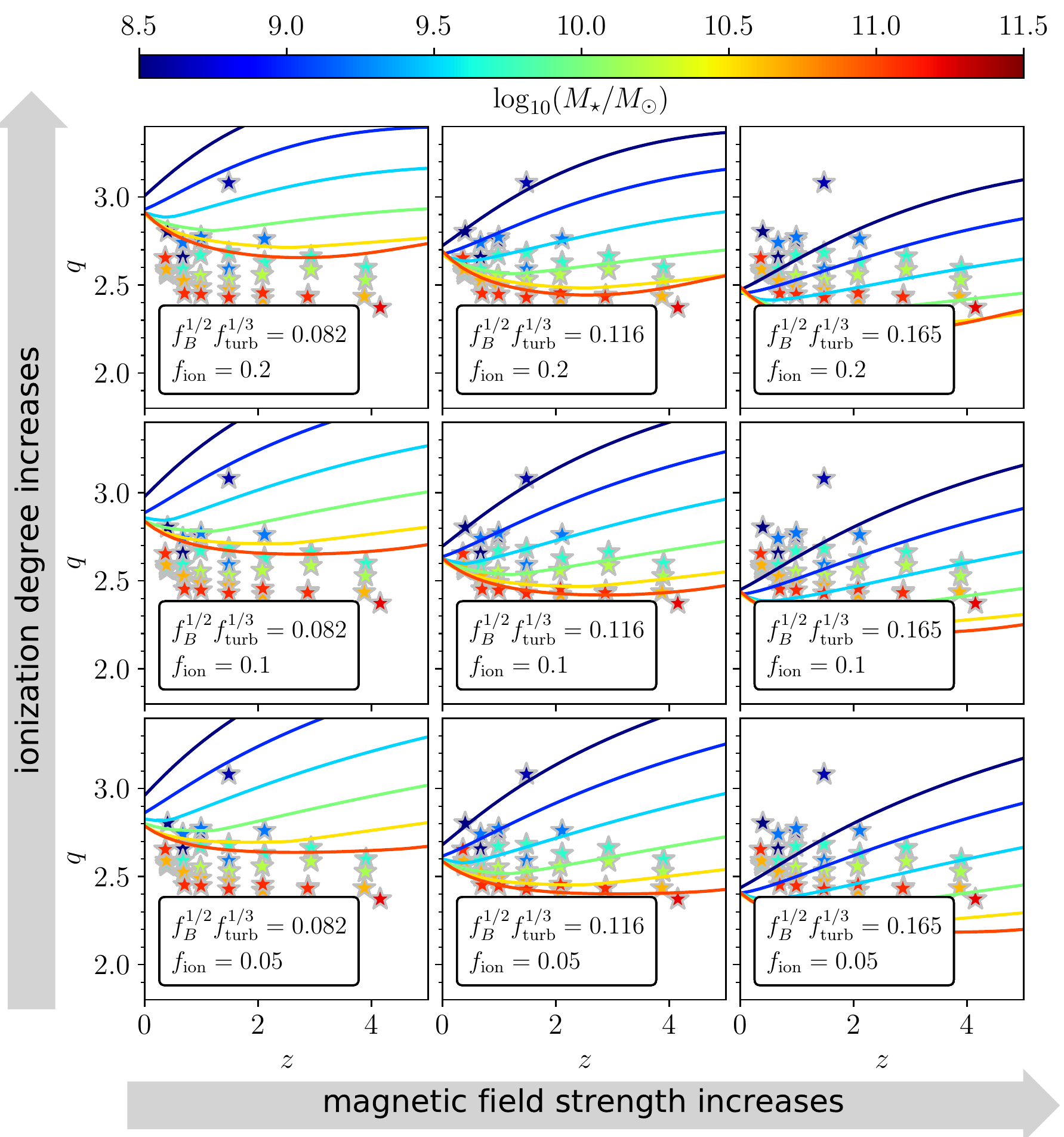}
\caption{
 $q$ parameter as a function of redshift, $z$, for galaxies
with different stellar mass, $M_\star$, as given in the colorbar. 
The reference model compared to the observed data set of
\citet{DelvecchioEtAl2020} is presented in the middle panel.
In panels from left to right, the magnetic field strength in the model
is increased; see the discussion around Eq.~\ref{eq_B} for the parameters 
$f_B$ and $f_\mathrm{turb}$. 
From bottom to top, the ionization degree is increased via
the parameter $f_\mathrm{ion}$ which affects the 
free-free emission; see Sect.~\ref{sec_freefree}.
}
\label{fig_q_z__fB_fion}
\end{figure} 

In Fig.~\ref{fig_q_z}, we show that in the reference model, 
the spread of $q$
with respect to $M_\star$ increases with increasing $z$.
This might be caused by the transition from the dominance of 
nonthermal to thermal emission (see Figs.~\ref{fig_spec} 
and~\ref{fig_slope14_z}).
With model series A (see Table~\ref{tab_models}), we tested the
role of the ratio of free-free to synchrotron 
emission in the dependence of $q$ on $M_\star$ This series includes the reference model plus eight models in 
which the magnetic field strength and the ionization degree
are changed systematically. 
The magnetic field strength determines the
synchrotron flux and is controlled by the parameters
$f_B$ and $f_\mathrm{turb}$ (see Eq.~\ref{eq_B})
and the ionization degree $f_\mathrm{ion}$ 
determines the free-free emission (see Eq.~\ref{eq_Lff}).

In the individual panels of Fig.~\ref{fig_q_z__fB_fion}, 
we present $q(z,M_\star)$ for different 
values of $f_B$, $f_\mathrm{turb}$, and $f_\mathrm{ion}$.
The reference model is presented in the center panel. 
From left to right the magnetic field strength increases
as we increase the value of $f_B^{1/2} f_\mathrm{turb}^{1/3}$.
As expected, with smaller $B$ the $q(z)$ curves move up, because 
$L_{1.4}$ decreases. 
Additionally, we find that the dependence of 
$q(z,M_\star)$ on $M_\star$ becomes stronger as  
$f_B^{1/2} f_\mathrm{turb}^{1/3}$, namely,~the magnetic field strength 
decreases. 

From bottom to top, we show models with different values
of the ionization degree. 
Lowering $f_\mathrm{ion}$ affects the $q(z,M_\star)$
mostly at higher redshifts as expected. 
Generally, we find that the $q(z,M_\star)$ curves
tend to become closer again at high $z$, if the value of 
 $f_\mathrm{ion}$ is higher.
In conclusion, the dependence of $q(z,M_\star) $
on $M_\star$ seems to be stronger if the radio flux is dominated by
synchrotron emission.

\subsubsection{CR model}
\label{sec_CR}

\begin{figure}[t]
\centering
  \includegraphics[width=0.5\textwidth]{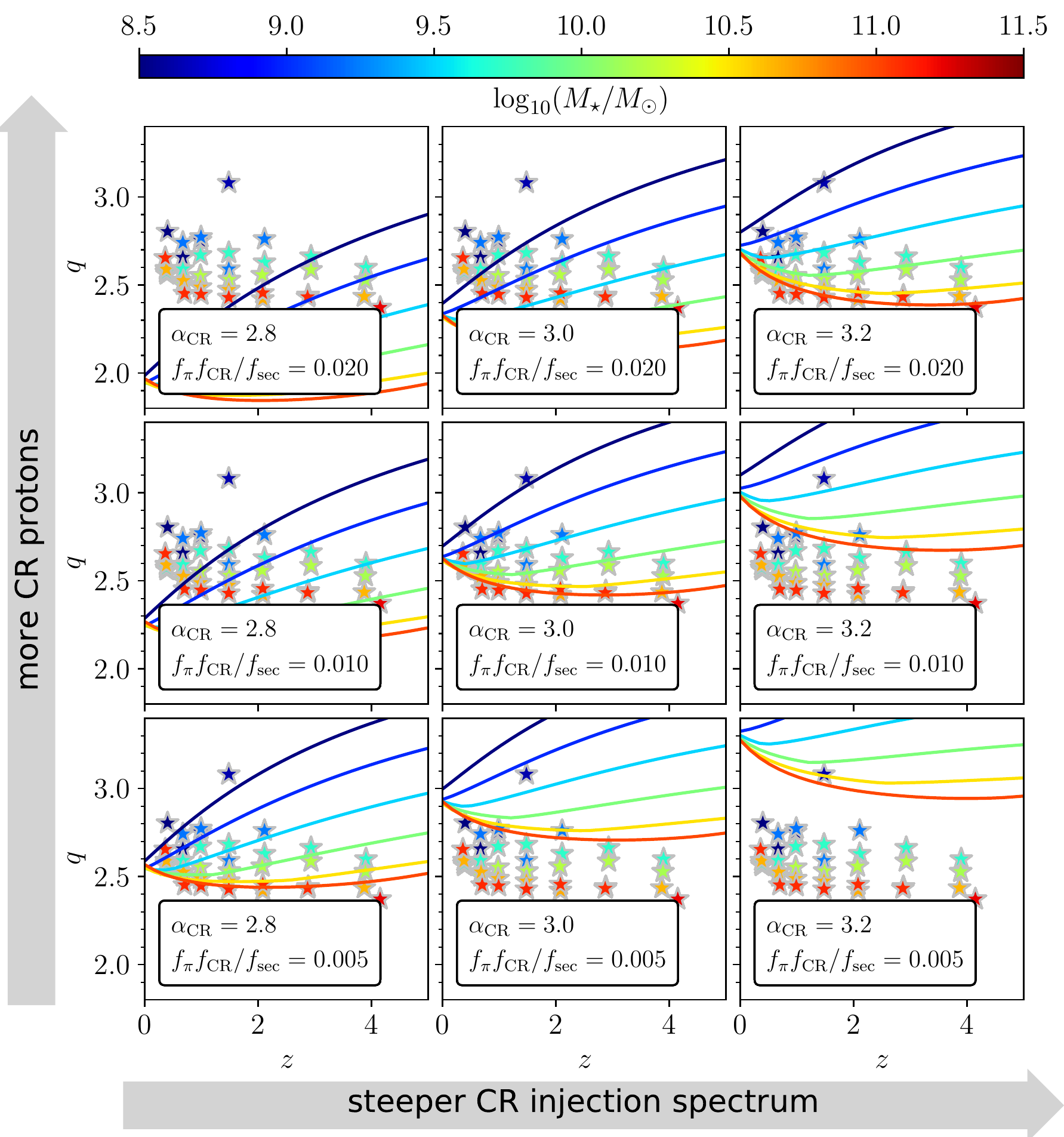}
\caption{
 $q$ parameter as a function of redshift, $z$, for galaxies
with different stellar mass, $M_\star$, as given in the colorbar. 
The reference model compared to the observed data set of
\citet{DelvecchioEtAl2020} is presented in the middle panel.
From left to right, 
the slope of the CR injection spectrum gets steeper
via the parameter $\alpha_\mathrm{CR}$;
see Eq.~\ref{eq_Q}. 
From bottom to top, the number of CR protons in the model is increased
via the parameters $f_\pi$, $f_\mathrm{CR}$, and $f_\mathrm{sec}$;
see Eq.~\ref{eq_Ne}.
}
\label{fig_q_z__CR_sISRF}
\end{figure} 

Our cosmic ray model includes several assumptions and parameters 
that are not very well constrained. 
Therefore, we explored the dependence of our model on these parameters
in series B (see Table~\ref{tab_models}).
In particular, we changed the value of $f_\pi f_\mathrm{CR}/f_\mathrm{sec}$,
which determines the number of CR protons $Q_{p,0}$ (see Eq.~\ref{eq_Qp0}).
The number of CR electrons which ultimately produce the
synchrotron emission depends on $Q_{p,0}$ but also 
on the cooling time
scale, and the slope of the injection spectrum $\alpha_\mathrm{CR}$ (see Eq.~\ref{eq_Q}).
As discussed before, the slope $\alpha_\mathrm{CR}$ 
is strongly connected
to the slope of the radio spectrum. 
In series B, we changed both $f_\pi f_\mathrm{CR}/f_\mathrm{sec}$ and
$\alpha_\mathrm{CR}$ systematically to explore the effect of the 
key parameters of the CR model.

The dependence of $q(z,M_\star)$ on the number of 
CR protons and the slope of the injection spectrum 
is presented in Fig.~\ref{fig_q_z__CR_sISRF}.
Again, the reference model is shown in the center panel of the 
figure.
For both, decreasing the slope of the CR injection spectrum 
and the total number of CR protons leads to a 
shift of the $q(z,M_\star)$
curves toward lower values.
The origin of the decrease in  $q(z,M_\star)$ for an increase in the
total number of CR protons is trivial since more CR protons imply more 
CR electrons and, therefore, more synchrotron emission. 
For a steeper $N_e(\gamma_e)$ spectrum, 
$N_e(\gamma_e^\mathrm{max})$ is smaller, which implies that
there are fewer electrons 
to emit synchrotron
at $\nu_\mathrm{obs}=1.4~\mathrm{GHz}$. 
Hence, the radio emission,~$L_{1.4}$, decreases for an increasing
$\alpha_\mathrm{CR}$ and consequently $q$ increases. 
We do not find significant changes in the overall shape of the
$q(z,M_\star)$ curves; namely,~there is no change in the $M_\star$ 
and $z$ trend and only the absolute values are shifted.
We note that Fig.~\ref{fig_q_z__CR_sISRF} indicates that in
our model, increasing the total number of CR protons has a 
very similar effect to flattening the CR injection spectrum.

\subsubsection{Galactic radius and scale height models}
\label{sec_Hgal}

In this section, we present the dependence of the 
resulting IRRC on our models of the galactic radius
$R_\mathrm{gal}$
and scale height $H_\mathrm{gal}$. 
The default evolution is given in Eqs.~\ref{eq_Rgal}
and~\ref{eq_Hgal}, however, in
series C and D we test alternative scenarios.

\begin{figure}
\centering
\subfigure{\includegraphics[width=0.45\textwidth]{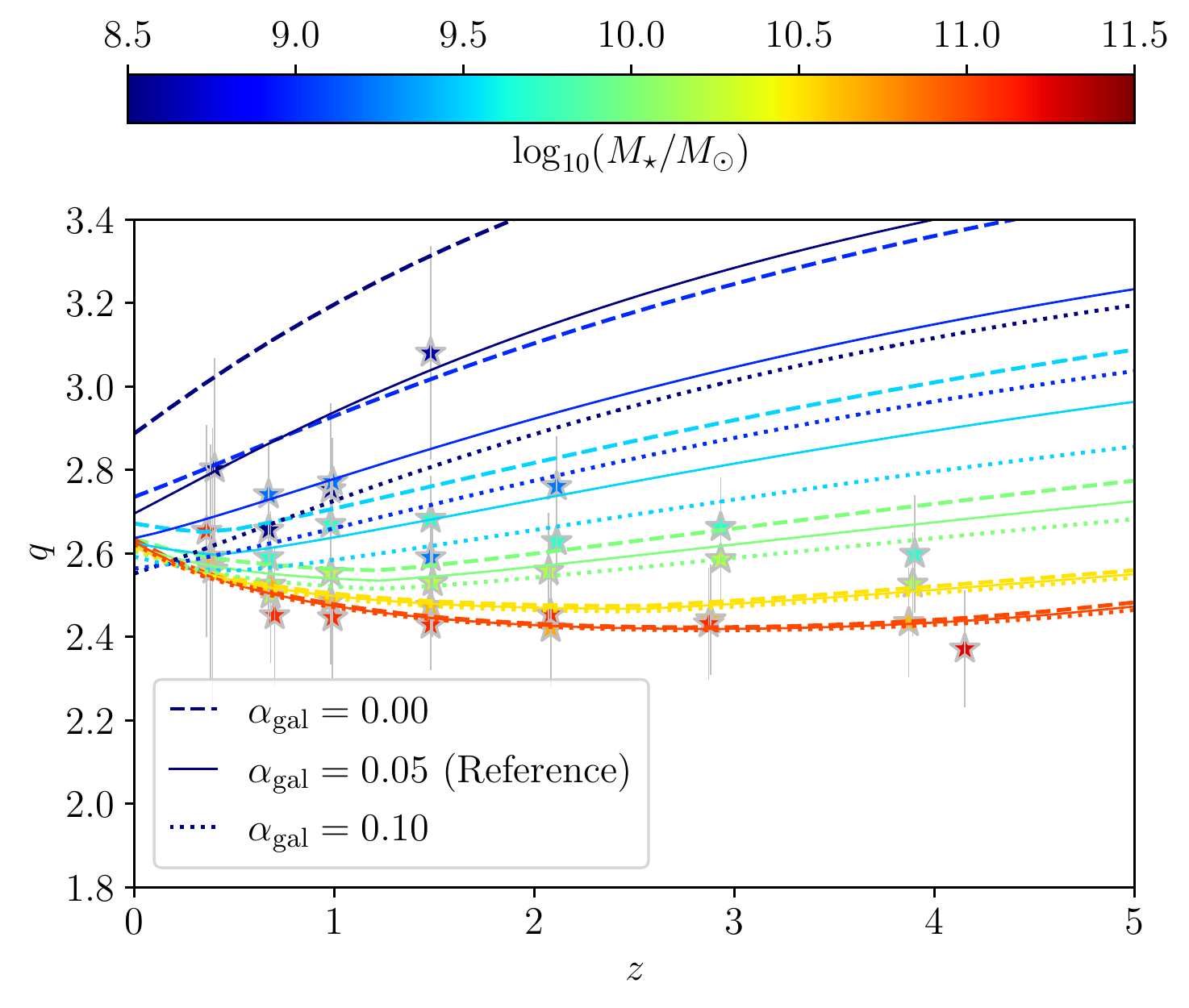}}
\caption{
Different dependence of the galactic radius on stellar mass
(by variation of $\alpha_\mathrm{gal}$, see Eq.~\ref{eq_Rgal})
compared to the reference model ($\alpha_\mathrm{gal}=0.05$).}
\label{fig_q_z__chi}
\end{figure} 

The results for series C are presented in Fig.~\ref{fig_q_z__chi},
where the strong effect of $\alpha_\mathrm{gal}$ on the spread of
the $q(z,M_\star)$ is evident. 
While our reference model ($\alpha_\mathrm{gal}=0.05$) reproduces
the observed data best
within the uncertainty regime,
lower values of $\alpha_\mathrm{gal}$ 
result in a stronger $M_\star$ dependence of $q$. 
For stronger scaling of $R_\mathrm{gal}$ with $M_\star$,
for instance,~ for $\alpha_\mathrm{gal}=0.1$ the $q(z,M_\star)$ curves 
are closer together.

\begin{figure}
\centering
\subfigure{\includegraphics[width=0.45\textwidth]{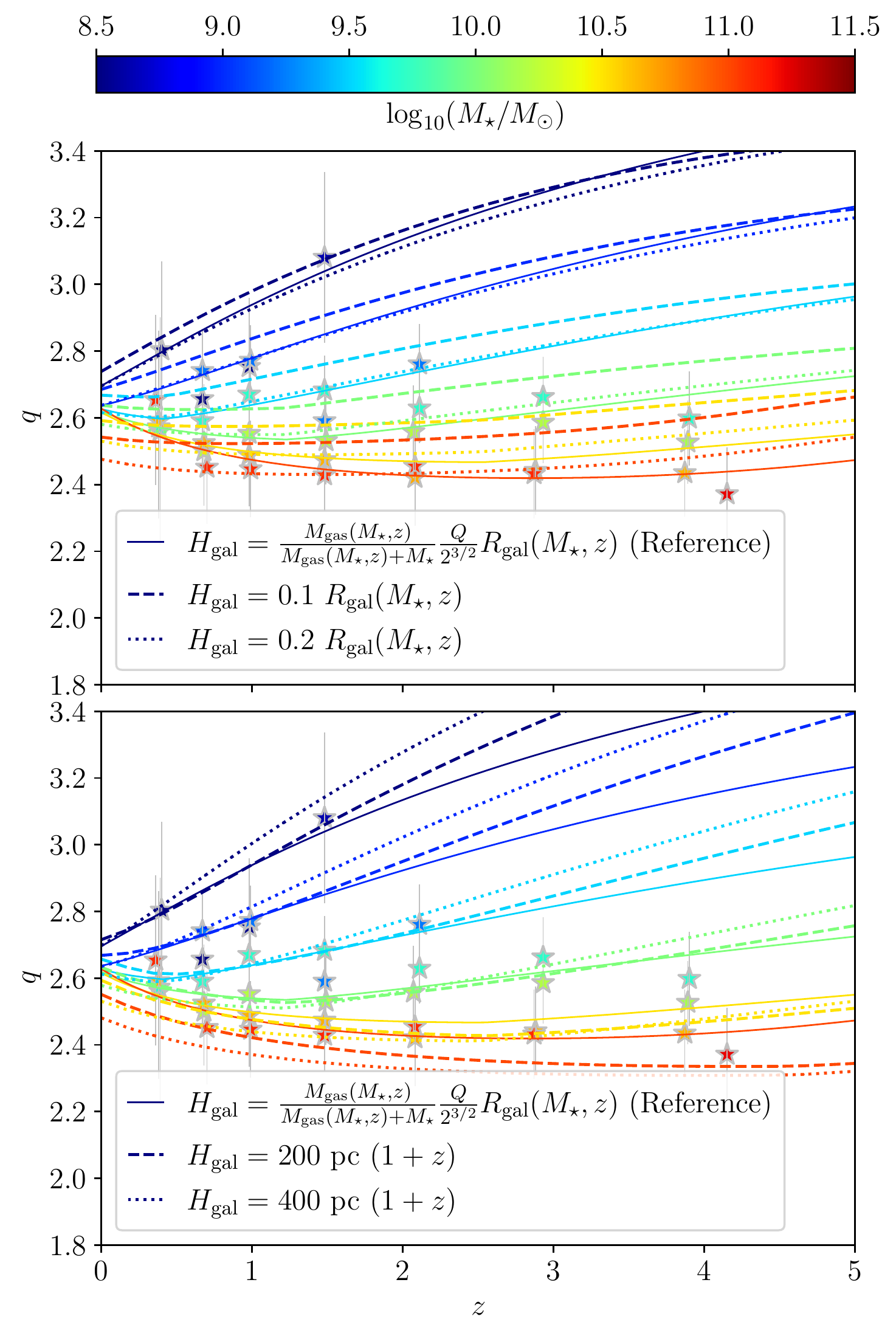}}
\caption{
Different models of the galactic scale height $H_\mathrm{gal}$ compared to the reference model.
\textit{Top:} $H_\mathrm{gal}$ scales with the characteristic galactic
radius $R_\mathrm{gal}$ as modeled in Eq.~\ref{eq_Rgal}.
\textit{Bottom:} 
Model in which $H_\mathrm{gal}$ 
increases with redshift.  
}
\label{fig_q_z__Hgal}
\end{figure} 

The model for $H_\mathrm{gal}$ seems to affect the
resulting $q(z,M_\star)$ less, at least for the scenarios we present
in Fig.~\ref{fig_q_z__Hgal}.
In the upper panel of Fig.~\ref{fig_q_z__Hgal}, a scaling of 
$H_\mathrm{gal}$ with the galactic radius is assumed and appears
to result is a similar scaling of $q(z,M_\star)$ as the 
reference model, even though the $M_\star$ dependence at $z\approx 0$ 
is slightly stronger for the models with $H_\mathrm{gal}\propto R_\mathrm{gal}$.
Therefore, at low $z$, the reference model is closer to the 
observed data.
A scaling of $H_\mathrm{gal}\propto (1+z)$ is presented in the 
lower panel of Fig.~\ref{fig_q_z__Hgal}, results in a good
agreement with observations at intermediate redshifts. 
At high $z$, however, the $q(z,M_\star)$ curves increase faster 
for galaxies with small $M_\star$ if $H_\mathrm{gal}\propto (1+z)$.

\subsection{Extrapolation to high redshift}
\label{sec_highz}

\begin{figure}
\centering
  \subfigure{\includegraphics[width=0.45\textwidth]{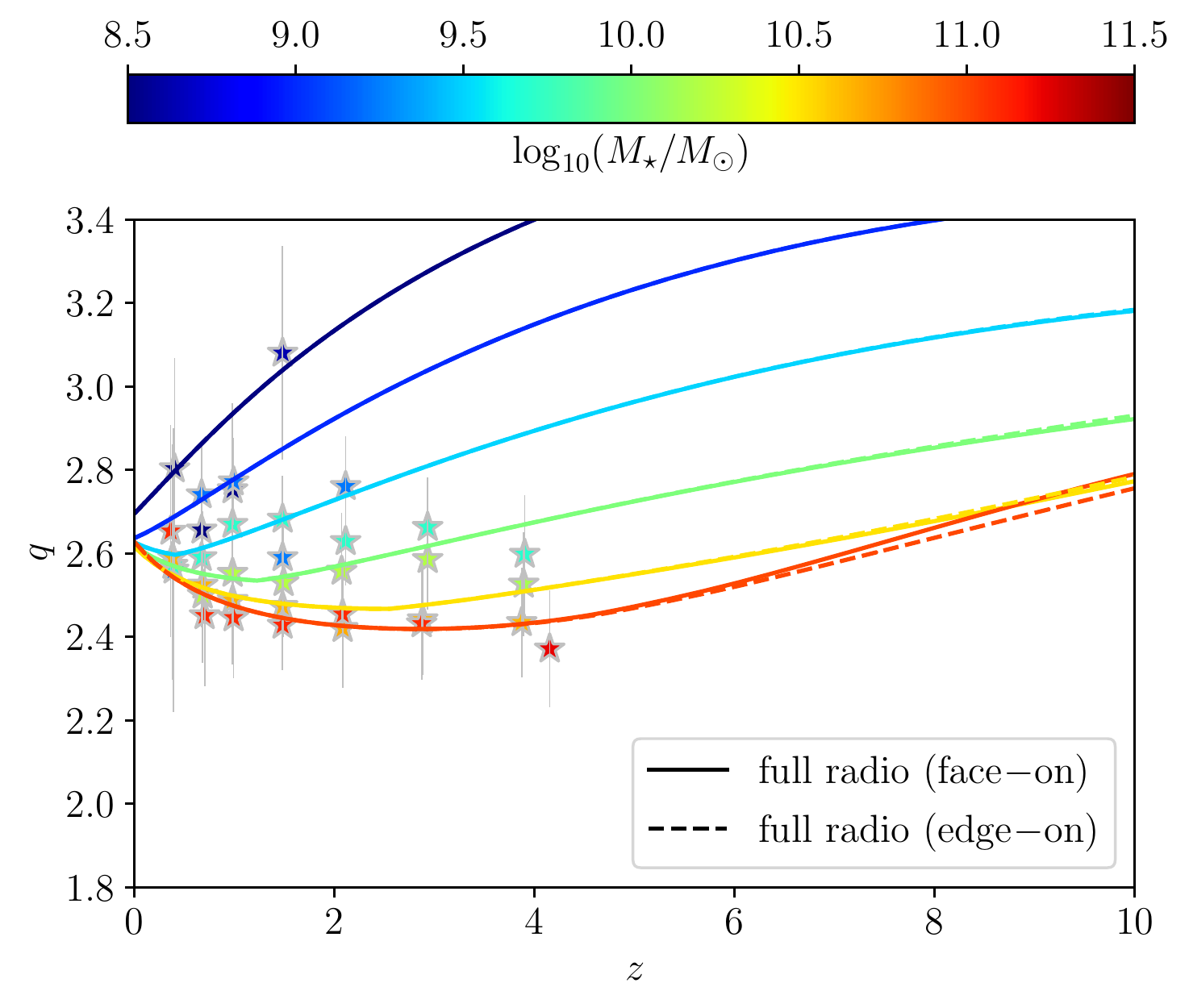}}
\caption{
Extension of the reference model to high $z$.
}
\label{fig_q_z_highz}
\end{figure} 

The reference model describes the observed IR-radio correlation
well up to the maximum redshift of $z=4$. 
In principle, the model can be extrapolated to redshifts beyond that. 
However, we note that some of the input scaling relations for the 
model as well as the assumption of well-settled, double-exponential disks 
might break down at extreme $z$ and therefore the results
of this section are rather speculative. 
We note, however, that early observations with the James Webb Space Telescope (JWST) 
reveal that disk galaxies
dominate the population even at $z > 1.5$
\citep{FerreiraEtAl2022}.

Another assumption that will break down at high redshifts is the saturation of the small-scale dynamo. 
While the kinematic phase is very short
\citep{Kazantsev1968,KulsrudAnderson1992,RogachevskiiKleeorin1997,SchoberEtAl2012.1}, at $z\gg4$ the dynamo 
could still be in the nonlinear stage. 
In this phase \citep{Beresnyak2012}, saturation is reached at the smallest length
scales of the turbulent inertial range and there is an inverse cascade
of magnetic energy to larger scales.
Semi-analytical models by \citet{SchoberEtAl2013} predict
that the dynamo saturates on time scales of
$4-270$ Myr, depending on the 
driving mechanism of turbulence. 

In Fig.~\ref{fig_q_z_highz}, the $q(z, M_\star)$ curves are 
shown up to $z=10$. 
The redshift and mass dependence predicted by our model does not 
significantly change at higher $z$.
However, with absorption processes becoming more relevant in 
high-$z$ galaxies, the observed radio flux depends
on whether the galaxy is face-on or edge-on. 
For galaxies with high $M_\star$, this orientation effect becomes important 
at $z\gtrsim 6$.
We note, however, that our model only includes orientation effects for the radio emission. 
A more detailed model of the IR emission that takes into account the
dust absorption/emission with respect to the orientation of the disk
might affect the high-$z$ evolution of $q$ \citep[see also][for an observational study of the difference between the IR-to-radio 
luminosity ratio in local face-on and edge-on galaxies]{LiEtAl2016}.

The interesting trends predicted by the model
at $6 \gtrsim z\gtrsim 8$ will be testable in the near
future.
Based on the currently proposed SFR-$L_{1.4}$ calibrations for 
galaxies at intermediate and high redshift, objects with SFR in 
the range ${\sim}5-10\,M_{\odot}$/yr are expected to be 
detectable at $z\,{\sim}$\,6 in the deepest future Square Kilometre Array (SKA) 
continuum surveys (using the parameters of the SKA continuum 
reference surveys as outlined in \citealt{PrandoniSeymour2015}). 
For $z\,{\sim}$\,6 main sequence galaxies \citep[e.g.,][]{DuncanEtaAl2014,SalmonEtAl2015},
this would imply an accessible stellar mass 
range of ${\gtrsim} 5{\times}10^9\,M_{\odot}$.

\section{Discussion}
\label{sec_discussion}

\subsection{Novel interpretations of the IRRC with our semi-analytical model}

The strength of a simplified galaxy model as presented in this paper is
that it allows
for an exploration of a large parameter range and tests of
different scenarios of galaxy evolution. 
This makes it possible to identify the underlying physical processes
behind observed correlations, such as the IRRC.

The dependence of the IRRC on stellar mass, $M_\star$, for instance arises
from the empirical input relations (see Sect.~\ref{sec_global}) 
that make up the foundation of our model.
The SFR rate in Eq.~\ref{eq_SFR} that is used as an input is not a 
simple power-law, but the $M_\star$ dependence can be roughly 
approximated as $\dot{M}_\star \propto M_\star^{0.8}$.
Therefore the gas mass, as given in Eq.~\ref{eq_Mgas} scales as ${M}_\mathrm{gas} \propto M_\star^{0.65}$.
Together with $R_\mathrm{gal}\propto M_\star^{\alpha_\mathrm{gal}}$ (Eq.~\ref{eq_Rgal}) and 
$H_\mathrm{gal}\propto R_\mathrm{gal}$ (Eq.~\ref{eq_Hgal}),
the gas density (Eq.~\ref{eq_ngas}) scales 
as $n_\mathrm{gas} \approx M_\mathrm{gas}/(R_\mathrm{gal}^2 H_\mathrm{gal})\propto M_\star^{0.65-3\alpha_\mathrm{gal}}$. 
This results in a scaling of the magnetic field strength of
$B\propto(\dot{n}_\mathrm{SN} H_\mathrm{gal})^{1/3}n_\mathrm{gas}^{1/6} \propto M_\star^{(2.25-\alpha_\mathrm{gal})/6}$, 
where we use $\dot{n}_\mathrm{SN}\propto\dot{M}_\star/V_\mathrm{gal}$.
The number of cosmic ray electrons as a function of the
Lorentz factor is given in Eq.~\ref{eq_Ne}.
For synchrotron emission, the number of cosmic rays with a
Lorentz factor $\gamma_e^\mathrm{max}$ is relevant. 
This number is $N_e^\mathrm{max} \propto Q_{p,0} (\gamma_e^\mathrm{max})^{-\alpha_\mathrm{CR}}\tau_e$.
For our model, we find that $Q_{p,0} \propto M_\star^{0.8}$ and
$\gamma_e^\mathrm{max}\propto M_\star^{-0.19 +0.59\alpha_\mathrm{gal}}$, 
while the CR cooling time, $\tau_e$, is dominated by different processes 
for different stellar masses.
In low-mass galaxies and low-to-moderate redshifts,
typically $\tau_e \approx \tau_\mathrm{synch} \propto M_\star^{-0.57+1.75\alpha_\mathrm{gal}}$;
whereas for high-mass galaxies, $\tau_e \approx \tau_\mathrm{ion} \propto M_\star^{-0.84+3.59\alpha_\mathrm{gal}}$.
Therefore, we find
\begin{equation}
  N_e^\mathrm{max} \propto 
  \begin{cases} 
     M_\star^{0.23 + 0.19\alpha_\mathrm{CR}+1.75\alpha_\mathrm{gal} -0.59\alpha_\mathrm{CR}\alpha_\mathrm{gal}} & \mathrm{for~low~mass~galaxies,}  \\
     M_\star^{-0.04 + 0.19\alpha_\mathrm{CR}+3.59\alpha_\mathrm{gal} -0.59\alpha_\mathrm{CR}\alpha_\mathrm{gal}}  &  \mathrm{for~high~mass~galaxies.}
  \end{cases}
\end{equation}
The $1.4$ GHz emission can then be estimated as 
$L_{1.4}\propto j_\nu^\mathrm{synch} R_\mathrm{gal}^2 H_\mathrm{gal}$, 
assuming that it is dominated by synchrotron emission. 
According to Eq.~\ref{eq_jsynchnu}, we have $j_\nu^\mathrm{synch}\propto B~ N_e^\mathrm{max}$
and we find the following for the regimes with different CR cooling channels:
\begin{equation}
  L_{1.4}\propto 
  \begin{cases} 
     M_\star^{0.61 + 0.19\alpha_\mathrm{CR} + 3.58\alpha_\mathrm{gal} -0.59\alpha_\mathrm{CR}\alpha_\mathrm{gal}} & \mathrm{for~low~mass~galaxies,}  \\
     M_\star^{0.34 + 0.19\alpha_\mathrm{CR}+5.42\alpha_\mathrm{gal} -0.59\alpha_\mathrm{CR}\alpha_\mathrm{gal}}  &  \mathrm{for~high~mass~galaxies.}
  \end{cases}
\end{equation}
For our reference model with $\alpha_\mathrm{gal}=0.05$ and $\alpha_\mathrm{CR}=3$, we find
$L_{1.4}\propto M_\star^{1.27}$ and $L_{1.4}\propto M_\star^{1.09}$ for low- and high-mass galaxies respectively. 
Therefore, in both cases, radio emission increases more strongly 
with $M_\star$ than the IR emission that scales as
$L_{IR}\propto \dot{M}_\star \propto M_\star^{0.8}$ (see Eq.~\ref{eq_LFIR}).
Hence, we expect $q$ to decrease with increasing $M_\star$.
This trend is consistent with the results of the full model presented in 
this work and is, for instance, presented in Fig.~\ref{fig_q_z}.
Additionally, we expect the dependence of $q$ on $M_\star$ to be less 
if $\alpha_\mathrm{gal}$ is larger,
which can also be seen in Fig.~\ref{fig_q_z__chi}.

Calculations with the full model, which are presented in Sect.~\ref{sec_results},
are needed to understand the detailed dependence of $q$ on $M_\star$ and $z$.
In particular, the contributions of free-free emission and different cooling
mechanisms of cosmic ray electrons cannot be included in a simple estimate as
discussed in the previous paragraph.

\subsection{Comparison to other models for the IRRC}

The first attempts at understanding the observed IRRC
were one-zone models. 
The so-called calorimeter theory by \citet{Voelk1989} was based on the assumption that CR electrons lose their energy via synchrotron emission before escaping the galaxy
and that the galaxy is optically thick in the UV, such that emission by young stars is absorbed by dust
and re-radiated in the FIR.
This calorimeter model results in a tight IRRC but
the slope of the radio spectra is too steep in comparison to observations.
Non-calorimeter one-zone models were subsequently proposed 
\citep{NiklasBeck1997,LackiEtAl2010a,SchleicherBeck2013,SchoberEtAl2016}, but they
have generally needed to be finetuned to maintain the
tightness of the IRRC.
For example, in the non-calorimeter model proposed by \citet{LackiEtAl2010a},
the efficient cooling of CR electrons decreases radio emission.
To maintain a tight IRRC, the decreasing radio emission is 
either balanced by the decreasing FIR emission in lower surface density
galaxies or by secondary CRs and the dependence of synchrotron frequency on energy in starbursts.
However, the predicted spectral slopes by \citet{LackiEtAl2010a}
are still in conflict with observations, even if such one-zone models
include a large number of free parameters. \\
To model more realistic galaxies and to better understand the 
underlying physics of the IRRC, 
\citet{VollmerEtAl2022}  developed a 1D model of
turbulent clumpy star-forming galactic disks, including a 
recipe for the nonthermal radio continuum emission.
They conclude that the observed SFR–radio correlation can be
reproduced by their fiducial model in a reasonable way (within
$\approx4\sigma$).
However, the mass range of the model galaxies in \citet{VollmerEtAl2022}
is not broad enough to explore a mass dependence of the IRRC. 
This is a central advantage of the model we present in this 
study and ultimately allows us to interpret the 
observations by \citet{DelvecchioEtAl2020}.

1D models are still significantly simplified
in comparison to real galaxies and therefore other authors have chosen
to simulate individual galaxies in detail, with the goal of identifying the 
origin of the IRRC. 
\citet{WerhahnEtAl2021} is one the first studies that used 3D
magneto-hydrodynamical (MHD) simulations of isolated galaxies 
including a self-consistent evolution of the CR proton energy density.
In post-processing, they calculate the steady-state
spectra of CR electrons to estimate the radio luminosity.
\citet{WerhahnEtAl2021} confirmed the findings of \citet{LackiEtAl2010a} and, 
in particular, revealed that the increasing bremsstrahlung and Coulomb losses at high gas surface densities is almost exactly counteracted by an increasing
contribution of secondary radio emission with increasing $\dot{M}_\star$.
In a companion study, \citet{PfrommerEtAl2022} showed that the 
global IRRC is indeed connected to the saturated stage of the small-scale
dynamo.
The approach via full 3D MHD-CR simulations 
exhibits fewer free parameters and thus can be
considered to be more predictive than one-zone models. 
However, such simulations are computationally 
expensive and are not suitable for exploring a large parameter 
space, unlike models such as the one presented in this work.

\subsection{Weaknesses of the semi-analytical approach}

With the model proposed in this paper, we can explore
the parameter space at low computational costs. 
However, the galaxies are modelled in a very idealized way.
They are characterized by
exponential density profiles, but lack 
spiral arms, bulges, and a multi-phase interstellar medium.
In the evolution of the galaxies, neither accretion nor merging
events are taken into account directly.
Full 3D simulations of galaxies as presented by \citet{WerhahnEtAl2021} and \citet{PfrommerEtAl2022}
are much more suitable for
exploring the role of detailed galaxy properties. 

Instead, we used observed scaling relations to model the global 
properties of the galaxies.
However, these might break down at extreme redshift.
This might be also the case for the assumed
scaling relation in Eq.~\ref{eq_LFIR}), 
which we used to estimate the IR luminosity; 
in particular, if the dust properties in the early Universe differ
from those of the local galaxies. 
We note that IR emission may  be modified not only with a change in the 
dust density, but also through temperature variations, 
changes in metallicity, and a change in sources for the ISM heating
as well as the details of radiative transfer physics. 
An expansion of our model by a description of the IR emission that goes beyond an observed scaling 
relation could therefore be very insightful.
On the other hand, such an extension of the model would imply a larger number of 
free parameters.

Another central ingredient of synchrotron emission is 
the cosmic ray model,
in which we assume steady-state and neglect diffusion and 
propagation effects.
Since the distribution of CRs is homogenous and 
follows the distribution of the gas 
and the SFR, face-on and edge-on views
result in similar radio luminosities.
This effect is enhanced by our assumption of 
isotropic magnetic fields that are not 
composed of small-scale and large-scale components.

\begin{figure}
\centering
\subfigure{\includegraphics[width=0.45\textwidth]{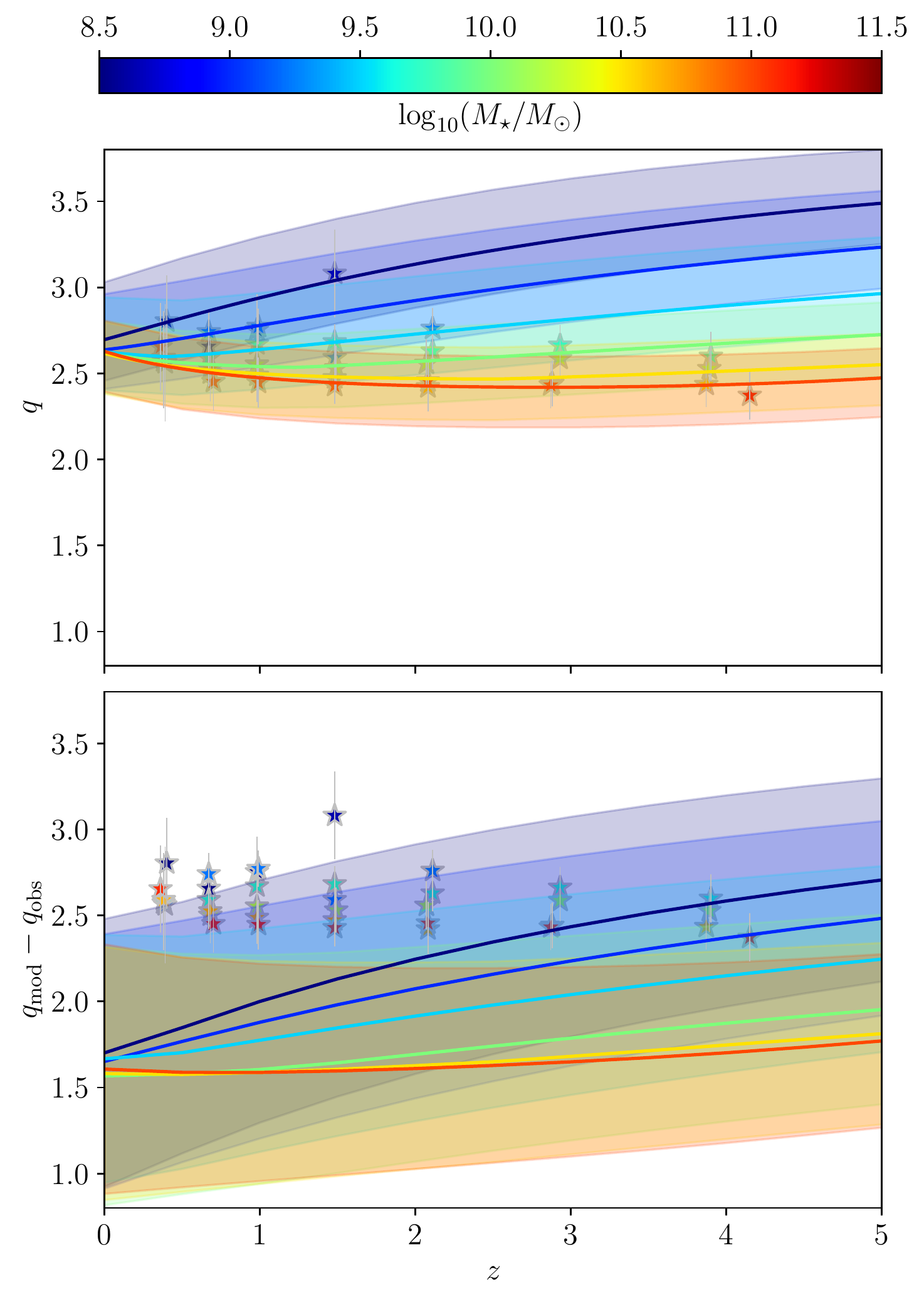}}
\caption{
Model predictions for galaxy populations in which the 
free parameters have a range of values.
\textit{Top:} 
Model predictions for a galaxy population, which on average
has the parameters of our reference model, but in which 
$f_\mathrm{turb}$, $f_\mathrm{CR}$,
$f_B$, $f_\pi$, $f_\mathrm{sec}$, $f_\mathrm{ion}$ and 
$f_\mathrm{sISRF}$
vary randomly with up to $\pm25\%$ scatter around the default values.
\textit{Bottom:} 
Model predictions for a galaxy population with parameters
that match the ones of
of the reference model, except for the value of 
$\alpha_\mathrm{CR}$.
Here, a galaxy population with 
$1.75 \leq \alpha_\mathrm{CR} \leq 3.25$ is considered which results in $0.4 \lesssim \alpha_{1.4} \lesssim 1.3$.
}
\label{fig_q_z_error_fchi}
\end{figure} 

Our galaxy model is based on a simplified model
of a star-forming galaxy that is described by a set of 
free parameters (see the yellow boxes in Fig.~\ref{fig_Model_anatomy2}).
A combination of free parameters that results in $q$ parameters
that are comparable to the observed values by \citet{DelvecchioEtAl2020} 
is listed in Table~\ref{table_model_parameters} and makes up our 
reference model. In
Figs.~\ref{fig_q_z__fB_fion}, ~\ref{fig_q_z__CR_sISRF}, and~\ref{fig_q_z__Hgal}, we show how the model curves change 
when the respective free parameters are changed individually.
We go on to consider a galaxy population in which the free parameters
vary.
In particular, we consider $f_\mathrm{turb}$, $f_\mathrm{CR}$,
$f_B$, $f_\pi$, $f_\mathrm{sec}$, $f_\mathrm{ion}$, and 
$f_\mathrm{sISRF}$, which have average values that match the one of 
the reference model but are assumed to vary up to $\pm 25~\%$ within a 
population of galaxies. 
The resulting error bands are presented in the top panel
of Fig.~\ref{fig_q_z_error_fchi}. 
The one-sigma error bands are overlapping, nearly washing out the 
dependence of $q$ on stellar mass $M_\star$. 
We note, however, that a covariance 
between some of the free parameters in our model can be expected and 
therefore, the error bands in Fig.~\ref{fig_q_z_error_fchi} 
are likely to be overestimated. 
Nevertheless, this analysis, in addition to the results presented in Figs.~\ref{fig_q_z__fB_fion}, ~\ref{fig_q_z__CR_sISRF}, and~\ref{fig_q_z__Hgal}, demonstrates that a fine-tuning of our 
model is required to describe the observation by \citet{DelvecchioEtAl2020}.

A final obstacle when comparing with the 
observational data from \citet{DelvecchioEtAl2020}, is
the fact that they used a constant spectral slope of $\alpha_{1.4}=0.75$ in their analysis. 
In our model, however, $\alpha_{1.4}$ is not an input parameter, 
but an output that mostly depends on the slope of injection
spectrum, $\alpha_\mathrm{CR}$, but also on the thermal to non-thermal 
ratio.
If we use our reference model, but lower the value of the 
$\alpha_\mathrm{CR}$ such that $\alpha_{1.4}\approx0.75$, the $q(z)$ 
curves decrease significantly (see the trend in Fig.~\ref{fig_q_z__CR_sISRF}).
The decrease in $q(z)$ with decreasing $\alpha_\mathrm{CR}$ 
is caused by an increase in the 
number of cosmic ray electrons with
energies $E_{e}^\mathrm{max}$ (see Eq.~\ref{eq_Emax})
that lead to synchrotron emission at $1.4~\mathrm{GHz}$.
If, in addition, a distribution of $\alpha_{1.4}$ is 
considered, an additional uncertainty arises. 
The strong dependence of the radio luminosity
on $\alpha_\mathrm{CR}$ can be seen in the top panel of Fig.~\ref{fig_Lradio_LFIR}, where $\alpha_\mathrm{CR}$ is
only varied between $2.8$ and $3.2$, implying $0.9 \lesssim \alpha_{1.4} \lesssim 1.1$,
but the resulting $L_{1.4}$ captures the entire 
observed scatter. 
Therefore, the input parameter $\alpha_\mathrm{CR}$, 
alongside the efficiency of the dynamo, $f_B$, and
the efficiency of cosmic ray production $f_\mathrm{CR}$, 
are the key parameters that determine the value of $q(z)$.
To approximately reproduce an observed scatter of 
$0.4 \lesssim \alpha_{1.4} \lesssim 1.3$ \citep[see e.g.,][]{TabatabaeiEtAl2017},
we then used our reference model with a galaxy population that
includes members with parameters $1.75 \leq \alpha_\mathrm{CR} \leq 3.25$.
The resulting uncertainty in the $q(z)$ 
curves is large, as can be seen in the lower panel of
Fig.~\ref{fig_q_z_error_fchi}, and the $q(z)$ curves are 
lower than the observed values.

\subsection{Potential extensions of the model}

There are several potential extensions of the galaxy model. 
For instance, a model for clumping of the density field could be implemented
or a redshift evolution of various galaxy properties, such as the
small-scale dynamo efficiency, the ionization degree, or the 
dust temperature. 
Furthermore, it would be possible to consider distributions of 
galactic radii and scale height at a given $z$ to obtain an estimate for 
errors of $q(z, M_\star)$. 
However, since our current idealized galaxy model is successful in 
explaining the observed IRRC, adding further free parameters 
to the model is not necessary at present.

\section{Conclusion}
\label{sec_conclusion}

In this study, we  modeled the radio and infrared 
emission of idealized galaxies with the aim of understanding the 
origin of the IRRC and, in particular, its evolution across cosmic history. 
While our semi-analytical galaxy model includes a list of free 
parameters, the focus lies on how these underlying model parameters 
vary with two physical galaxy properties (assuming that the model 
galaxies follow the scaling relations for main sequence galaxies): 
stellar mass, $M_\star$, and redshift, $z$.
Our approach is summarized as follows. 
In the first step, global aspects of the galaxies, namely,~the scale height, radius, gas mass, and star formation rate, are estimated from observational correlations. 
Then a 3D distribution of the interstellar gas, the SFR, as well as the 
magnetic field, the thermal and non-thermal electron population is constructed 
that allows us to calculate the galactic radio emission. 
Finally, absorption processes that depend 
on the line of sight of the observation are included. 

The model reproduces the observed dependence of the IRRC on $M_\star$ 
and $z$ by \citet{DelvecchioEtAl2020} well. 
The best working model requires a fraction of $5~\%$ of the SN
energy to be converted into turbulence 
and $10~\%$ of that being 
converted to magnetic energy by a turbulent dynamo. 
This results in a magnetic field strength in the core of the galaxy of 
$10^{0.5}-10^{1.5}~\mu\mathrm{G}$ at $z=0$ up to 
$10^{1.2}-10^{2.3}~\mu\mathrm{G}$ at $z=5$ for galaxies 
with $M_\star=10^{8.5}-10^{11}~M_\odot$, respectively. 
Then the observed trend of $q$, with values of $q\approx2.8$ for 
$M_\star=10^{8.5}~M_\odot$ and $q\approx2.4$ for $M_\star=10^{11}~M_\odot$ 
and vanishing dependence on $z$ can be reproduced,
if $\approx 4\%$ of the SN energy is converted into CR protons; see Fig.~\ref{fig_q_z}.

We used our model to explore the effect of
varying galaxy properties on the evolution of $q$ with $z$ systematically.
In particular, we find that increasing the ionization degree,
which implies an increase in the free-free emission, only affects galaxies with high stellar mass.
Minor modifications of the $q(z,M_\star)$ curves when changing the ionization degree between $f_\mathrm{ion}=0.05-0.2$ can only be seen for galaxies with $M_\star=10^{11}~\mathrm{M}_\odot$.
Increasing the magnetic field strength, 
by varying the efficiency of the dynamo with the parameters $f_B$ and $f_\mathrm{turb}$, shifts the $q(z,M_\star)$ curves to lower values while weakening the 
$M_\star$ dependence (see Fig.~\ref{fig_q_z__fB_fion}).
Naturally, the $q(z,M_\star)$ curves also depend on the free parameters of the cosmic ray model. 
More cosmic rays, namely,~larger values of $f_\mathrm{CR}$, shift the $q(z,M_\star)$ curves to lower values since the synchrotron luminosity increases (see Fig.~\ref{fig_q_z__CR_sISRF}).
The $q(z,M_\star)$ curves also move down when
the slope of the CR injection spectrum $\alpha_\mathrm{CR}$ is decreased. 
The latter effect is caused by the fact that 
the CR population consists of fewer electrons with the characteristic energy for synchrotron emission at $\nu=1.4~\mathrm{GHz}$ and the typical galactic magnetic field strengths when the injection spectrum is steeper.  

The semi-analytical galaxy model presented in this paper 
is a powerful tool for testing different scenarios of 
galaxy evolution and, in particular, for efficient 
exploration of the free parameter space. This is the 
central benefit of such a phenomenological model compared 
to high-resolution (magneto)hydrodynamical simulations. 
Comparison between the model predictions with currently 
available data sets allows us to constrain parameters of 
galaxy evolution and the dynamics of the magnetized ISM. 
For instance, it can be used to test turbulent dynamo 
theory across galaxy evolution. 
Finally, our model allows for the construction of a robust and
physically motivated radio-calibration of the SFR that can 
be employed for current and future ultra-deep radio 
surveys with the Square Kilometer Array and its 
pathfinders.


\begin{acknowledgements} 
We thank Ivan Delvecchio for providing the observational data set 
and for useful discussions. 
The detailed comments by 
the referee Aritra Basu are 
highly appreciated and have improved the manuscript significantly. 
JS~acknowledges the support by the Swiss National Science 
Foundation under Grant No.\ 185863. 
MTS acknowledges support from a Scientific Exchanges visitor 
fellowship (IZSEZO\_202357) from the Swiss National Science Foundation.
RSK acknowledges financial support from the European 
Research Council via the ERC Synergy Grant ``ECOGAL'' (project ID 855130), 
from the Deutsche Forschungsgemeinschaft (DFG) via the Collaborative 
Research Center ``The Milky Way System''  (SFB 881 -- funding ID 138713538 
-- subprojects A1, B1, B2 and B8), from the Heidelberg Cluster of 
Excellence (EXC 2181 - 390900948) ``STRUCTURES'', funded by the German 
Excellence Strategy, and from the German Ministry for Economic 
Affairs and Climate Action in project ``MAINN'' (funding ID 50OO2206). 
RSK also thanks for computing resources provided by the Ministry of 
Science, Research and the Arts (MWK) of the State of Baden-W\"{u}rttemberg 
through bwHPC and DFG through grant INST 35/1134-1 FUGG and for data 
storage at SDS@hd through grant INST 35/1314-1 FUGG.
DRGS gratefully acknowledges support by the ANID BASAL projects 
ACE210002 and FB210003, as well as via the Millenium Nucleus 
NCN19-058 (TITANs). 
DRGS thanks for funding via Fondecyt Regular (project code 1201280).
\end{acknowledgements}



\newpage
\appendix

\section{Radial profiles}
\label{app_RadProf}

\begin{figure*}
\centering
  \subfigure{\includegraphics[width=0.85\textwidth]{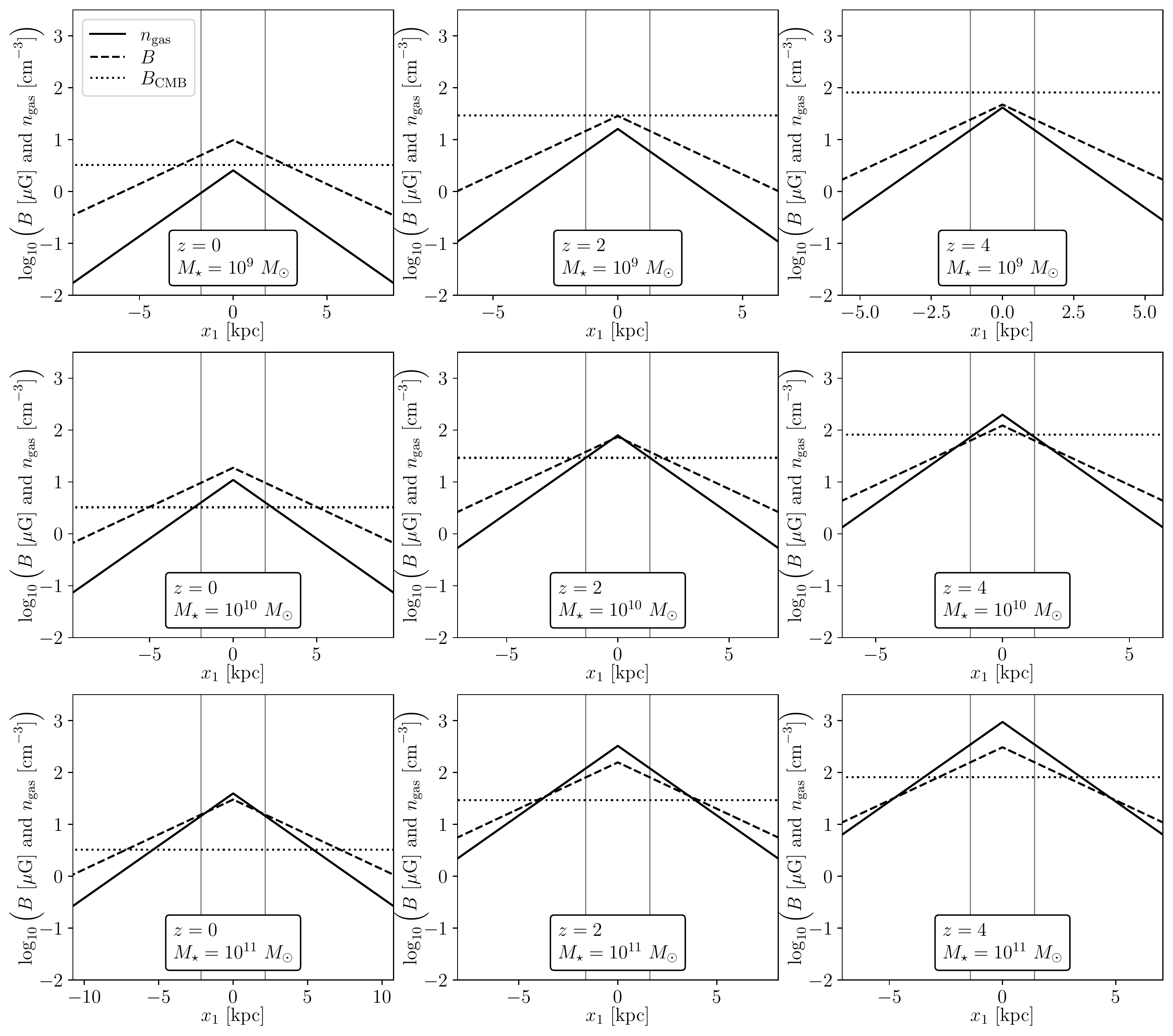}}
\caption{Radial profiles of the gas
density $n_\mathrm{gas}$ and the magnetic field strength
$B$.
For comparison the CMB equivalent field strength $B_\mathrm{CMB}$
is shown.
Grey vertical lines indicate the characteristic radius $R_\mathrm{gal}$,
i.e.,~they enclose $2R_\mathrm{gal}$.
}
\label{fig_various_x}
\end{figure*}

In Fig.~\ref{fig_various_x}, the radial profiles of the gas
density $n_\mathrm{gas}$ and the magnetic field strength
$B$ are presented.
The figure shows the result for the reference galaxy model.

We note that the radial profiles of the magnetic field
strength in our model show certain dissimilarities 
compared to observations of non-thermal radio
emission in star-forming galaxies. 
Observed profiles are naturally more complex than 
an exponential profile. 
For example, \citet{BeckEtAl2020} reported that the radial variations of radio intensities peaks of M51 between a radius of $8$ and $12$ kpc, 
but their scale lengths differ and are on the order of $3-4$ kpc,
depending on the fitting range.

For a better agreement with the observed profiles, our model
would need to be extended. 
First of all, the transport of cosmic ray electrons needs
to be implemented, which is challenging.
Also, an extension of the magnetic
field model would be needed, potentially in the form
of a dynamo efficiency factor, $f_B$, that depends on the 
distance from the center of the galaxy.

\section{Model for the interstellar radiation field}
\label{app_ISRF}

For the thermal interstellar radiation field, we 
employed a model that includes both a stellar contribution as well as
the cosmic microwave background (CMB).
To this end, we follow \citet{SchleicherBeck2013} and
scale the energy density of the ISRF with 
the SFR in a given galaxy:
\begin{eqnarray}
   u_\mathrm{ISRF}(M_\star,z) &=&  
   \frac{8~\pi^5 k_\mathrm{B}^4}{15~c^3 h^3}~T_\mathrm{CMB}(z)^4 \nonumber \\
   && + 10^{-12} \frac{\mathrm{erg}}{\mathrm{cm}^3}  \frac{\dot{M}_\star(M_\star,z) R_\mathrm{gal}(M_\star,z)^{-2}}{0.3~M_\odot~\mathrm{kpc^{-2}}\mathrm{yr}^{-1} }.
\label{eq_uISRF_Schleicher}
\end{eqnarray}
We note that we have made the assumption that the SFR surface density is given by $\dot{M}_\star(M_\star,z) R_\mathrm{gal}(M_\star,z)^{-2}$.
The CMB temperature evolves with redshift as  $T_\mathrm{CMB} = 2.73(1+z)~\mathrm{K}$.

To mimic a stronger radiation field in regions where the SFR is higher, 
we multiply equation \ref{eq_uISRF_Schleicher} by a dimensionless weighting factor :
\begin{eqnarray}
   u_\mathrm{ISRF}(M_\star,z, x_1,x_2,x_3) &=& 
     \frac{8~\pi^5 k^4}{15~c^3 h^3}~T_\mathrm{CMB}(z)^4 \nonumber \\
   && + 10^{-12} \frac{\mathrm{erg}}{\mathrm{cm}^3}  \frac{\dot{M}_\star(M_\star,z) R_\mathrm{gal}(M_\star,z)^{-2}}{\dot{\Sigma}_{\star,0} } \nonumber \\
   &&\times  \omega(x_1,x_2,x_3).
\end{eqnarray}
For the typical SFR surface density of a MW-like galaxy we
use $\dot{\Sigma}_{\star,0} = 0.3~M_\odot~\mathrm{kpc^{-2}}\mathrm{yr}^{-1}$.
The function $\omega(x_1,x_2,x_3)$ has the same spatial dependence as
the SFR density:
\begin{eqnarray}
  \omega(x_1,x_2,x_3) =  f_\mathrm{sISRF} \left[ \mathrm{exp}\left(-\frac{(x_1^2+x_2^2)^{1/2}}{R_\mathrm{gal}} \right) \mathrm{exp}\left(-\frac{x_3}{H_\mathrm{gal}} \right)\right]^{n_\mathrm{SFR}}
\label{eq_sISRF}
,\end{eqnarray}
and the default value of $f_\mathrm{sISRF}$ is chosen such that the $\omega(x_1,x_2,x_3)$ has a value of $1$
in the center of the galaxy, namely,~$f_\mathrm{sISRF}=1$.

\section{Dependence on resolution and box size}
\label{app_res}

\begin{figure}[t]
\centering
  \subfigure{\includegraphics[width=0.5\textwidth]{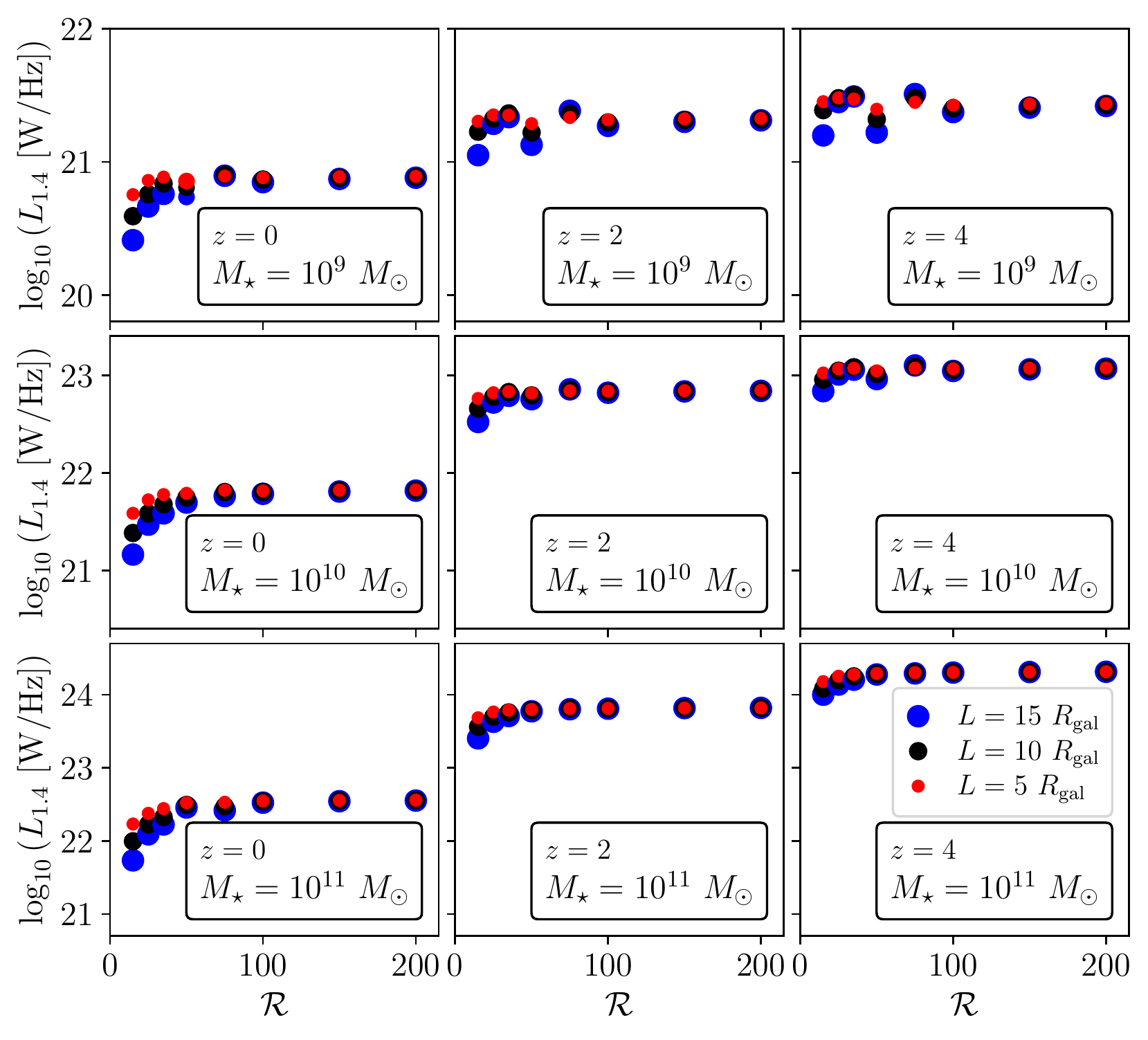}}
\caption{
Rest-frame 1.4 GHz luminosity $L_{1.4}$ for different box sizes
$L$ and resolution $\mathcal{R}$. 
Here, the default model (see Table~\ref{table_model_parameters}) of face-on galaxies 
is used and a convergence analysis at different 
redshift $z$ and stellar mass $M_\star$ is present as given in the labels.
}
\label{fig_convergence}
\end{figure}

In this section, we discuss the numerical convergence of 
our results and justify the default values of spatial 
resolution $\mathcal{R}$ and box size $L$.

The 3D model of the galaxy is set up on a grid and we
assume that the gas density has a double exponential profile (see Eq.~\ref{eq_rhogas}). 
The quantities relevant to the radio emission, such as the magnetic
field strength and the cosmic ray density, are coupled to the 
gas density and therefore also decrease exponentially 
with increasing distance from the center of the box. 
For results that are independent of the numerical model,
the value of $L$ needs to be at least larger than the 
effective radius and scale height, $R_\mathrm{gal}$ and $H_\mathrm{gal}$,
of the galaxy. 
There can be, however, significant contributions coming from radio emission at 
radii larger than $R_\mathrm{gal}$. 
The default value in our study is $L = 10 R_\mathrm{gal}$.

The result of the convergence study is presented in 
Fig.~\ref{fig_convergence}. 
There, we plot the value of $L_{1.4}$ for face-on galaxies
with different stellar masses and redshifts. 
The resolution is increased from $\mathcal{R}=15$ to 
$\mathcal{R}=200$. 
Different sizes of the numerical domain, 
namely,~$L = 5 R_\mathrm{gal}$, $L = 10 R_\mathrm{gal}$,
and $L = 15 R_\mathrm{gal}$, are presented by different 
colors. 
The small box sizes converge at the lowest resolution,
but generally converges can be expected for 
$\mathcal{R}\gtrsim100$.

\section{Nonlinearity of the IRRC}
\label{app_nonlin}

\begin{figure}[t]
\centering
  \subfigure{\includegraphics[width=0.45\textwidth]{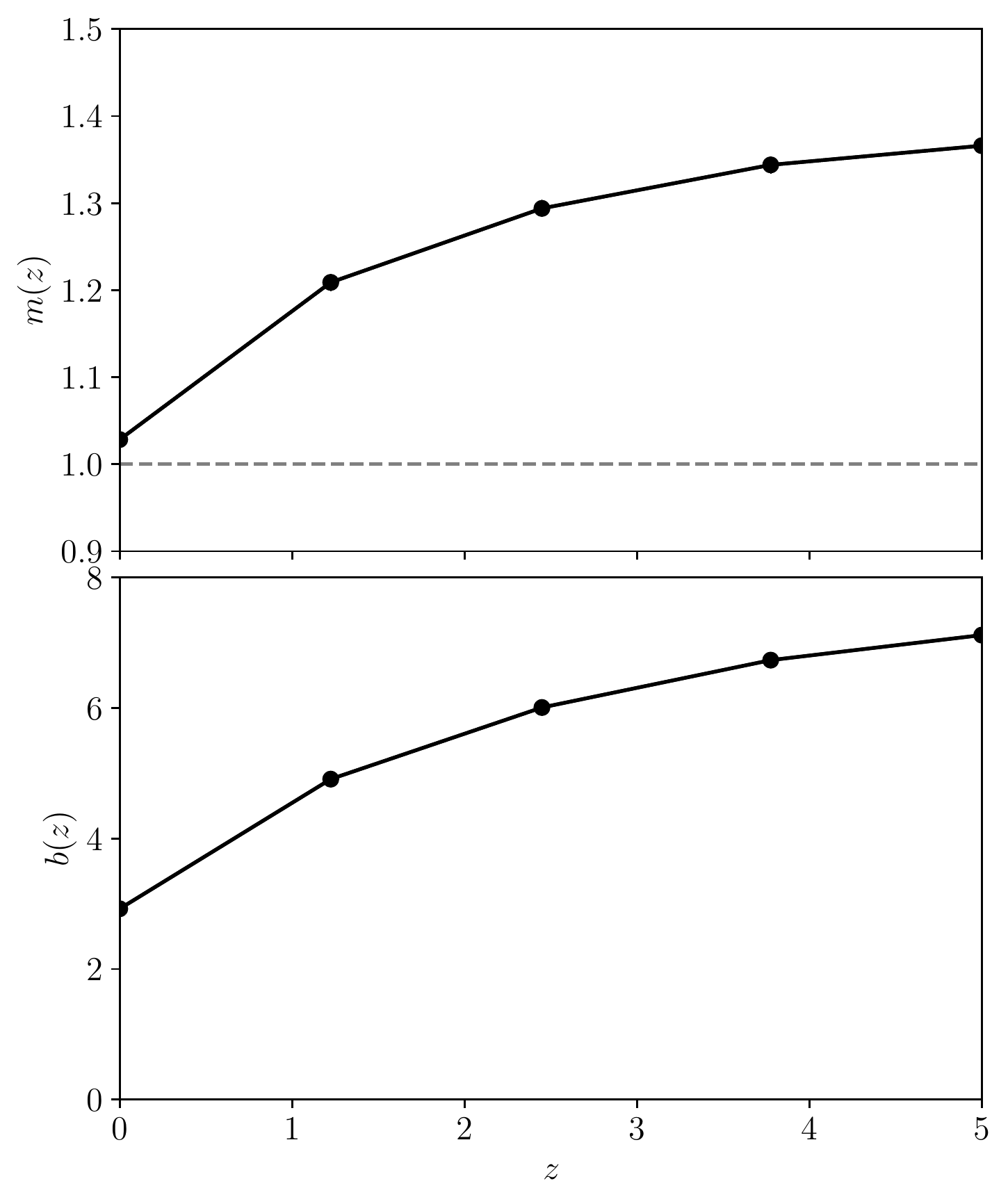}}
\caption{
Results for the fits in the IRRC obtained from
the reference model to the function~(\ref{eq_fit}).
The fitting parameters, $m$ and $b,$ are presented 
as a function of redshift $z$.}
\label{fig_fit}
\end{figure} 

The result for fitting the modeled IRRC at 
different redshifts
with the function given in Eq.~\ref{eq_fit} 
is presented in Fig.~\ref{fig_fit} 
for the reference model.
In agreement with the observational study of \citet{MolnarEtAl2021}, 
we find that the IRRC is slightly nonlinear at $z=0$.
The nonlinearity, that is,~the value of the slope $m$, 
increases slightly up to $z\approx1$ but decreases 
at $z\gtrsim1$. 
Strictly speaking, a linear IRRC is never obtained
from our reference model for $0\leq z \leq5$.

The fitting parameter $b$ is found to increase up to
 $z\approx2.5$ and decreases for higher $z$.
This parameter would correspond to $q$, in the case of
a linear IRRC, i.e.~for $m=1$.

\section{Comparison to $M_\star$-agnostic observations of the IRRC}
\label{sec_otherobs}
~\\
\begin{figure}[t]
\centering
  \includegraphics[width=0.5\textwidth]{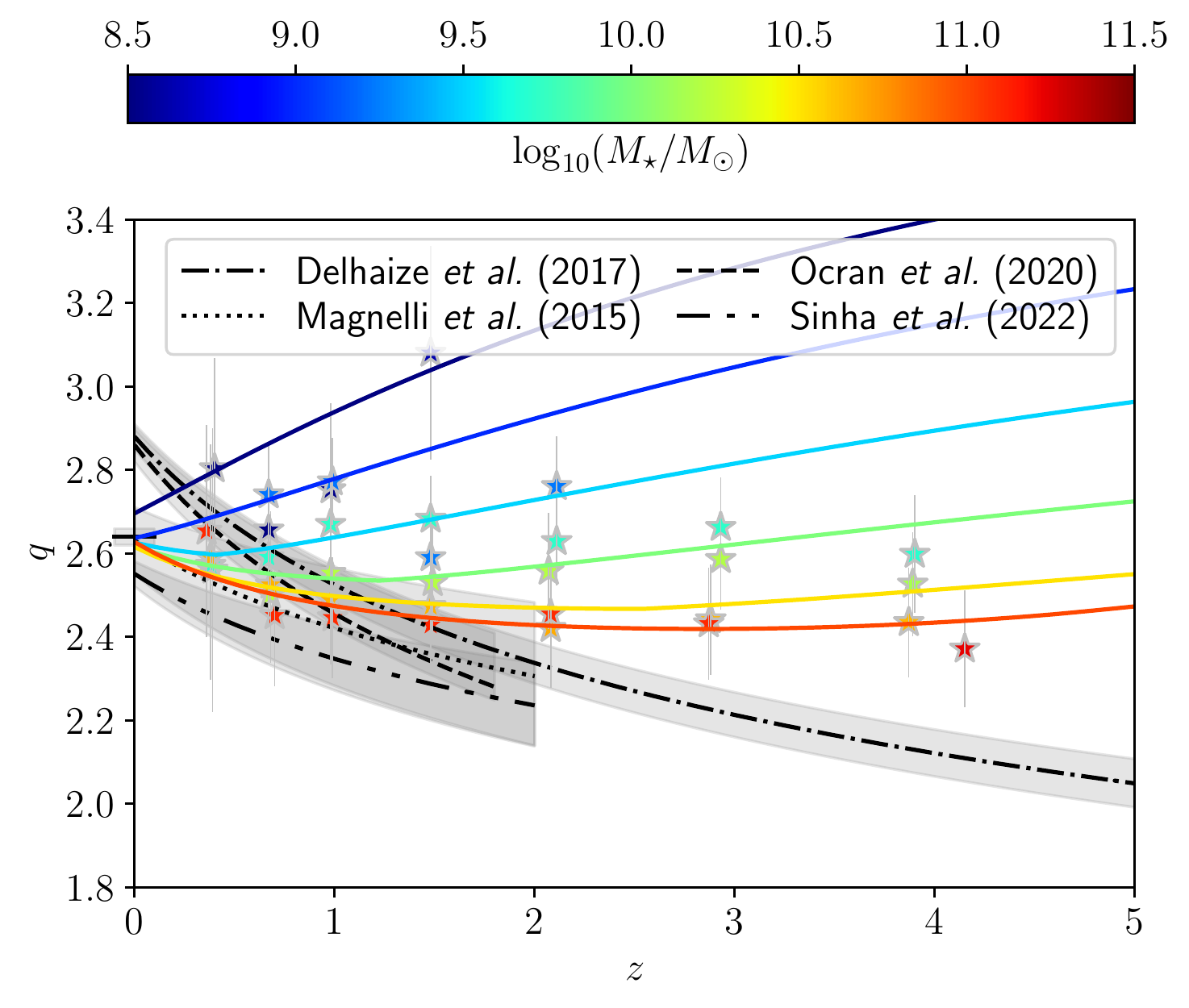}
\caption{
Comparison of the (face-on) reference model with other
observational results that do not include the dependence 
of $q$ on the stellar mass.
The lines of the observational relations are shown over the 
redshift ranges of the individual data sets. 
The short vertical line at the $q$ axis indicates the value of $q = 2.64 \pm 0.02$ obtained by \citet{Bell2003}
from observations of local star-forming galaxies.
}
\label{fig_q_otherobs}
\end{figure} 
~\\
The aim of this study is to understand the origin
of the dependence of the IRRC on both
redshift $z$ and stellar mass $M_\star$. 
To our knowledge, the first observational study on these different
dependencies was presented by \citet{DelvecchioEtAl2020}
and we compare our phenomenological model to their
data set.
However, an extensive amount of literature on the evolution of the IRRC on
$z$ exists and we present the comparison between other,
$M_\star$ agnostic,
observational studies with our model in this appendix. 

In Fig.~\ref{fig_q_otherobs}, our reference model is compared to
results from different studies in which the stellar mass dependence is
not explicitly considered
\citep{Bell2003,MagnelliEtAl2015,DelhaizeEtAl2017,OcranEtAl2020,SinhaEtAl2022}
as well as with the \citet{DelvecchioEtAl2020} data. 
At $z=0$, observational results presented in Fig.~\ref{fig_q_otherobs}
agree well with both the \citet{DelvecchioEtAl2020} data and
our reference model. 
For increasing $z$, the majority of observational studies predict a decrease in
$q$. 
This can be interpreted as an observational bias caused by 
the fact that galaxies with smaller $M_\star$ fall below
the sensitivity limit at lower $z$ compared to more 
massive galaxies.
This leads to an increase in the average $M_\star$ in a given 
galaxy sample with $z$ and, according to our model, to
a systematic decrease in the average $q$ value.


\begin{thebibliography}{110}
\expandafter\ifx\csname natexlab\endcsname\relax\def\natexlab#1{#1}\fi

\bibitem[{{Aharonian} {et~al.}(2010){Aharonian}, {Kelner}, \&
  {Prosekin}}]{AharonianEtAl2010}
{Aharonian}, F.~A., {Kelner}, S.~R., \& {Prosekin}, A.~Y. 2010, \prd, 82,
  043002

\bibitem[{{An} {et~al.}(2021){An}, {Vaccari}, {Smail}, {Jarvis}, {Whittam},
  {Hale}, {Jin}, {Collier}, {Daddi}, {Delhaize}, {Frank}, {Murphy}, {Prescott},
  {Sekhar}, {Taylor}, {Ao}, {Knowles}, {Marchetti}, {Randriamampandry}, \&
  {Randriamanakoto}}]{AnEtAl2021}
{An}, F., {Vaccari}, M., {Smail}, I., {et~al.} 2021, \mnras, 507, 2643

\bibitem[{{Appleton} {et~al.}(2004){Appleton}, {Fadda}, {Marleau}, {Frayer},
  {Helou}, {Condon}, {Choi}, {Yan}, {Lacy}, {Wilson}, {Armus}, {Chapman},
  {Fang}, {Heinrichson}, {Im}, {Jannuzi}, {Storrie-Lombardi}, {Shupe},
  {Soifer}, {Squires}, \& {Teplitz}}]{AppletonEtAl2004}
{Appleton}, P.~N., {Fadda}, D.~T., {Marleau}, F.~R., {et~al.} 2004, \apjs, 154,
  147

\bibitem[{{Basu} {et~al.}(2017){Basu}, {Roychowdhury}, {Heesen}, {Beck},
  {Brinks}, {Westcott}, \& {Hindson}}]{BasuEtAl2017}
{Basu}, A., {Roychowdhury}, S., {Heesen}, V., {et~al.} 2017, \mnras, 471, 337

\bibitem[{{Basu} {et~al.}(2015){Basu}, {Wadadekar}, {Beelen}, {Singh},
  {Archana}, {Sirothia}, \& {Ishwara-Chandra}}]{BasuEtAl2015}
{Basu}, A., {Wadadekar}, Y., {Beelen}, A., {et~al.} 2015, \apj, 803, 51

\bibitem[{{Beck} {et~al.}(2020){Beck}, {Berkhuijsen}, {Gie{\ss}{\"u}bel}, \&
  {Mulcahy}}]{BeckEtAl2020}
{Beck}, R., {Berkhuijsen}, E.~M., {Gie{\ss}{\"u}bel}, R., \& {Mulcahy}, D.~D.
  2020, \aap, 633, A5

\bibitem[{{Bell}(1978{\natexlab{a}})}]{Bell1978a}
{Bell}, A.~R. 1978{\natexlab{a}}, \mnras, 182, 147

\bibitem[{{Bell}(1978{\natexlab{b}})}]{Bell1978b}
{Bell}, A.~R. 1978{\natexlab{b}}, \mnras, 182, 443

\bibitem[{{Bell}(2003)}]{Bell2003}
{Bell}, E.~F. 2003, \apj, 586, 794

\bibitem[{{Beresnyak}(2012)}]{Beresnyak2012}
{Beresnyak}, A. 2012, \prl, 108, 035002

\bibitem[{{Bernhard} {et~al.}(2014){Bernhard}, {B{\'e}thermin}, {Sargent},
  {Buat}, {Mullaney}, {Pannella}, {Heinis}, \& {Daddi}}]{BernhardEtAl2014}
{Bernhard}, E., {B{\'e}thermin}, M., {Sargent}, M., {et~al.} 2014, \mnras, 442,
  509

\bibitem[{{Blandford} \& {Eichler}(1987)}]{BlandfordEichler1987}
{Blandford}, R. \& {Eichler}, D. 1987, \physrep, 154, 1

\bibitem[{{Blumenthal} \& {Gould}(1970)}]{BlumenthalGould1970}
{Blumenthal}, G.~R. \& {Gould}, R.~J. 1970, Reviews of Modern Physics, 42, 237

\bibitem[{{Bogdan} \& {V{\"o}lk}(1983)}]{BogdanVolk1983}
{Bogdan}, T.~J. \& {V{\"o}lk}, H.~J. 1983, \aap, 122, 129

\bibitem[{{Bonato} {et~al.}(2021){Bonato}, {Prandoni}, {De Zotti}, {Best},
  {Bondi}, {Calistro Rivera}, {Cochrane}, {G{\"u}rkan}, {Haskell},
  {Kondapally}, {Magliocchetti}, {Leslie}, {Malek}, {R{\"o}ttgering}, {Smith},
  {Tasse}, \& {Wang}}]{BonatoEtAl2021}
{Bonato}, M., {Prandoni}, I., {De Zotti}, G., {et~al.} 2021, \aap, 656, A48

\bibitem[{{Bourne} {et~al.}(2011){Bourne}, {Dunne}, {Ivison}, {Maddox},
  {Dickinson}, \& {Frayer}}]{BourneEtAl2011}
{Bourne}, N., {Dunne}, L., {Ivison}, R.~J., {et~al.} 2011, \mnras, 410, 1155

\bibitem[{{Brandenburg} \& {Subramanian}(2005)}]{BrandenburgSubramanian2005}
{Brandenburg}, A. \& {Subramanian}, K. 2005, \physrep, 417, 1

\bibitem[{{Burgarella} {et~al.}(2013){Burgarella}, {Buat}, {Gruppioni},
  {Cucciati}, {Heinis}, {Berta}, {B{\'e}thermin}, {Bock}, {Cooray}, {Dunlop},
  {Farrah}, {Franceschini}, {Le Floc'h}, {Lutz}, {Magnelli}, {Nordon},
  {Oliver}, {Page}, {Popesso}, {Pozzi}, {Riguccini}, {Vaccari}, \&
  {Viero}}]{BurgarellaEtAl2013}
{Burgarella}, D., {Buat}, V., {Gruppioni}, C., {et~al.} 2013, \aap, 554, A70

\bibitem[{{Burkhart} {et~al.}(2010){Burkhart}, {Stanimirovi{\'c}}, {Lazarian},
  \& {Kowal}}]{BurkhartEtAl2010}
{Burkhart}, B., {Stanimirovi{\'c}}, S., {Lazarian}, A., \& {Kowal}, G. 2010,
  \apj, 708, 1204

\bibitem[{Calistro~Rivera {et~al.}(2017)Calistro~Rivera, Williams, Hardcastle,
  Duncan, Röttgering, Best, Brüggen, Chyży, Conselice, de~Gasperin, Engels,
  Gürkan, Intema, Jarvis, Mahony, Miley, Morabito, Prandoni, Sabater, Smith,
  Tasse, van~der Werf, \& White}]{CalistroRiveraEtAl2017}
Calistro~Rivera, G., Williams, W.~L., Hardcastle, M.~J., {et~al.} 2017, Monthly
  Notices of the Royal Astronomical Society, 469, 3468

\bibitem[{{Caprioli} \& {Spitkovsky}(2014)}]{CaprioliSpitkovsky2014}
{Caprioli}, D. \& {Spitkovsky}, A. 2014, \apj, 783, 91

\bibitem[{{Chy{\.z}y} {et~al.}(2011){Chy{\.z}y}, {We{\.z}gowiec}, {Beck}, \&
  {Bomans}}]{ChyzyEtAl2011}
{Chy{\.z}y}, K.~T., {We{\.z}gowiec}, M., {Beck}, R., \& {Bomans}, D.~J. 2011,
  \aap, 529, A94

\bibitem[{{Condon}(1992)}]{Condon1992}
{Condon}, J.~J. 1992, \araa, 30, 575

\bibitem[{{Daddi} {et~al.}(2010){Daddi}, {Elbaz}, {Walter}, {Bournaud},
  {Salmi}, {Carilli}, {Dannerbauer}, {Dickinson}, {Monaco}, \&
  {Riechers}}]{DaddiEtAl2010}
{Daddi}, E., {Elbaz}, D., {Walter}, F., {et~al.} 2010, \apjl, 714, L118

\bibitem[{{de Jong} {et~al.}(1985){de Jong}, {Klein}, {Wielebinski}, \&
  {Wunderlich}}]{deJongEtAl1985}
{de Jong}, T., {Klein}, U., {Wielebinski}, R., \& {Wunderlich}, E. 1985, \aap,
  147, L6

\bibitem[{{Delhaize} {et~al.}(2017){Delhaize}, {Smol{\v{c}}i{\'c}},
  {Delvecchio}, {Novak}, {Sargent}, {Baran}, {Magnelli}, {Zamorani},
  {Schinnerer}, {Murphy}, {Aravena}, {Berta}, {Bondi}, {Capak}, {Carilli},
  {Ciliegi}, {Civano}, {Ilbert}, {Karim}, {Laigle}, {Le F{\`e}vre}, {Marchesi},
  {McCracken}, {Salvato}, {Seymour}, \& {Tasca}}]{DelhaizeEtAl2017}
{Delhaize}, J., {Smol{\v{c}}i{\'c}}, V., {Delvecchio}, I., {et~al.} 2017, \aap,
  602, A4

\bibitem[{{Delvecchio} {et~al.}(2021){Delvecchio}, {Daddi}, {Sargent},
  {Jarvis}, {Elbaz}, {Jin}, {Liu}, {Whittam}, {Algera}, {Carraro}, {D'Eugenio},
  {Delhaize}, {Kalita}, {Leslie}, {Moln{\'a}r}, {Novak}, {Prandoni},
  {Smol{\v{c}}i{\'c}}, {Ao}, {Aravena}, {Bournaud}, {Collier},
  {Randriamampandry}, {Randriamanakoto}, {Rodighiero}, {Schober}, {White}, \&
  {Zamorani}}]{DelvecchioEtAl2020}
{Delvecchio}, I., {Daddi}, E., {Sargent}, M.~T., {et~al.} 2021, \aap, 647, A123

\bibitem[{{Dey} {et~al.}(2022){Dey}, {Goyal}, {Ma{\l}ek}, {Galvin}, {Seymour},
  {D{\'\i}az Santos}, {Piotrowska}, \& {Charmandaris}}]{DeyEtAl2022}
{Dey}, S., {Goyal}, A., {Ma{\l}ek}, K., {et~al.} 2022, arXiv e-prints,
  arXiv:2207.08929

\bibitem[{{Draine}(2003)}]{Draine2003}
{Draine}, B.~T. 2003, \araa, 41, 241

\bibitem[{{Draine}(2011)}]{Draine2011}
{Draine}, B.~T. 2011, {Physics of the Interstellar and Intergalactic Medium}

\bibitem[{{Duncan} {et~al.}(2014){Duncan}, {Conselice}, {Mortlock}, {Hartley},
  {Guo}, {Ferguson}, {Dav{\'e}}, {Lu}, {Ownsworth}, {Ashby}, {Dekel},
  {Dickinson}, {Faber}, {Giavalisco}, {Grogin}, {Kocevski}, {Koekemoer},
  {Somerville}, \& {White}}]{DuncanEtaAl2014}
{Duncan}, K., {Conselice}, C.~J., {Mortlock}, A., {et~al.} 2014, \mnras, 444,
  2960

\bibitem[{Federrath {et~al.}(2011)Federrath, Chabrier, Schober, Banerjee,
  Klessen, \& Schleicher}]{FederrathEtAl2011b}
Federrath, C., Chabrier, G., Schober, J., {et~al.} 2011, \prl, 107, 114504

\bibitem[{{Federrath} {et~al.}(2014){Federrath}, {Schober}, {Bovino}, \&
  {Schleicher}}]{FederrathEtAl2014b}
{Federrath}, C., {Schober}, J., {Bovino}, S., \& {Schleicher}, D.~R.~G. 2014,
  \apjl, 797, L19

\bibitem[{{Ferreira} {et~al.}(2022){Ferreira}, {Adams}, {Conselice},
  {Sazonova}, {Austin}, {Caruana}, {Ferrari}, {Verma}, {Trussler},
  {Broadhurst}, {Diego}, {Frye}, {Pascale}, {Wilkins}, {Windhorst}, \&
  {Zitrin}}]{FerreiraEtAl2022}
{Ferreira}, L., {Adams}, N., {Conselice}, C.~J., {et~al.} 2022, \apjl, 938, L2

\bibitem[{{Gaensler} {et~al.}(2011){Gaensler}, {Haverkorn}, {Burkhart},
  {Newton-McGee}, {Ekers}, {Lazarian}, {McClure-Griffiths}, {Robishaw},
  {Dickey}, \& {Green}}]{GaenslerEtAl2011}
{Gaensler}, B.~M., {Haverkorn}, M., {Burkhart}, B., {et~al.} 2011, \nat, 478,
  214

\bibitem[{{Garrett}(2002)}]{Garrett2002}
{Garrett}, M.~A. 2002, \aap, 384, L19

\bibitem[{{Gent} {et~al.}(2021){Gent}, {Mac Low}, {K{\"a}pyl{\"a}}, \&
  {Singh}}]{GentEtAl2021}
{Gent}, F.~A., {Mac Low}, M.-M., {K{\"a}pyl{\"a}}, M.~J., \& {Singh}, N.~K.
  2021, \apjl, 910, L15

\bibitem[{{Graziani} {et~al.}(2020){Graziani}, {Schneider}, {Ginolfi}, {Hunt},
  {Maio}, {Glatzle}, \& {Ciardi}}]{GrazianiEtAl2020}
{Graziani}, L., {Schneider}, R., {Ginolfi}, M., {et~al.} 2020, \mnras, 494,
  1071

\bibitem[{{Heesen} {et~al.}(2022){Heesen}, {Staffehl}, {Basu}, {Beck}, {Stein},
  {Tabatabaei}, {Hardcastle}, {Chy{\.z}y}, {Shimwell}, {Adebahr}, {Beswick},
  {Bomans}, {Botteon}, {Brinks}, {Br{\"u}ggen}, {Dettmar}, {Drabent}, {de
  Gasperin}, {G{\"u}rkan}, {Heald}, {Horellou}, {Nikiel-Wroczynski},
  {Paladino}, {Piotrowska}, {R{\"o}ttgering}, {Smith}, \&
  {Tasse}}]{HeesenEtAl2022}
{Heesen}, V., {Staffehl}, M., {Basu}, A., {et~al.} 2022, \aap, 664, A83

\bibitem[{{Helou} {et~al.}(1985){Helou}, {Soifer}, \&
  {Rowan-Robinson}}]{HelouEtAl1985}
{Helou}, G., {Soifer}, B.~T., \& {Rowan-Robinson}, M. 1985, \apjl, 298, L7

\bibitem[{{Inami} {et~al.}(2022){Inami}, {Algera}, {Schouws}, {Sommovigo},
  {Bouwens}, {Smit}, {Stefanon}, {Bowler}, {Endsley}, {Ferrara}, {Oesch},
  {Stark}, {Aravena}, {Barrufet}, {da Cunha}, {Dayal}, {De Looze}, {Fudamoto},
  {Gonzalez}, {Graziani}, {Hodge}, {Hygate}, {Nanayakkara}, {Pallottini},
  {Riechers}, {Schneider}, {Topping}, \& {van der Werf}}]{InamiEtAl2022}
{Inami}, H., {Algera}, H. S.~B., {Schouws}, S., {et~al.} 2022, arXiv e-prints,
  arXiv:2203.15136

\bibitem[{{Ivison} {et~al.}(2010{\natexlab{a}}){Ivison}, {Alexander}, {Biggs},
  {Brandt}, {Chapin}, {Coppin}, {Devlin}, {Dickinson}, {Dunlop}, {Dye},
  {Eales}, {Frayer}, {Halpern}, {Hughes}, {Ibar}, {Kov{\'a}cs}, {Marsden},
  {Moncelsi}, {Netterfield}, {Pascale}, {Patanchon}, {Rafferty}, {Rex},
  {Schinnerer}, {Scott}, {Semisch}, {Smail}, {Swinbank}, {Truch}, {Tucker},
  {Viero}, {Walter}, {Wei{\ss}}, {Wiebe}, \& {Xue}}]{IvisonEtAl2010b}
{Ivison}, R.~J., {Alexander}, D.~M., {Biggs}, A.~D., {et~al.}
  2010{\natexlab{a}}, \mnras, 402, 245

\bibitem[{{Ivison} {et~al.}(2010{\natexlab{b}}){Ivison}, {Magnelli}, {Ibar},
  {Andreani}, {Elbaz}, {Altieri}, {Amblard}, {Arumugam}, {Auld}, {Aussel},
  {Babbedge}, {Berta}, {Blain}, {Bock}, {Bongiovanni}, {Boselli}, {Buat},
  {Burgarella}, {Castro-Rodr{\'\i}guez}, {Cava}, {Cepa}, {Chanial}, {Cimatti},
  {Cirasuolo}, {Clements}, {Conley}, {Conversi}, {Cooray}, {Daddi},
  {Dominguez}, {Dowell}, {Dwek}, {Eales}, {Farrah}, {F{\"o}rster Schreiber},
  {Fox}, {Franceschini}, {Gear}, {Genzel}, {Glenn}, {Griffin}, {Gruppioni},
  {Halpern}, {Hatziminaoglou}, {Isaak}, {Lagache}, {Levenson}, {Lu}, {Lutz},
  {Madden}, {Maffei}, {Magdis}, {Mainetti}, {Maiolino}, {Marchetti},
  {Morrison}, {Mortier}, {Nguyen}, {Nordon}, {O'Halloran}, {Oliver}, {Omont},
  {Owen}, {Page}, {Panuzzo}, {Papageorgiou}, {Pearson}, {P{\'e}rez-Fournon},
  {P{\'e}rez Garc{\'\i}a}, {Poglitsch}, {Pohlen}, {Popesso}, {Pozzi},
  {Rawlings}, {Raymond}, {Rigopoulou}, {Riguccini}, {Rizzo}, {Rodighiero},
  {Roseboom}, {Rowan-Robinson}, {Saintonge}, {Sanchez Portal}, {Santini},
  {Schulz}, {Scott}, {Seymour}, {Shao}, {Shupe}, {Smith}, {Stevens}, {Sturm},
  {Symeonidis}, {Tacconi}, {Trichas}, {Tugwell}, {Vaccari}, {Valtchanov},
  {Vieira}, {Vigroux}, {Wang}, {Ward}, {Wright}, {Xu}, \&
  {Zemcov}}]{IvisonEtAl2010a}
{Ivison}, R.~J., {Magnelli}, B., {Ibar}, E., {et~al.} 2010{\natexlab{b}}, \aap,
  518, L31

\bibitem[{{Jarvis} {et~al.}(2010){Jarvis}, {Smith}, {Bonfield}, {Hardcastle},
  {Falder}, {Stevens}, {Ivison}, {Auld}, {Baes}, {Baldry}, {Bamford}, {Bourne},
  {Buttiglione}, {Cava}, {Cooray}, {Dariush}, {de Zotti}, {Dunlop}, {Dunne},
  {Dye}, {Eales}, {Fritz}, {Hill}, {Hopwood}, {Hughes}, {Ibar}, {Jones},
  {Kelvin}, {Lawrence}, {Leeuw}, {Loveday}, {Maddox}, {Micha{\l}owski},
  {Negrello}, {Norberg}, {Pohlen}, {Prescott}, {Rigby}, {Robotham},
  {Rodighiero}, {Scott}, {Sharp}, {Temi}, {Thompson}, {van der Werf}, {van
  Kampen}, {Vlahakis}, \& {White}}]{JarvisEtAl2010}
{Jarvis}, M.~J., {Smith}, D.~J.~B., {Bonfield}, D.~G., {et~al.} 2010, \mnras,
  409, 92

\bibitem[{{Jim{\'e}nez-Andrade} {et~al.}(2019){Jim{\'e}nez-Andrade},
  {Magnelli}, {Karim}, {Zamorani}, {Bondi}, {Schinnerer}, {Sargent},
  {Romano-D{\'\i}az}, {Novak}, {Lang}, {Bertoldi}, {Vardoulaki}, {Toft},
  {Smol{\v{c}}i{\'c}}, {Harrington}, {Leslie}, {Delhaize}, {Liu}, {Karoumpis},
  {Kartaltepe}, \& {Koekemoer}}]{Jimenez-AndradeEtAl2019}
{Jim{\'e}nez-Andrade}, E.~F., {Magnelli}, B., {Karim}, A., {et~al.} 2019, \aap,
  625, A114

\bibitem[{{Jim{\'e}nez-Andrade} {et~al.}(2021){Jim{\'e}nez-Andrade}, {Murphy},
  {Heywood}, {Smail}, {Penner}, {Momjian}, {Dickinson}, {Armus}, \&
  {Lazio}}]{Jimenez-AndradeEtAl2021}
{Jim{\'e}nez-Andrade}, E.~F., {Murphy}, E.~J., {Heywood}, I., {et~al.} 2021,
  \apj, 910, 106

\bibitem[{{Kazantsev}(1968)}]{Kazantsev1968}
{Kazantsev}, A.~P. 1968, Soviet Journal of Experimental and Theoretical
  Physics, 26, 1031

\bibitem[{{Kennicutt} \& {Evans}(2012)}]{KennicuttEvans2012}
{Kennicutt}, R.~C. \& {Evans}, N.~J. 2012, \araa, 50, 531

\bibitem[{{Kennicutt}(1998)}]{Kennicutt1998}
{Kennicutt}, Jr., R.~C. 1998, \apj, 498, 541

\bibitem[{{Klein} {et~al.}(2018){Klein}, {Lisenfeld}, \&
  {Verley}}]{KleinEtAl2018}
{Klein}, U., {Lisenfeld}, U., \& {Verley}, S. 2018, \aap, 611, A55

\bibitem[{{Klessen} \& {Glover}(2016)}]{KlessenGlover2016}
{Klessen}, R.~S. \& {Glover}, S. C.~O. 2016, Saas-Fee Advanced Course, 43, 85

\bibitem[{{Koley} \& {Roy}(2019)}]{KoleyRoy2019}
{Koley}, A. \& {Roy}, N. 2019, \mnras, 483, 593

\bibitem[{{Kulsrud} \& {Anderson}(1992)}]{KulsrudAnderson1992}
{Kulsrud}, R.~M. \& {Anderson}, S.~W. 1992, \apj, 396, 606

\bibitem[{{Lacki} \& {Beck}(2013)}]{LackiBeck2013}
{Lacki}, B.~C. \& {Beck}, R. 2013, \mnras, 430, 3171

\bibitem[{{Lacki} {et~al.}(2010){Lacki}, {Thompson}, \&
  {Quataert}}]{LackiEtAl2010a}
{Lacki}, B.~C., {Thompson}, T.~A., \& {Quataert}, E. 2010, \apj, 717, 1

\bibitem[{{Li} {et~al.}(2016){Li}, {Beck}, {Dettmar}, {Heald}, {Irwin},
  {Johnson}, {Kepley}, {Krause}, {Murphy}, {Orlando}, {Rand}, {Strong},
  {Vargas}, {Walterbos}, {Wang}, \& {Wiegert}}]{LiEtAl2016}
{Li}, J.-T., {Beck}, R., {Dettmar}, R.-J., {et~al.} 2016, \mnras, 456, 1723

\bibitem[{{Licquia} \& {Newman}(2015)}]{LicquiaNewman2015}
{Licquia}, T.~C. \& {Newman}, J.~A. 2015, \apj, 806, 96

\bibitem[{{Liu} {et~al.}(2022){Liu}, {Kretschmer}, \& {Teyssier}}]{LiuEtAl2022}
{Liu}, Y., {Kretschmer}, M., \& {Teyssier}, R. 2022, \mnras, 513, 6028

\bibitem[{{Mac Low} \& {Klessen}(2004)}]{MacLowKlessen2004}
{Mac Low}, M.-M. \& {Klessen}, R.~S. 2004, Rev. Mod. Phys., 76, 125

\bibitem[{{Magnelli} {et~al.}(2015){Magnelli}, {Ivison}, {Lutz}, {Valtchanov},
  {Farrah}, {Berta}, {Bertoldi}, {Bock}, {Cooray}, {Ibar}, {Karim}, {Le
  Floc'h}, {Nordon}, {Oliver}, {Page}, {Popesso}, {Pozzi}, {Rigopoulou},
  {Riguccini}, {Rodighiero}, {Rosario}, {Roseboom}, {Wang}, \&
  {Wuyts}}]{MagnelliEtAl2015}
{Magnelli}, B., {Ivison}, R.~J., {Lutz}, D., {et~al.} 2015, \aap, 573, A45

\bibitem[{{Mancini} {et~al.}(2015){Mancini}, {Schneider}, {Graziani},
  {Valiante}, {Dayal}, {Maio}, {Ciardi}, \& {Hunt}}]{ManciniEtAl2015}
{Mancini}, M., {Schneider}, R., {Graziani}, L., {et~al.} 2015, \mnras, 451, L70

\bibitem[{{Marassi} {et~al.}(2019){Marassi}, {Schneider}, {Limongi}, {Chieffi},
  {Graziani}, \& {Bianchi}}]{MarassiEtAl2019}
{Marassi}, S., {Schneider}, R., {Limongi}, M., {et~al.} 2019, \mnras, 484, 2587

\bibitem[{{Martin-Alvarez} {et~al.}(2018){Martin-Alvarez}, {Devriendt}, {Slyz},
  \& {Teyssier}}]{Martin-AlvarezEtAl2018}
{Martin-Alvarez}, S., {Devriendt}, J., {Slyz}, A., \& {Teyssier}, R. 2018,
  \mnras, 479, 3343

\bibitem[{{Martin-Alvarez} {et~al.}(2021){Martin-Alvarez}, {Katz}, {Sijacki},
  {Devriendt}, \& {Slyz}}]{Martin-AlvarezEtAl2021}
{Martin-Alvarez}, S., {Katz}, H., {Sijacki}, D., {Devriendt}, J., \& {Slyz}, A.
  2021, \mnras, 504, 2517

\bibitem[{{McCheyne} {et~al.}(2022){McCheyne}, {Oliver}, {Sargent},
  {Kondapally}, {Smith}, {Haskell}, {Duncan}, {Best}, {Sabater}, {Bonato},
  {Calistro Rivera}, {Cochrane}, {Campos Varillas}, {Hurley}, {Leslie},
  {Ma{\l}ek}, {Magliocchetti}, {Prandoni}, {Read}, {Rottgering}, {Tasse},
  {Vaccari}, \& {Wang}}]{McCheyneEtAl2022}
{McCheyne}, I., {Oliver}, S., {Sargent}, M., {et~al.} 2022, \aap, 662, A100

\bibitem[{{Micha{\l}owski} {et~al.}(2010{\natexlab{a}}){Micha{\l}owski},
  {Hjorth}, \& {Watson}}]{MichalowskiEtAl2010b}
{Micha{\l}owski}, M., {Hjorth}, J., \& {Watson}, D. 2010{\natexlab{a}}, \aap,
  514, A67

\bibitem[{{Micha{\l}owski}(2015)}]{Michalowski2015}
{Micha{\l}owski}, M.~J. 2015, \aap, 577, A80

\bibitem[{{Micha{\l}owski} {et~al.}(2010{\natexlab{b}}){Micha{\l}owski},
  {Watson}, \& {Hjorth}}]{MichalowskiEtAl2010a}
{Micha{\l}owski}, M.~J., {Watson}, D., \& {Hjorth}, J. 2010{\natexlab{b}},
  \apj, 712, 942

\bibitem[{{Moln{\'a}r} {et~al.}(2021){Moln{\'a}r}, {Sargent}, {Leslie},
  {Magnelli}, {Schinnerer}, {Zamorani}, {Delhaize}, {Smol{\v{c}}i{\'c}},
  {Tisani{\'c}}, \& {Vardoulaki}}]{MolnarEtAl2021}
{Moln{\'a}r}, D.~C., {Sargent}, M.~T., {Leslie}, S., {et~al.} 2021, \mnras,
  504, 118

\bibitem[{{Mowla} {et~al.}(2019){Mowla}, {van Dokkum}, {Brammer}, {Momcheva},
  {van der Wel}, {Whitaker}, {Nelson}, {Bezanson}, {Muzzin}, {Franx},
  {MacKenty}, {Leja}, {Kriek}, \& {Marchesini}}]{MowlaEtAl2019}
{Mowla}, L.~A., {van Dokkum}, P., {Brammer}, G.~B., {et~al.} 2019, \apj, 880,
  57

\bibitem[{{Murphy} {et~al.}(2011){Murphy}, {Condon}, {Schinnerer}, {Kennicutt},
  {Calzetti}, {Armus}, {Helou}, {Turner}, {Aniano}, {Beir{\~a}o}, {Bolatto},
  {Brandl}, {Croxall}, {Dale}, {Donovan Meyer}, {Draine}, {Engelbracht},
  {Hunt}, {Hao}, {Koda}, {Roussel}, {Skibba}, \& {Smith}}]{MurphyEtAl2011}
{Murphy}, E.~J., {Condon}, J.~J., {Schinnerer}, E., {et~al.} 2011, \apj, 737,
  67

\bibitem[{Murphy {et~al.}(2006)Murphy, Helou, Braun, Kenney, Armus, Calzetti,
  Draine, R.~C.~Kennicutt, Roussel, Walter, Bendo, Buckalew, Dale, Engelbracht,
  Smith, \& Thornley}]{MurphyEtAl2006}
Murphy, E.~J., Helou, G., Braun, R., {et~al.} 2006, The Astrophysical Journal,
  651, L111

\bibitem[{{Niklas} \& {Beck}(1997)}]{NiklasBeck1997}
{Niklas}, S. \& {Beck}, R. 1997, \aap, 320, 54

\bibitem[{{Novikov} \& {Thorne}(1973)}]{NovikovThorne1973}
{Novikov}, I.~D. \& {Thorne}, K.~S. 1973, in Black Holes (Les Astres Occlus),
  343--450

\bibitem[{{Ocran} {et~al.}(2020){Ocran}, {Taylor}, {Vaccari},
  {Ishwara-Chandra}, {Prandoni}, {Prescott}, \& {Mancuso}}]{OcranEtAl2020}
{Ocran}, E.~F., {Taylor}, A.~R., {Vaccari}, M., {et~al.} 2020, \mnras, 491,
  5911

\bibitem[{{Pfrommer} {et~al.}(2022){Pfrommer}, {Werhahn}, {Pakmor},
  {Girichidis}, \& {Simpson}}]{PfrommerEtAl2022}
{Pfrommer}, C., {Werhahn}, M., {Pakmor}, R., {Girichidis}, P., \& {Simpson},
  C.~M. 2022, \mnras, 515, 4229

\bibitem[{{Prandoni} \& {Seymour}(2015)}]{PrandoniSeymour2015}
{Prandoni}, I. \& {Seymour}, N. 2015, in Advancing Astrophysics with the Square
  Kilometre Array (AASKA14), 67

\bibitem[{{Rieder} \& {Teyssier}(2016)}]{RiederTeyssier2016}
{Rieder}, M. \& {Teyssier}, R. 2016, \mnras, 457, 1722

\bibitem[{{Rieder} \& {Teyssier}(2017)}]{RiederTeyssier2017}
{Rieder}, M. \& {Teyssier}, R. 2017, \mnras, 471, 2674

\bibitem[{{Rogachevskii} \& {Kleeorin}(1997)}]{RogachevskiiKleeorin1997}
{Rogachevskii}, I. \& {Kleeorin}, N. 1997, \pre, 56, 417

\bibitem[{{Rybicki} \& {Lightman}(1986)}]{RybickiLightman1986}
{Rybicki}, G.~B. \& {Lightman}, A.~P. 1986, {Radiative Processes in
  Astrophysics}

\bibitem[{{Salmon} {et~al.}(2015){Salmon}, {Papovich}, {Finkelstein}, {Tilvi},
  {Finlator}, {Behroozi}, {Dahlen}, {Dav{\'e}}, {Dekel}, {Dickinson},
  {Ferguson}, {Giavalisco}, {Long}, {Lu}, {Mobasher}, {Reddy}, {Somerville}, \&
  {Wechsler}}]{SalmonEtAl2015}
{Salmon}, B., {Papovich}, C., {Finkelstein}, S.~L., {et~al.} 2015, \apj, 799,
  183

\bibitem[{{Sargent} {et~al.}(2012){Sargent}, {B{\'e}thermin}, {Daddi}, \&
  {Elbaz}}]{SargentEtAl2012}
{Sargent}, M.~T., {B{\'e}thermin}, M., {Daddi}, E., \& {Elbaz}, D. 2012, \apjl,
  747, L31

\bibitem[{{Sargent} {et~al.}(2014){Sargent}, {Daddi}, {B{\'e}thermin},
  {Aussel}, {Magdis}, {Hwang}, {Juneau}, {Elbaz}, \& {da
  Cunha}}]{SargentEtAl2014}
{Sargent}, M.~T., {Daddi}, E., {B{\'e}thermin}, M., {et~al.} 2014, \apj, 793,
  19

\bibitem[{{Sargent} {et~al.}(2010){Sargent}, {Schinnerer}, {Murphy}, {Carilli},
  {Helou}, {Aussel}, {Le Floc'h}, {Frayer}, {Ilbert}, {Oesch}, {Salvato},
  {Smol{\v c}i{\'c}}, {Kartaltepe}, \& {Sanders}}]{SargentEtAl2010}
{Sargent}, M.~T., {Schinnerer}, E., {Murphy}, E., {et~al.} 2010, \apjl, 714,
  L190

\bibitem[{{Schaye} \& {Dalla Vecchia}(2008)}]{SchayeDallaVecchia2008}
{Schaye}, J. \& {Dalla Vecchia}, C. 2008, \mnras, 383, 1210

\bibitem[{{Schlei\-cher} {et~al.}(2013){Schlei\-cher}, {Schober}, {Federrath},
  {Bovino}, \& {Schmidt}}]{SchleicherEtAl2013}
{Schlei\-cher}, D.~R.~G., {Schober}, J., {Federrath}, C., {Bovino}, S., \&
  {Schmidt}, W. 2013, \njop, 15, 023017

\bibitem[{{Schleicher} \& {Beck}(2013)}]{SchleicherBeck2013}
{Schleicher}, D.~R.~G. \& {Beck}, R. 2013, \aap, 556, A142

\bibitem[{{Schleicher} \& {Beck}(2016)}]{SchleicherBeck2016}
{Schleicher}, D. R.~G. \& {Beck}, R. 2016, \aap, 593, A77

\bibitem[{{Schlickeiser}(2002)}]{Schlickeiser2002}
{Schlickeiser}, R. 2002, {Cosmic Ray Astrophysics}

\bibitem[{{Schober} {et~al.}(2012{\natexlab{a}}){Schober}, {Schleicher},
  {Bovino}, \& {Klessen}}]{SchoberEtAl2012.3}
{Schober}, J., {Schleicher}, D., {Bovino}, S., \& {Klessen}, R.~S.
  2012{\natexlab{a}}, \pre, 86, 066412

\bibitem[{{Schober} {et~al.}(2012{\natexlab{b}}){Schober}, {Schleicher},
  {Federrath}, {Klessen}, \& {Banerjee}}]{SchoberEtAl2012.1}
{Schober}, J., {Schleicher}, D., {Federrath}, C., {Klessen}, R., \& {Banerjee},
  R. 2012{\natexlab{b}}, \pre, 85, 026303

\bibitem[{{Schober} {et~al.}(2013){Schober}, {Schleicher}, \&
  {Klessen}}]{SchoberEtAl2013}
{Schober}, J., {Schleicher}, D.~R.~G., \& {Klessen}, R.~S. 2013, \aap, 560, A87

\bibitem[{{Schober} {et~al.}(2016){Schober}, {Schleicher}, \&
  {Klessen}}]{SchoberEtAl2016}
{Schober}, J., {Schleicher}, D.~R.~G., \& {Klessen}, R.~S. 2016, \apj, 827, 109

\bibitem[{Schober {et~al.}(2017)Schober, Schleicher, \&
  Klessen}]{SchoberEtAl2017}
Schober, J., Schleicher, D. R.~G., \& Klessen, R.~S. 2017, Monthly Notices of
  the Royal Astronomical Society, 468, 946

\bibitem[{{Schreiber} {et~al.}(2015){Schreiber}, {Pannella, M.}, {Elbaz, D.},
  {B\'ethermin, M.}, {Inami, H.}, {Dickinson, M.}, {Magnelli, B.}, {Wang, T.},
  {Aussel, H.}, {Daddi, E.}, {Juneau, S.}, {Shu, X.}, {Sargent, M. T.}, {Buat,
  V.}, {Faber, S. M.}, {Ferguson, H. C.}, {Giavalisco, M.}, {Koekemoer, A. M.},
  {Magdis, G.}, {Morrison, G. E.}, {Papovich, C.}, {Santini, P.}, \& {Scott,
  D.}}]{SchreiberEtAl2015}
{Schreiber}, C., {Pannella, M.}, {Elbaz, D.}, {et~al.} 2015, A\&A, 575, A74

\bibitem[{Seta \& Federrath(2021)}]{SetaFederrath2021}
Seta, A. \& Federrath, C. 2021, Phys. Rev. Fluids, 6, 103701

\bibitem[{{Seymour} {et~al.}(2009){Seymour}, {Huynh}, {Dwelly}, {Symeonidis},
  {Hopkins}, {McHardy}, {Page}, \& {Rieke}}]{SeymourEtAl2009}
{Seymour}, N., {Huynh}, M., {Dwelly}, T., {et~al.} 2009, \mnras, 398, 1573

\bibitem[{{Sinha} {et~al.}(2022){Sinha}, {Basu}, {Datta}, \&
  {Chakraborty}}]{SinhaEtAl2022}
{Sinha}, A., {Basu}, A., {Datta}, A., \& {Chakraborty}, A. 2022, \mnras, 514,
  4343

\bibitem[{{Smith} {et~al.}(2021){Smith}, {Haskell}, {G{\"u}rkan}, {Best},
  {Hardcastle}, {Kondapally}, {Williams}, {Duncan}, {Cochrane}, {McCheyne},
  {R{\"o}ttgering}, {Sabater}, {Shimwell}, {Tasse}, {Bonato}, {Bondi},
  {Jarvis}, {Leslie}, {Prandoni}, \& {Wang}}]{SmithEtAl2021}
{Smith}, D.~J.~B., {Haskell}, P., {G{\"u}rkan}, G., {et~al.} 2021, \aap, 648,
  A6

\bibitem[{{Smith} {et~al.}(2014){Smith}, {Jarvis}, {Hardcastle}, {Vaccari},
  {Bourne}, {Dunne}, {Ibar}, {Maddox}, {Prescott}, {Vlahakis}, {Eales},
  {Maddox}, {Smith}, {Valiante}, \& {de Zotti}}]{SmithEtAl2014}
{Smith}, D.~J.~B., {Jarvis}, M.~J., {Hardcastle}, M.~J., {et~al.} 2014, \mnras,
  445, 2232

\bibitem[{{Sommovigo} {et~al.}(2022){Sommovigo}, {Ferrara}, {Pallottini},
  {Dayal}, {Bouwens}, {Smit}, {da Cunha}, {De Looze}, {Bowler}, {Hodge},
  {Inami}, {Oesch}, {Endsley}, {Gonzalez}, {Schouws}, {Stark}, {Stefanon},
  {Aravena}, {Graziani}, {Riechers}, {Schneider}, {van der Werf}, {Algera},
  {Barrufet}, {Fudamoto}, {Hygate}, {Labb{\'e}}, {Li}, {Nanayakkara}, \&
  {Topping}}]{SommovigoEtAl2022}
{Sommovigo}, L., {Ferrara}, A., {Pallottini}, A., {et~al.} 2022, \mnras, 513,
  3122

\bibitem[{{Tabatabaei} {et~al.}(2017){Tabatabaei}, {Schinnerer}, {Krause},
  {Dumas}, {Meidt}, {Damas-Segovia}, {Beck}, {Murphy}, {Mulcahy}, {Groves},
  {Bolatto}, {Dale}, {Galametz}, {Sandstrom}, {Boquien}, {Calzetti},
  {Kennicutt}, {Hunt}, {De Looze}, \& {Pellegrini}}]{TabatabaeiEtAl2017}
{Tabatabaei}, F.~S., {Schinnerer}, E., {Krause}, M., {et~al.} 2017, \apj, 836,
  185

\bibitem[{{Thompson} {et~al.}(2005){Thompson}, {Quataert}, \&
  {Murray}}]{ThompsonEtAl2005}
{Thompson}, T.~A., {Quataert}, E., \& {Murray}, N. 2005, \apj, 630, 167

\bibitem[{{van der Kruit}(1971)}]{VanDerKruit1971}
{van der Kruit}, P.~C. 1971, \aap, 15, 110

\bibitem[{{van der Kruit}(1973)}]{VanDerKruit1973}
{van der Kruit}, P.~C. 1973, \aap, 29, 263

\bibitem[{{Voelk}(1989)}]{Voelk1989}
{Voelk}, H.~J. 1989, \aap, 218, 67

\bibitem[{{Vollmer} {et~al.}(2022){Vollmer}, {Soida}, \&
  {Dallant}}]{VollmerEtAl2022}
{Vollmer}, B., {Soida}, M., \& {Dallant}, J. 2022, arXiv e-prints,
  arXiv:2207.06173

\bibitem[{{Werhahn} {et~al.}(2021){Werhahn}, {Pfrommer}, \&
  {Girichidis}}]{WerhahnEtAl2021}
{Werhahn}, M., {Pfrommer}, C., \& {Girichidis}, P. 2021, \mnras, 508, 4072

\bibitem[{{Yun} {et~al.}(2001){Yun}, {Reddy}, \& {Condon}}]{YunEtAl2001}
{Yun}, M.~S., {Reddy}, N.~A., \& {Condon}, J.~J. 2001, \apj, 554, 803

\end{thebibliography}
\end{document}